\documentclass[
 reprint, 
 amsmath, 
 aps, 
 pra, 
 showkeys, 
 superscriptaddress 
]{revtex4-2}

\usepackage{hyperref} 
\usepackage{graphicx} 
\usepackage{orcidlink} 
\usepackage[caption=false]{subfig} 
\usepackage{array} 
\usepackage{multirow} 
\usepackage{enumitem} 
\usepackage[ruled, linesnumbered]{algorithm2e} 
\usepackage{braket} 
\usepackage{amsthm} 
\usepackage{setspace} 

\hypersetup{
    colorlinks,
    linkcolor=blue,
    citecolor=blue,
    urlcolor=blue
} 

\makeatletter
\def\switch@array{\let\@array@sw\@array@sw@LaTeX} 
\makeatother

\newcolumntype{C}[1]{>{\centering\arraybackslash}p{#1}} 

\setlist{topsep=5pt, itemsep=5pt, parsep=0pt, wide=0pt, listparindent=\parindent} 
\setlist[enumerate]{label=\arabic*)} 

\renewcommand{\thetable}{\arabic{table}} 

\providecommand{\proj}[1]{\ket{#1}\!\bra{#1}}
\providecommand{\tu}{\tilde{u}}
\providecommand{\tf}{\tilde{f}}
\providecommand{\td}{\tilde{d}}
\providecommand{\Hsym}{H_{\mathrm{sym}}}
\providecommand{\Msym}{M_N^{\mathrm{sym}}}
\providecommand{\Hneg}{H_{\mathrm{neg}}}
\providecommand{\Mneg}{M_N^{\mathrm{neg}}}
\newtheorem{theorem}{Theorem}
\newtheorem{claim}{Claim}[theorem]
\newtheorem{subclaim}{Claim}[claim]

\begin{document}

\title{Quantum Circuits for Exact Preparation of Colored Motzkin and Fredkin States}

\author{Igor Gaidai}
\affiliation{Department of Physics and Astronomy, University of Tennessee at Chattanooga, Chattanooga, TN 37403, USA}
\affiliation{UTC Quantum Center, University of Tennessee at Chattanooga, Chattanooga, TN 37403, USA}

\author{Kaustav Mukherjee}
\affiliation{Department of Physics and Astronomy, University of Tennessee at Chattanooga, Chattanooga, TN 37403, USA}
\affiliation{UTC Quantum Center, University of Tennessee at Chattanooga, Chattanooga, TN 37403, USA}

\author{Tomasz Szo\l{}dra}
\affiliation{Zentrum für Optische Quantentechnologien, Universität Hamburg, Luruper Chaussee 149, 22761 Hamburg, Germany}

\author{Hatem Barghathi}
\affiliation{Department of Physics and Astronomy, University of Tennessee, Knoxville, Tennessee 37996, USA}
\affiliation{Institute for Advanced Materials and Manufacturing, University of Tennessee, Knoxville, Tennessee 37996, USA}

\author{Peter Schmelcher}
\affiliation{Zentrum für Optische Quantentechnologien, Universität Hamburg, Luruper Chaussee 149, 22761 Hamburg, Germany}

\author{Adrian Del Maestro}
\affiliation{Department of Physics and Astronomy, University of Tennessee, Knoxville, Tennessee 37996, USA}
\affiliation{Institute for Advanced Materials and Manufacturing, University of Tennessee, Knoxville, Tennessee 37996, USA}
\affiliation{Min H. Kao Department of Electrical Engineering and Computer Science, University of Tennessee, Knoxville, Tennessee 37996, USA}

\author{Rick Mukherjee}
\affiliation{Department of Physics and Astronomy, University of Tennessee at Chattanooga, Chattanooga, TN 37403, USA}
\affiliation{UTC Quantum Center, University of Tennessee at Chattanooga, Chattanooga, TN 37403, USA}

\begin{abstract}
    Motzkin and Fredkin states are ground states of local, frustration-free spin chains, yet they violate the area law of entanglement: their half-chain entanglement entropy grows as $\log N$, $\sqrt{N}$ or $N$ with the number of sites $N$ for the uncolored, colored and area-weighted colored variants.
    This entanglement places the dynamics of the colored variants beyond the reach of classical tensor-network simulations.
    In this work, we construct unitary quantum circuits that prepare these states exactly and deterministically, without post-selection or mid-circuit measurements, using either a logarithmic ancilla register or no ancillas at all.
    Their CX gate count and depth grow as $O(N^2)$, with leading-order prefactors that we derive analytically.
    Interestingly, these states have been connected to the Haldane phase and the AdS/CFT correspondence.
    Our circuits can be used to explore these connections on quantum computers.
\end{abstract}

\maketitle

\section{Introduction}

Ground states of gapped local Hamiltonians in one dimension obey an area law of entanglement~\cite{hastings2007area}, and gapless ones typically violate it only logarithmically~\cite{eisert2010colloquium}.
However, certain model Hamiltonians violate it much more strongly, yielding ground states whose entanglement entropy grows polynomially with system size.
The Motzkin and Fredkin spin chains are prominent examples of such constructions, exhibiting entanglement entropy that grows as $\log N$, $\sqrt{N}$, and $N$ (for $t > 1$), for uncolored, colored, and area-weighted colored states with $N$ spins, respectively, despite being frustration-free and local one-dimensional systems~\cite{bravyi2012criticality, movassagh2016supercritical, zhang2017novel, salberger2017deformed, salberger2018fredkin}.
Beyond their entanglement scaling, these states have been connected to a variety of broader physical phenomena, including symmetry-protected topological order such as the Haldane phase \cite{barbiero2017haldane}, connections to the AdS/CFT correspondence \cite{alexander2019exact, alexander2021exact}, and long-distance entanglement relevant to quantum communication and repeater-type resource applications \cite{dell2019long}.

The same entanglement growth that makes the Motzkin and Fredkin states interesting also makes their colored variants challenging to simulate using classical methods such as matrix product states (MPS)~\cite{schuch2008entropy} and more general tensor-network approaches~\cite{orus2014practical}.
Exact representations of these states have been constructed with tensor networks~\cite{alexander2019exact, alexander2021exact} and neural networks~\cite{zha_exact_2026}, but they describe only the states themselves and not the dynamics that follows when these states are perturbed.

Motzkin and Fredkin states therefore belong to the class of problems that motivate quantum simulation~\cite{feynman1982simulating, lloyd1996universal, georgescu2014quantum, altman2021quantum, daley2022practical}, which aims to reproduce quantum phenomena that are intractable for classical computers.
Over the past decade, quantum simulation has emerged as a major frontier in quantum science, with applications spanning a broad range of physical settings, including condensed-matter systems~\cite{scholl2021quantum, mukherjee2024excitons, zhang2017observation, kim2023evidence}, high-energy physics Hamiltonians~\cite{vovrosh2022dynamical, martinez2016real, wang2024quantum, klco2018quantum}, exotic quantum phases and ground states~\cite{zeybek2025formation, satzinger2021realizing, zeybek2024bond, iqbal2024non}, frustrated spin models~\cite{glaetzle2015designing, tian2025engineering, kim2010quantum, king2018observation}, and disordered spin systems~\cite{signoles2021glassy, mukherjee2024influence, smith2016many, mukherjee2026tuning, roushan2017spectroscopic}.
A diverse range of experimental platforms, including trapped ions~\cite{blatt2012quantum, monroe2021programmable}, Rydberg atom arrays~\cite{browaeys2020many}, superconducting circuits~\cite{kjaergaard2020superconducting}, and photonic processors, are now being developed for both analog and digital quantum simulation~\cite{moses2023race, wright2019benchmarking, kim2023evidence, google2025quantum, reagor2018demonstration, henriet2020quantum, bluvstein2024logical, norcia2023midcircuit, reichardt2024logical, psiquantum2025manufacturable, madsen2022quantum}, as well as for other quantum algorithms, such as variational quantum algorithms~\cite{cerezo2021variational, gaidai2026benchmarking}, including the quantum approximate optimization algorithm~\cite{farhi2014quantum, wilkie2024quantum, zhou2020quantum}.

Before we can study the properties of Motzkin and Fredkin states, we need to understand how to prepare them on quantum devices.
Reference~\cite{mukherjee2026quantum} recently proposed a Rydberg-atom-based protocol that uses spatially varying light shifts and Rydberg interactions to prepare uncolored Motzkin states.
However, the states are prepared with a fidelity that decreases with system size, and since the protocol is specified in terms of particular physical controls, it is challenging to extend it to devices where these controls may not be available.

Another protocol based on quantum push-down automata~\cite{gopalakrishnan2025pushdown} uses the radiation emitted by an open quantum system to generate superpositions of uncolored and 2-colored Motzkin paths, but with unequal weights.
Additionally, the protocol relies on post-selection on the final state of the emitter, so it is non-unitary as an overall state preparation protocol.

References~\cite{feldmeier_digital_2026, dorstel2026frustration} proposed protocols that prepare ground states of arbitrary frustration-free Hamiltonians (a class that includes all Motzkin and Fredkin state variants considered here) using repeated rounds of mid-circuit measurements and feedback.
However, in these protocols the average number of rounds needed to reach a target fidelity is bounded below by the inverse of the spectral gap, which scales at least as $N^{8/3}$ for the uncolored Motzkin and Fredkin chains~\cite{chen2017gapless}, and exponentially so for the $t > 1$ area-weighted variants~\cite{levine2017gap, zhang2017entropy, andrei2022spin}.

In this work, we overcome these limitations by developing quantum circuits that deterministically prepare the exact uncolored, colored and area-weighted Motzkin and Fredkin states, as well as other similar states.
Compared to some of the other approaches considered above, our construction has the following advantages:
\begin{enumerate}
    \item It is efficient: its circuit depth and CX gate count grow as $O(N^2)$ for all state variants, and their leading-order terms are derived analytically.
    \item It is hardware-agnostic: it is specified as a quantum circuit, so any gate-based quantum computer can run it after compiling the circuit into its native controls.
    \item It is unitary: it applies a fixed sequence of gates without post-selection or mid-circuit measurements, which not all hardware supports, so every run yields the target state, and it can be inverted, controlled and embedded as a subroutine in larger quantum algorithms.
    \item It is exact: under ideal gates, its unit fidelity at any system size is certified by construction, so any deviation observed on hardware is attributable to noise alone, which also makes it a clean hardware benchmark.
\end{enumerate}

\section{Motzkin and Fredkin States}
\label{sec:motzkin-fredkin}
Before presenting the sequential preparation protocol, we first review the mathematical construction and entanglement properties of Motzkin and Fredkin states, highlighting the features that make these states particularly interesting for quantum simulation.

An uncolored Motzkin state on $N$ sites is a uniform superposition of all uncolored Motzkin paths of length $N$.
An uncolored Motzkin path of length $N$ is a sequence of $N$ steps, where each step can be one of \{up ($+1$ height), flat ($+0$ height) or down ($-1$ height)\}, such that its start, minimum and end heights are all 0.
Each site carries a local state $\ket{u}$, $\ket{f}$ or $\ket{d}$ that encodes its step, so every sequence of $N$ steps (i.e. path $p$) labels the corresponding basis state $\ket{p}$ of the chain.
The first $\ell$ steps of a path form its $\ell$-prefix, and the height of a prefix is the height after its last step.

\begin{figure}
    \includegraphics[width=\columnwidth]{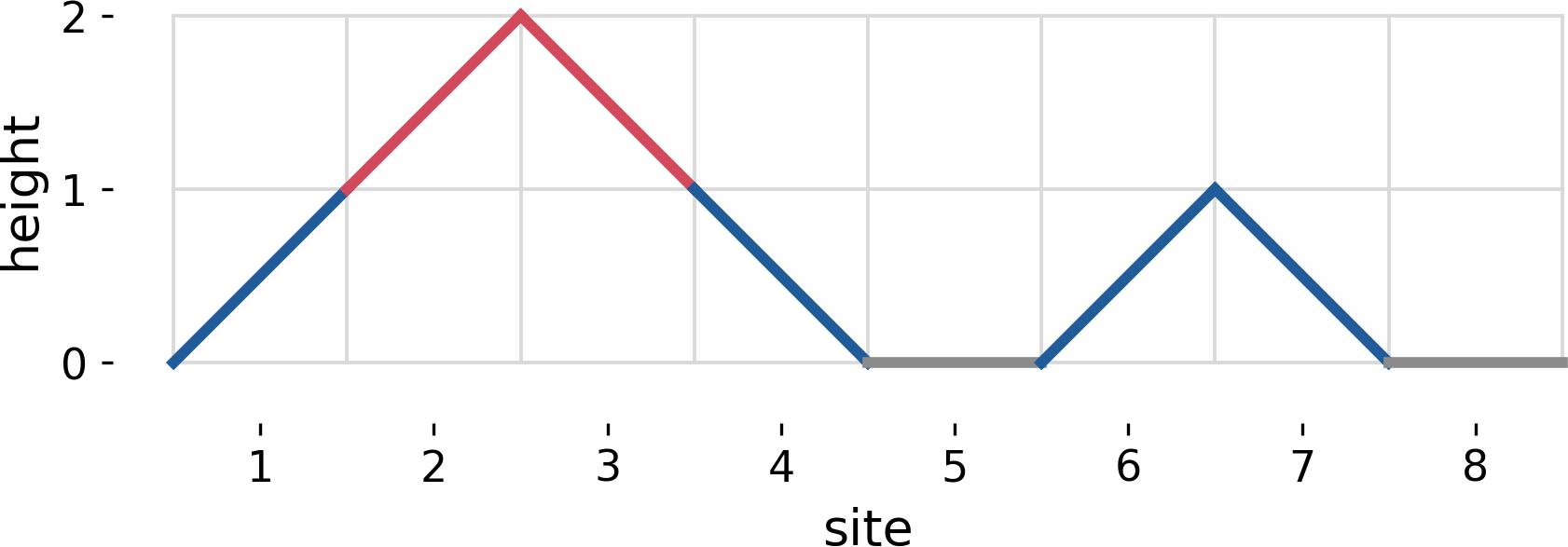}

    \includegraphics[width=\columnwidth]{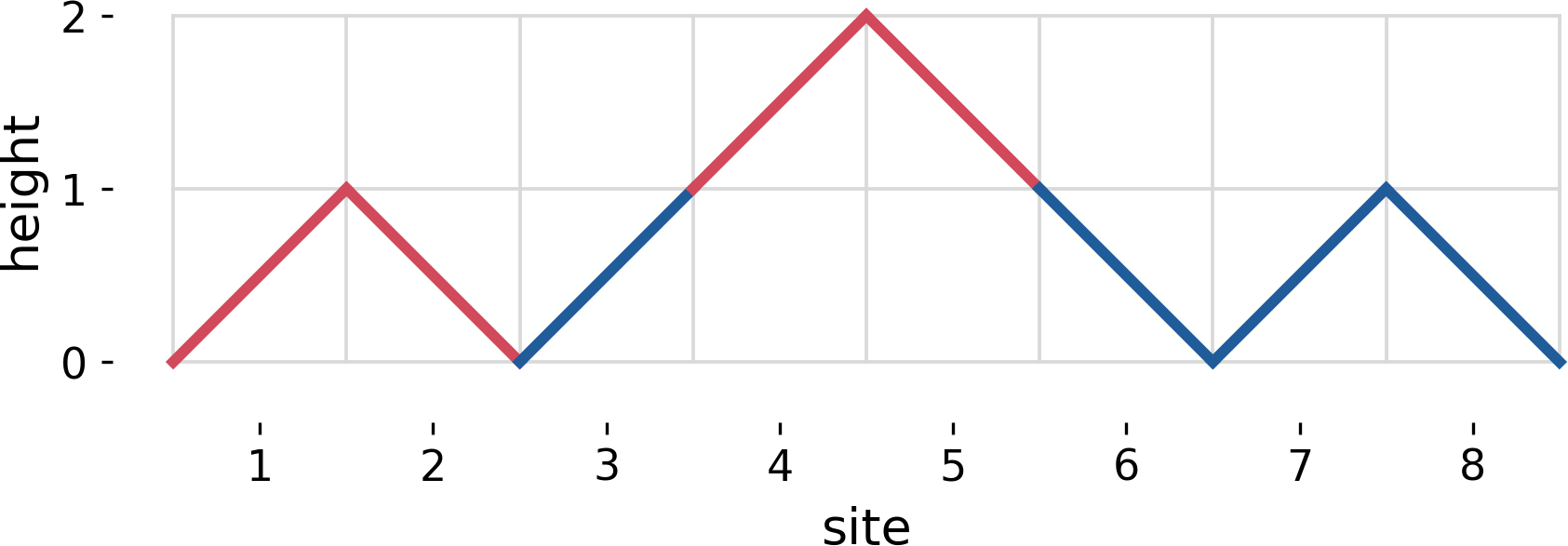}

    \caption{Example of 2-colored Motzkin~\cite{bravyi2012criticality, movassagh2016supercritical} (top) and Fredkin~\cite{salberger2017deformed, salberger2018fredkin} (bottom) paths on 8 sites.}
    \label{fig:path-examples}
\end{figure}

One common generalization of uncolored Motzkin states is the $k$-colored Motzkin state, defined as a uniform superposition of all $k$-colored Motzkin paths of length $N$.
In the $k$-colored version of Motzkin paths, in addition to the rules of the uncolored Motzkin paths, each up step is colored with one of the $k$ colors.
For example, in Fig.~\ref{fig:path-examples}, two colors are used: red and blue.
Each down step has to have the same color as its matching up step, where steps match in the usual parenthesis sense.
For example, in Fig.~\ref{fig:path-examples} (top), the down step at site 3 matches the up step at site 2.
Similarly, the down step at site 4 matches the up step at site 1, and site 7 matches site 6.
The flat steps are uncolored (neutral).
In the case of $k = 1$, colored Motzkin states reduce to uncolored Motzkin states.

Yet another generalization is the area-weighted (or $t$-deformed) Motzkin state~\cite{levine2017gap, zhang2017novel}, where the superposition of the Motzkin paths ($k$-colored, in general) becomes non-uniform.
In particular, the relative weight of each path in the superposition becomes exponential in the area under the path, which, in the case of Motzkin (and Fredkin) paths, can be shown to be equal to
\begin{equation}
    A(p) = \sum_{j=1}^{N} h_j(p),
    \label{eq:area}
\end{equation}
where $A(p)$ is the area under path $p$ and $h_j(p)$ is the height after site $j$.

Specifically, the $t$-deformed $k$-colored Motzkin state on $N$ sites is given by
\begin{eqnarray}
    \ket{M_N^{(k)}; t}
    &=&\frac{1}{\sqrt{\mathcal{N}}}
    \sum_{p\in M_N^{(k)}} t^{A(p)}\,\ket{p},
    \label{eq:weighted-motzkin-state}
\end{eqnarray}
where $\mathcal{N} = \sum_{p\in M_N^{(k)}} t^{2A(p)}$ is the normalization factor, $M_N^{(k)}$ is the set of all $k$-colored Motzkin paths of length $N$, and $t > 0$ is a deformation parameter (see Table~\ref{tab:notation} for the notation).
For $t=1$, every path has equal weight and Eq.~\eqref{eq:weighted-motzkin-state} reduces to the usual uniform $k$-colored Motzkin state $\ket{M_N^{(k)}}$.
For $t<1$, the superposition is dominated by paths of small height, whereas for $t>1$ it is dominated by paths of large height, i.e. paths with $h_{N/2}\sim N/2$.
The deformation also changes the entanglement scaling: the half-chain entanglement entropy is bounded for $t<1$ and grows linearly with $N$ for $t>1$ and $k\geq2$~\cite{zhang2017novel} (see Sec.~\ref{subsec:EE}).

\begin{table}
    \centering
    \renewcommand{\arraystretch}{1.2}
    \begin{tabular}{| l | >{\raggedright\arraybackslash}p{0.72\columnwidth} |}
        \hline
        Symbol & Meaning \\
        \hline
        $N$ & Number of sites \\
        $k$ & Number of colors (default $1$) \\
        $t$ & Deformation parameter of Eq.~\eqref{eq:weighted-motzkin-state} (default $1$) \\
        $\ell$ & Number of prepared sites, see Sec.~\ref{sec:construction} (default $N$) \\
        $h_j(p)$ & Height of path $p$ after site $j$ \\
        $M_N^{(k)}$ & Set of $k$-colored Motzkin paths of length $N$ \\
        $\ket{S}$ & Uniform superposition of all paths in a set $S$ \\
        $\ket{S; t}$ & $t$-deformed superposition of the paths in $S$, as in Eq.~\eqref{eq:weighted-motzkin-state} \\
        $\ket{S; \ell}$ & $\ell$-prefix state of the target state $\ket{S}$, see Sec.~\ref{sec:construction} \\
        $\ket{S; \ell}'$ & $\ell$-prefix state before the height update, see Sec.~\ref{sec:uncolored-motzkin} \\
        \hline
    \end{tabular}
    \caption{Notation for paths, sets of paths and the states labeled by them. Qualifiers and parameters at their default values can be omitted. For example, $M_N = M_N^{(1)}$, i.e. the set of uncolored Motzkin paths of length $N$. Additional notation used only in the appendices is listed in Table~\ref{tab:notation-appendix}.}
    \label{tab:notation}
\end{table}

Fredkin states are very similar to Motzkin states, except that Fredkin paths have only up or down steps and no flat steps \cite{salberger2017deformed, salberger2018fredkin}.
An example of a 2-colored Fredkin path is shown in Fig.~\ref{fig:path-examples} (bottom).

\subsection{Characterization of Motzkin and Fredkin States in Terms of Entanglement Entropy and Reduced Density Matrix}
\label{subsec:EE}

\begin{figure}
    \centering
    \makebox[\linewidth][l]{\hspace{0.1549\linewidth}\includegraphics[width=0.8074\linewidth, trim=44 277.05 12.33 0, clip]{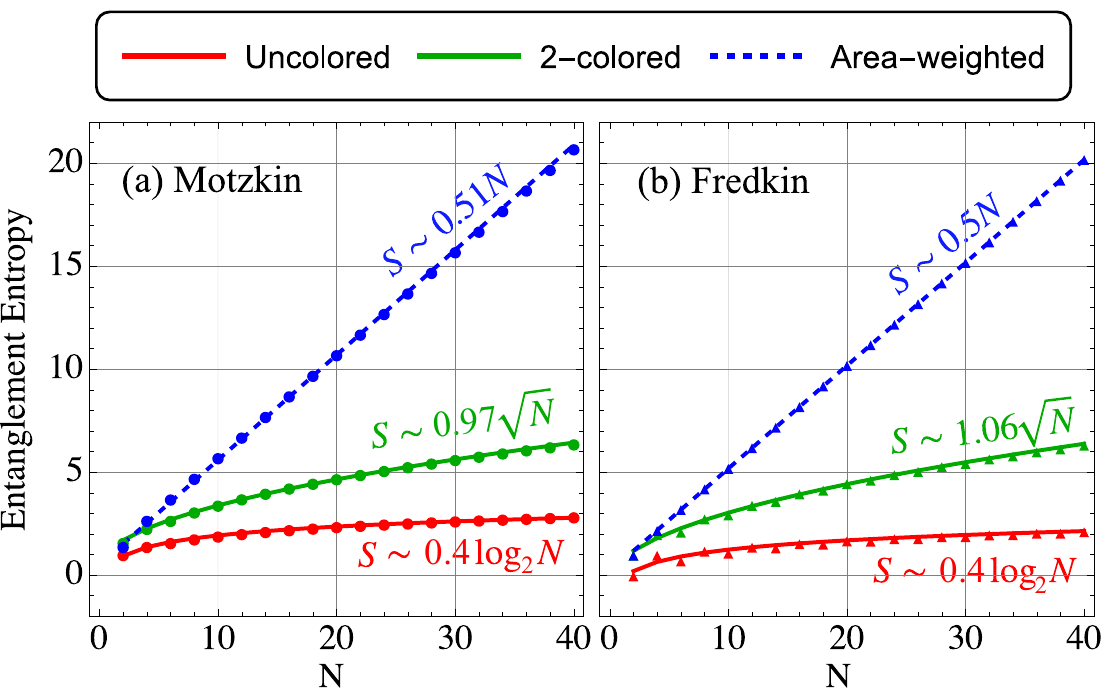}}\\
    \makebox[\linewidth][l]{\hspace{0.0494\linewidth}\includegraphics[width=0.4886\linewidth, trim=0 0 242.73 53, clip]{figures/rdm/RDM_plots_v7a.pdf}\hspace{0.0469\linewidth}\includegraphics[width=0.4152\linewidth, trim=285.6 0 0 53, clip]{figures/rdm/RDM_plots_v7a.pdf}}\\
    \includegraphics[width=\linewidth, trim=0 0 17.24 10, clip]{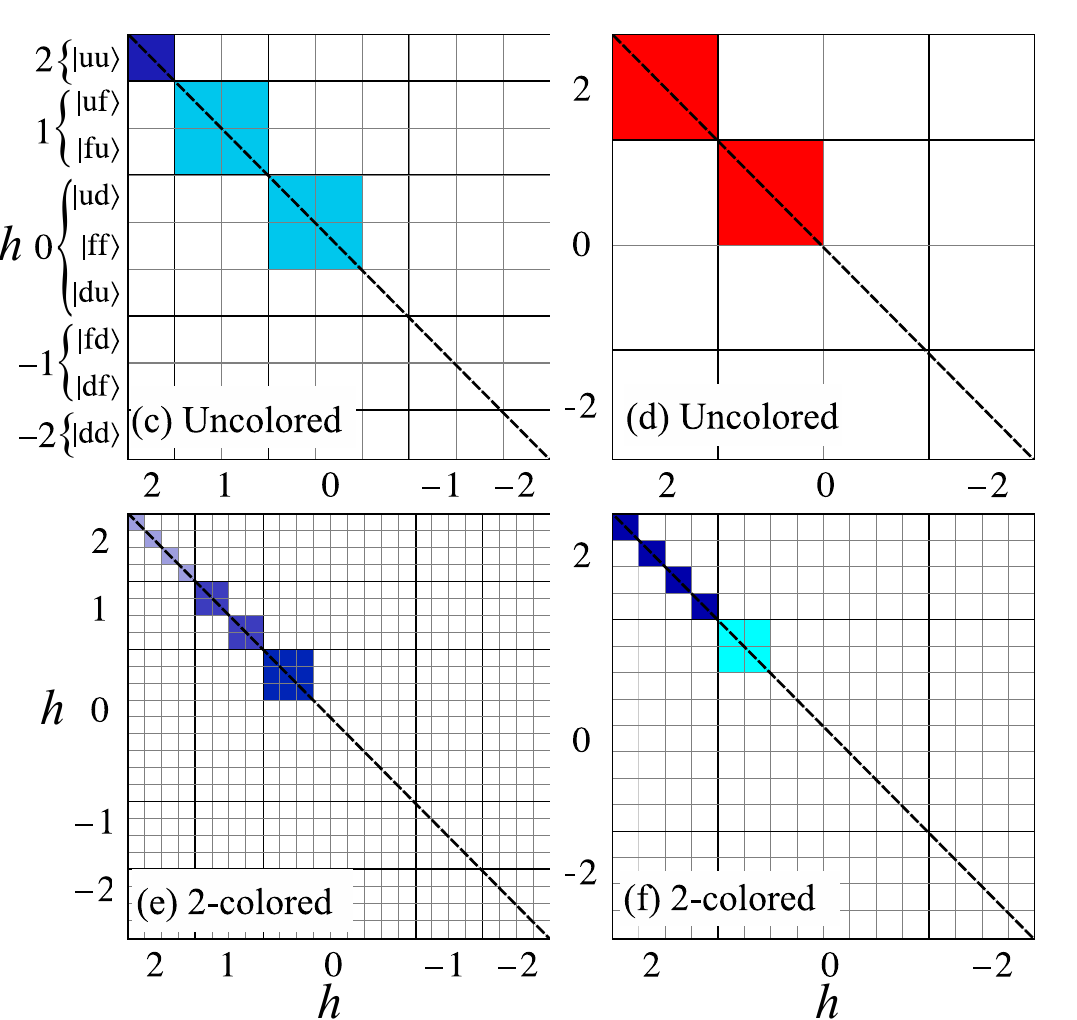}\\
    \makebox[\linewidth][l]{\hspace{0.3723\linewidth}\includegraphics[width=0.3866\linewidth]{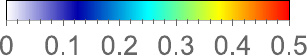}}
    \caption{Half-chain entanglement entropy $S_A$ (in bits) versus the number of sites $N$ for (a) the Motzkin and (b) Fredkin states. Red, green, and blue markers represent the uncolored, 2-colored, and area-weighted 2-colored ($t=2$) states, respectively. Dashed and solid lines are fits, labeled by their leading terms (with $S \equiv S_A$). Reduced density matrix (RDM) of the (c) uncolored Motzkin, (d) uncolored Fredkin, (e) 2-colored Motzkin, and (f) 2-colored Fredkin states for a chain of $N = 4$ sites, obtained by tracing out the right half of the chain. Rows and columns are labeled by the height $h$ of the left half ($N/2$-prefix), and the color scale indicates the magnitude of each matrix element.
    }
    \label{fig:EE_Motzkin_Fredkin}
\end{figure}

As mentioned in the Introduction, one of the reasons for interest in Motzkin and Fredkin states is that they violate the area law of entanglement despite being unique ground states of local, frustration-free Hamiltonians~\cite{bravyi2012criticality, movassagh2016supercritical, salberger2017deformed, salberger2018fredkin}.
Figures~\ref{fig:EE_Motzkin_Fredkin}(a) and (b) show the half-chain entanglement entropy $S_A$ of Motzkin and Fredkin states, respectively, as a function of system size $N$.
For a bipartition of the system $\ket{\psi}$ into subsystems $A$ and $B$, $S_A$ is the von Neumann entropy of the reduced density matrix $\rho_A$ of the left half of the chain (subsystem $A$),
\begin{equation}
    S_A = -{\rm Tr}\left(\rho_A\log_2\rho_A\right)
    = -\sum_i\lambda_i\log_2\lambda_i,
    \label{eq:entropy}
\end{equation}
where $ \rho_A = {\rm Tr}_B \left( \proj{\psi} \right)$ and \(\lambda_i\) are the eigenvalues of \(\rho_A\), so that $S_A$ is measured in bits.

In Figs.~\ref{fig:EE_Motzkin_Fredkin}(a) and (b), three scaling regimes can be observed.
For the uncolored states, the entropy grows logarithmically, $S_A \propto \log N$.
For the $k=2$ colored states, it grows as $S_A \propto \sqrt{N}$.
For the area-weighted 2-colored states with deformation parameter $t=2$, it grows linearly, $S_A \propto N$, which is a volume law of entanglement.

The distinct entanglement scaling can be understood from the structure of the reduced density matrix (RDM), which offers a direct visual representation of the entanglement structure and, as we discuss in Appendix~\ref{app:rdm_complexity}, also informs the circuit complexity.

Examples of RDMs for the uncolored and colored Motzkin and Fredkin states are shown in Figs.~\ref{fig:EE_Motzkin_Fredkin}(c)--(f).
Each cell corresponds to a pair of basis states $\ket{a},\ket{a'}$ of subsystem $A$, and its color gives the matrix element $\braket{a | \rho_A | a'}$.
Every path in the full state splits into a prefix on subsystem $A$ and a suffix on subsystem $B$.
For example, $\ket{uffd}$ has prefix $\ket{uf}$ and suffix $\ket{fd}$.
The rows and columns of the RDM therefore correspond to the prefixes.
The diagonal entries $\bra{a}\rho_A\ket{a}$, marked by the black dashed line, give the probabilities of the prefixes, while an off-diagonal entry $\bra{a}\rho_A\ket{a'}$ with $a\neq a'$ is proportional to the number of suffixes that $a$ and $a'$ share (for $t = 1$).
A detailed path-by-path construction of the RDM is provided in Appendix~\ref{app:rdm_construction}.

In Figs.~\ref{fig:EE_Motzkin_Fredkin}(c)--(f), the rows and columns of RDMs are organized by their height $h$ at the bipartition.
For the Motzkin and Fredkin states, we define the height operator of subsystem $A$ as
\begin{equation}
    \hat{h}_A=\sum_{i\in A}\hat{s}_i,\qquad \hat{s}_i=\sum_{c=1}^{k}\left(|u^c\rangle\langle u^c|-|d^c\rangle\langle d^c|\right)_i ,
\end{equation}
where $\hat s_i$ assigns $+1$, $-1$ and $0$ to up $\ket{u^c}$, down $\ket{d^c}$ and flat $\ket{f}$ steps, respectively, at site $i$ irrespective of color $c$.
Every prefix $|a\rangle$ is then an eigenstate, $\hat h_A|a\rangle=h(a)\,|a\rangle$, with eigenvalue equal to its height at the bipartition.
As an example, for the $N=4$ uncolored Motzkin state, the height operator yields $h=1$ on the prefixes $\ket{uf}$ and $\ket{fu}$, and $h=2$ on $\ket{uu}$.
Prefixes $\ket{uu}$, $\ket{uf}$, $\ket{fu}$, $\ket{ud}$ and $\ket{ff}$ are the only valid prefixes, i.e. the only ones that never go below height 0, and thus show non-zero matrix elements in Fig.~\ref{fig:EE_Motzkin_Fredkin}(c).

For the uncolored states, $\hat{h}_A$ reduces to the subsystem magnetization $\hat{S}^z_A = \sum_{i\in A} \hat{S}^z_i$ (up to the factor of $2$ for spin-$\tfrac{1}{2}$), where $\hat{S}^z_i$ is the $z$-component of the spin operator at site $i$, with eigenvalues $m=-l,\dots,l$ for spin $l$.
For colored states encoded in a higher spin, however, $\hat{S}^z_A$ weights each step by its magnetization $m$, which depends on its color (e.g. $m=\pm c$ for an up or down step of color $c$), and no longer equals the height operator.

Since the ground state consists of paths that end at height 0, it satisfies $\hat{h}_{\mathrm{tot}}\ket{\psi} = 0$ for the total height $\hat{h}_{\mathrm{tot}} = \hat{h}_A + \hat{h}_B$, therefore $\hat{h}_A\ket{\psi} = -\hat{h}_B\ket{\psi}$, and using the cyclicity of the partial trace over $B$,
\begin{equation}
    [\rho_A, \hat{h}_A] = 0 .
\end{equation}
The reduced density matrix is therefore block diagonal in the eigenvalues of $\hat{h}_A$, which provides an operator-level justification of the height-sector structure seen in Figs.~\ref{fig:EE_Motzkin_Fredkin}(c)--(f) and of the height-symmetry resolution of the entanglement spectrum.

While the entanglement entropy can be evaluated directly from the eigenvalues of the RDM using Eq.~\eqref{eq:entropy}, the matrix elements of the RDM provide a complementary route that connects the entanglement content directly to the Motzkin and Fredkin paths.
For the $N=4$ uncolored Motzkin state in Fig.~\ref{fig:EE_Motzkin_Fredkin}(c), the RDM consists of three blocks: one $1\times1$ block for $h=2$ and two $2\times2$ blocks for $h=1$ and $0$, where the shaded cells of each block correspond to the complete paths passing through height $h$ at the bipartition.
For example, the $h=2$ sector contains only the path $\ket{uudd}$, whereas the $h=1$ sector contains four paths $\{\ket{uffd},\ket{ufdf},\ket{fufd},\ket{fudf}\}$.
Since all paths carry equal amplitude, the weight $p_h$ of each height sector is the fraction of paths in that sector out of the $|M_4|=9$ paths in total, giving $p_h=1/9$, $4/9$ and $4/9$ for $h=2,1,0$, respectively.
Furthermore, since all entries within a block are identical, each block has rank one with the single non-zero eigenvalue $p_h$.
The entanglement entropy therefore reduces to the Shannon entropy of the height distribution, $S_A=-\sum_h p_h\log_2 p_h$.
For large $N$, the distribution $p_h$ has a standard deviation of order $\sqrt{N}$~\cite{movassagh2016supercritical}, and since the Shannon entropy grows as the logarithm of the standard deviation, $S_A\simeq\log_2\sqrt{N}=\frac{1}{2}\log_2 N$~\cite{bravyi2012criticality, movassagh2016supercritical}.

In the $k$-colored state, each height sector splits into $k^h$ subsectors, one for each coloring of the $h$ \emph{unmatched} up steps of the prefix.
A suffix can complete a prefix only if the down steps that match these up steps have the same colors, so prefixes whose unmatched up steps are colored differently share no suffixes.
For the RDM of the $N=4$, $2$-colored Motzkin state shown in Fig.~\ref{fig:EE_Motzkin_Fredkin}(e), the $h=2$ sector splits into $2^2$ subsectors and the $h=1$ sector splits further into $2^1$ subsectors, as compared to the uncolored state in panel (c).
For a uniform superposition, these subsectors carry equal weight, so each contributes an eigenvalue $p_h^{(k)}/k^h$, where $p_h^{(k)}$ is the height distribution of the $k$-colored state.
The RDM in Fig.~\ref{fig:EE_Motzkin_Fredkin}(e) consists of $7$ blocks: four $1\times1$ blocks for $h=2$, two $2\times2$ blocks for $h=1$, and one $3\times3$ block for $h=0$.
Out of $|M_4^{(2)}|=21$ paths, this gives $p_h^{(2)}=4/21$, $8/21$ and $9/21$ for $h=2,1,0$.
Thus, although the number of height sectors is the same as in the uncolored case, the color labeling increases the number of blocks and consequently the number of eigenvalues in Eq.~(\ref{eq:entropy}), raising the entanglement entropy.
The entanglement entropy for the $k$-colored state takes the form
\begin{equation}
    S_A = -\sum_h p_h^{(k)} \log_2 p_h^{(k)} + \braket{h}_k \log_2 k,
    \label{eq:EE_decomposition}
\end{equation}
with $\braket{h}_k=\sum_h h\,p_h^{(k)}\propto\sqrt{N}$ the mean height at the bipartition \cite{movassagh2016supercritical}.
The first term in Eq.~(\ref{eq:EE_decomposition}) is similar to that of the uncolored state and grows only logarithmically, whereas the second term grows as $\sqrt{N}$ for $k\geq2$ and dominates, so $S_A\propto\sqrt{N}$~\cite{Barghathi2026}.

Introducing area weighting ($t>1$) leaves the sector structure unchanged, but reshapes the height distribution by favoring paths with larger enclosed area, defined in Eq.~(\ref{eq:area}).
Such paths reach larger heights at the bipartition, so $p_h^{(k)}$ shifts toward $h=O(N)$.
Thus, $\braket{h}_k$ grows linearly with $N$, giving volume-law entanglement for $k\geq2$~\cite{zhang2017novel}.
For $k=1$, the second term in Eq.~(\ref{eq:EE_decomposition}) vanishes, and the entropy remains bounded.

Because Fredkin paths contain no flat steps, the height at the bipartition must have the same parity as the number of sites in subsystem $A$, so only every other height sector is possible.
Apart from this, the RDMs of the Fredkin states in Figs.~\ref{fig:EE_Motzkin_Fredkin}(d) and (f) have visually the same block-diagonal structure as their Motzkin counterparts: each height sector is a single rank-one block in the uncolored case and splits into $k^h$ equal-weight subsectors in the colored case.
Consequently, the uncolored, colored and area-weighted Fredkin states follow the same entanglement scaling as the corresponding Motzkin states~\cite{salberger2017deformed, salberger2018fredkin}.

\section{Circuit Construction}
\label{sec:construction}

The states considered here are prepared using a sequential generation (SG) approach.
In SG, the overall preparation process consists of $N$ iterations, where each iteration extends the existing $\ell$-prefix state into the $(\ell + 1)$-prefix state.
Here and below, the $\ell$-prefix state is defined as the state in which the probability of measuring a particular path on the first $\ell$ sites is equal to that in the target state.
We denote the $\ell$-prefix state of a target state $\ket{S}$ by $\ket{S; \ell}$, and omit the default value $\ell = N$, for which it is the target state itself.
All probability amplitudes prepared by SG are real and non-negative, which is also true for the target Motzkin and Fredkin states considered in this work.

Throughout the manuscript, individual sites may be represented by multilevel systems, i.e. qudits.
Of course, any qudit can be equivalently encoded with multiple qubits, so we do not require physical access to qudits to implement our algorithm.

We first introduce the construction for the uncolored Motzkin state (Sec.~\ref{sec:uncolored-motzkin}), then describe the changes needed for the colored Motzkin, area-weighted and Fredkin states (Secs.~\ref{sec:colored-motzkin}--\ref{sec:fredkin}), and finally discuss which other states can be prepared in the same way (Sec.~\ref{sec:other-states}).

\subsection{Uncolored Motzkin States}
\label{sec:uncolored-motzkin}

\begin{figure*}
    \includegraphics[width=\linewidth]{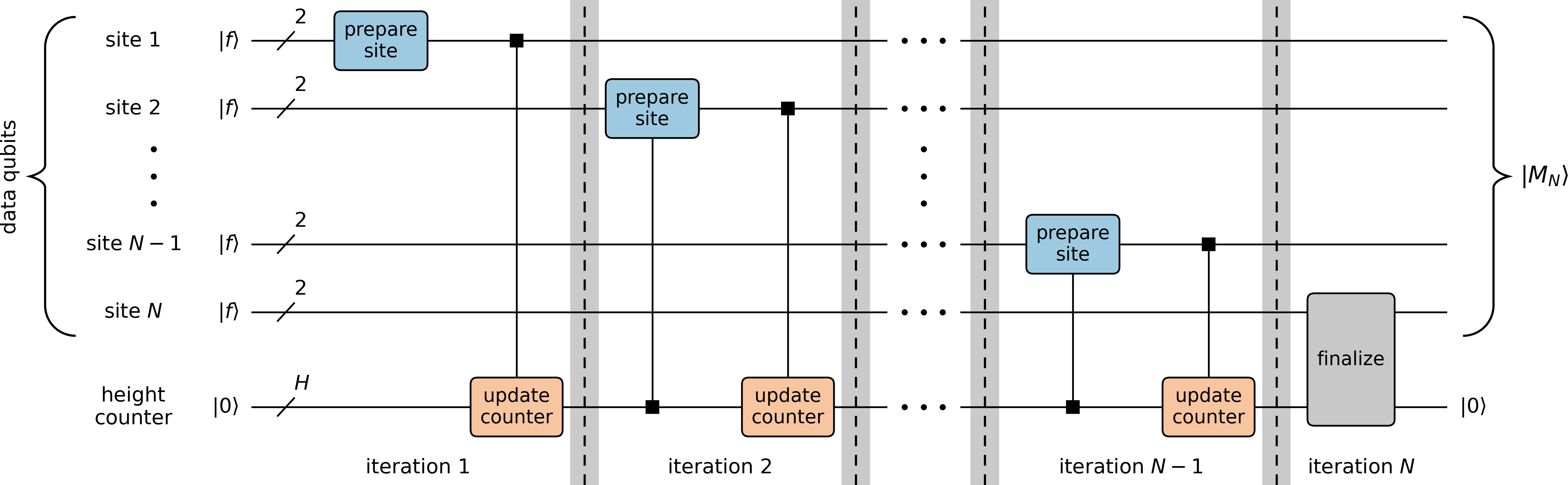}
    \caption{High-level overview of the SG-circuit preparing the uncolored Motzkin state $\ket{M_N}$. Each of the first $N - 1$ iterations prepares one site using the height counter as a control register, followed by an update of the counter. The last iteration prepares the last site and resets the height counter. Square controls indicate qubits used as controls within the corresponding operators. Kets on bundles of wires denote register states: each site starts in the flat state $\ket{f}$, and the height counter starts and ends in the state $\ket{0}$, i.e. holding the value 0.}
    \label{fig:overview}
\end{figure*}

\textbf{Overview.}
A high-level overview of the preparation circuit is shown in Fig.~\ref{fig:overview}.
Each site of an uncolored Motzkin path, i.e. an up ($u$), flat ($f$) or down ($d$) step, is encoded with 2 qubits as follows: $u \rightarrow 10$, $f \rightarrow 00$, $d \rightarrow 01$, where the first qubit is called the \textit{up} qubit, and the second qubit is called the \textit{down} qubit.
By construction, the up and down qubits cannot both be 1.
Below, we denote the up and down qubits of site $\ell$ by $\mathrm{up}_\ell$ and $\mathrm{down}_\ell$.

The circuit consists of $N$ structurally similar iterations, shown as circuit blocks separated by barriers (shaded dashed lines).
Internally, each of the first $N - 1$ iterations can be decomposed into 2 parts: preparation of the $\ell$-prefix state itself (``prepare site") and update of auxiliary information associated with that state, which will be needed for the next iteration.
The only such auxiliary information we have in this case is the height of each path in the currently prepared $\ell$-prefix state.
This information is stored in its own ancilla register, encoded with the usual binary encoding in LSB-first order (least significant bit).
The maximum height achieved by any prefix of any Motzkin path is limited by $\lfloor N/2 \rfloor$, so
\begin{equation}
    H = \lfloor \log_2 \lfloor N/2 \rfloor \rfloor + 1
\end{equation}
qubits in the height counter will be sufficient to store the height of any prefix.
The last iteration consists of a single ``finalize" operator, which prepares the last site and resets the height counter, so it ends up in the same all-0 state it started from.

The square controls in Fig.~\ref{fig:overview} (on ``prepare site" and ``update counter" operators) and below indicate multiplexing qubits, meaning that the effect of the corresponding gate depends on their value.

\begin{figure*}
    \centering
    \includegraphics[width=\textwidth]{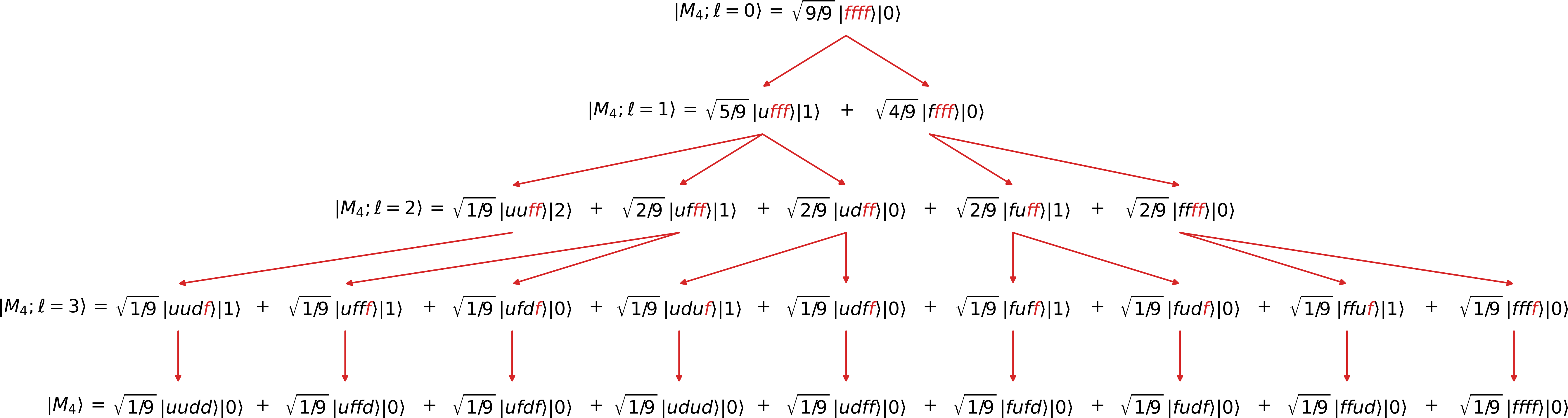}
    \caption{Example of the sequential generation procedure for the uncolored Motzkin state on 4 sites. Each row shows the intermediate $\ell$-prefix state $\ket{M_4; \ell}$ after preparing $\ell$ sites (qudits). In each term, the unprepared sites are shown in red, and the second ket is the height register, which stores the height of the prepared part. Arrows indicate how the probability of each prefix is distributed among its valid continuations. The final row corresponds to the target state $\ket{M_4}$.}
    \label{fig:state-tree}
\end{figure*}

As a specific example, we will use SG preparation of the uncolored Motzkin state on 4 sites
\begin{equation}
    \label{eq:target-state}
    \begin{aligned}
        \ket{M_4} &= \sqrt{1/9} \Bigl( \ket{uudd} + \ket{uffd} + \ket{ufdf} \\
        &\quad + \ket{udud} + \ket{udff} + \ket{fufd} \\
        &\quad + \ket{fudf} + \ket{ffud} + \ket{ffff} \Bigr),
    \end{aligned}
\end{equation}
i.e. a uniform superposition of all Motzkin paths of length 4.
Figure~\ref{fig:state-tree} shows all intermediate $\ell$-prefix states ($\ket{M_4; \ell}$, one in each row) obtained after each iteration of the state preparation circuit for $\ket{M_4}$.

Its preparation would use 8 data qubits and $H = 2$ qubits in the height counter (although Fig.~\ref{fig:state-tree} shows encoded values directly as qudits, instead of showing individual qubits).
It starts from a single computational basis state (CBS) $\ket{ffff}\ket{0}$ in the top row of Fig.~\ref{fig:state-tree}, where all sites are unprepared, i.e. in their default state $f \rightarrow 00$, and the height counter stores 0.

\begin{figure}
    \includegraphics[width=0.75\linewidth]{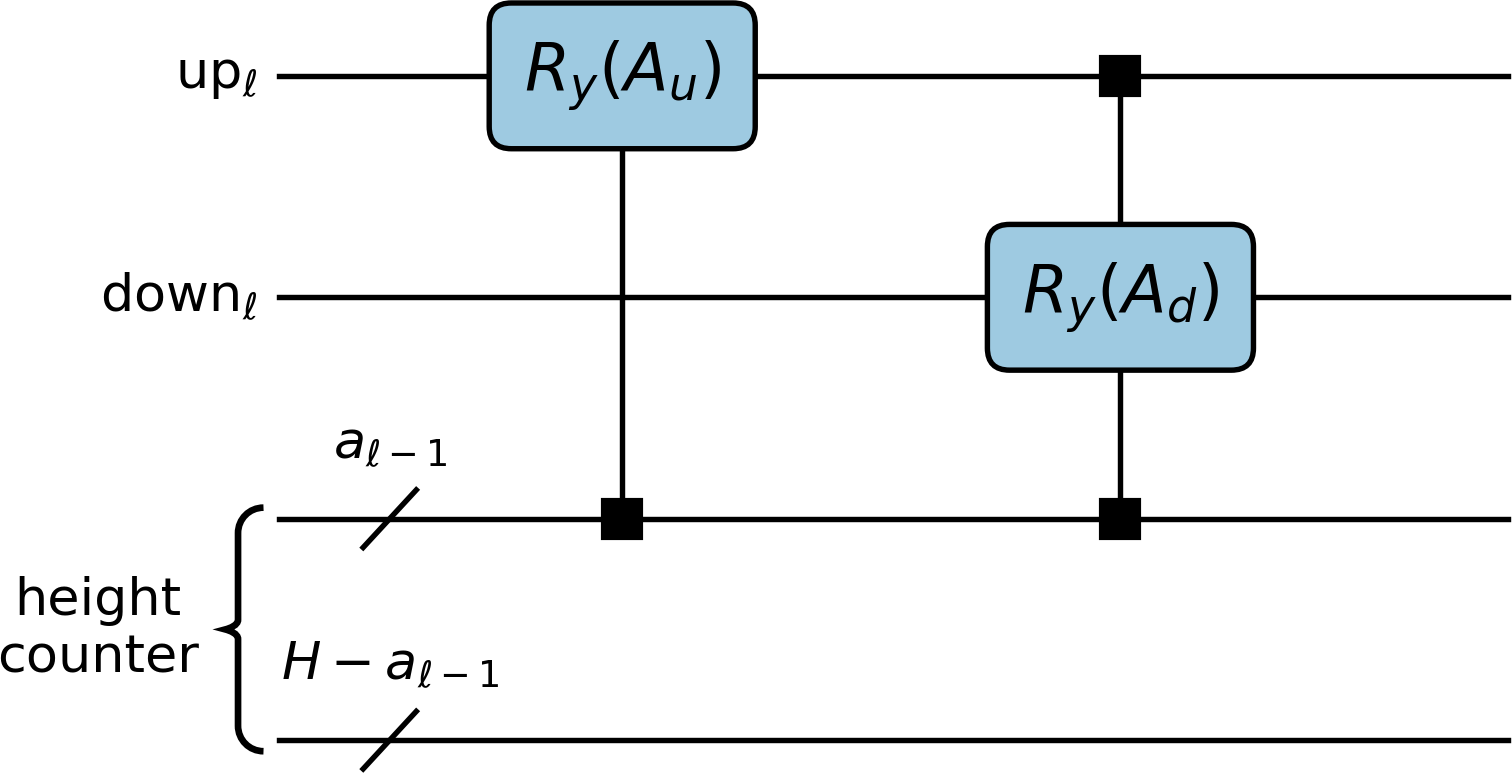}
    \caption{Implementation of the ``prepare site" operator. The $\mathrm{up}_\ell$ and $\mathrm{down}_\ell$ qubits are prepared using uniformly controlled $R_y$ rotations with angle arrays $A_u$ and $A_d$, respectively, multiplexed on the height counter. The down rotation is additionally multiplexed on the $\mathrm{up}_\ell$ qubit to prevent simultaneous up and down steps.}
    \label{fig:prepare-site}
\end{figure}

\addvspace{10pt}
\textbf{Site preparation.}
Each of the first $N - 1$ iterations begins with a ``prepare site" operator.
In iteration $\ell$, its input is the $(\ell - 1)$-prefix state prepared by the previous iterations, which it extends to the $\ell$-prefix state in the data qubits.
The heights in the counter are updated afterwards by the ``update counter" operator.
We have 2 qubits per site, so internally the ``prepare site" operation is implemented with 2 operators: the ``up rotation", splitting the $\mathrm{up}_\ell$ qubit, which creates prefixes with an up step in site $\ell$, and the ``down rotation", similarly creating prefixes with a down step in site $\ell$, as shown in Fig.~\ref{fig:prepare-site}.

For example, the second iteration starts from the 1-prefix state in Fig.~\ref{fig:state-tree}
\begin{equation*}
    \ket{M_4; \ell = 1} = \sqrt{5/9} \ket{ufff}\ket{1} + \sqrt{4/9} \ket{ffff}\ket{0}.
\end{equation*}
Its ``up rotation" would create the state
\begin{equation*}
    \begin{aligned} 
        & \sqrt{1/9} \ket{uuff}\ket{1} + \sqrt{4/9} \ket{ufff}\ket{1} \\
        & + \sqrt{2/9} \ket{fuff}\ket{0} + \sqrt{2/9} \ket{ffff}\ket{0}.
    \end{aligned}
\end{equation*}
The following ``down rotation" would then split off the new $\ket{udff}$ prefix from $\ket{ufff}$ (both underlined), creating the state
\begin{equation*}
    \begin{aligned}
        \ket{M_4; \ell = 2}' = {} & \sqrt{1/9} \ket{uuff}\ket{1} + \underline{\sqrt{2/9} \ket{udff}\ket{1}} \\
        & + \underline{\sqrt{2/9} \ket{ufff}\ket{1}} + \sqrt{2/9} \ket{fuff}\ket{0} \\
        & + \sqrt{2/9} \ket{ffff}\ket{0},
    \end{aligned}
\end{equation*}
which is the same as $\ket{M_4; \ell = 2}$, except that the heights are not yet updated at this point, which we mark with a prime.

Both operators are implemented as \textit{uniformly controlled rotations} (UCR)~\cite{mottonen2005transformation}, multiplexed on the prefix height.
The effect of UCR is to rotate the target qubit by an angle whose value depends on the value of the multiplexing register.
See Appendix~\ref{app:ucr} for its formal definition and the details of its implementation.

More specifically, the maximum height of an $\ell$-prefix is
\begin{equation}
    h_{\max}(\ell) = \min(\ell, N - \ell),
    \label{eq:hmax}
\end{equation}
so we do not need to multiplex rotations on \textit{all} qubits of the height register.
Multiplexing them on the first $a_{\ell - 1}$ qubits is sufficient, where
\begin{equation}
    a_\ell = \lceil \log_2 (h_{\max}(\ell) + 1) \rceil
\end{equation}
is the number of \emph{active} height register qubits needed to represent the value of $h_{\max}(\ell)$ after step $\ell$.
The rest of them are guaranteed to be in the $\ket{0}$ state.
The corresponding multiplexing angles $A_u(\ell - 1, :)$ (array of size $2^{a_{\ell - 1}}$) are calculated classically for every height (explained below).

Down qubit rotations are implemented similarly with their own angles $A_d$.
The main difference here is that the $\mathrm{up}_\ell$ qubit itself is now also a part of the multiplexing register.
This is because up and down steps cannot be taken simultaneously, so we want to exclude the prefixes that already took an up step by setting the down rotation angles corresponding to $\mathrm{up}_\ell = 1$ to 0.

Because the flat step is encoded as 00 (default qubit state), the probability amplitude remaining in the original $(\ell - 1)$-prefixes after the up and down rotations automatically encodes a flat step in site $\ell$, without any additional operations.

\addvspace{10pt}
\textbf{Rotation angles.}
The splitting angles $A_u$ and $A_d$ for the UCR during site preparation are calculated classically from combinatorial considerations.
We first evaluate a function $D_N(\ell, h)$, which returns the number of ways in which a given prefix of length=$\ell$ and height=$h$ can be completed into a Motzkin path on $N$ sites.

$D_N(\ell, h)$ can be evaluated via the following recurrence relation
\begin{equation}
    \label{eq:uncolored-recurrence}
    D_N(\ell, h) =
    \begin{cases}
        0, & \ell > N\ \mathrm{or}\ h < 0 \\
        1, & \ell = N\ \&\ h = 0 \\
        D_N(\ell + 1, h + 1) \\
        + D_N(\ell + 1, h) \\
        + D_N(\ell + 1, h - 1), &\mathrm{otherwise}.
    \end{cases}
\end{equation}

The length of a given prefix cannot exceed the total number of sites and its height cannot be less than $0$.
Therefore, in these cases the prefix is invalid and there are 0 ways to complete it to a Motzkin path.
If the length of a prefix is $= N$ and the height is $= 0$, then the prefix is already a Motzkin path and there is 1 trivial way to complete it (do nothing).
Otherwise, the prefix can be continued either into an $(\ell + 1, h + 1)$-prefix (via step up), $(\ell + 1, h)$-prefix (via flat step), or $(\ell + 1, h - 1)$-prefix (via step down), and the total number of ways to complete this prefix is given by the sum of those numbers from all its continuation branches.

For example, for $N = 4$, the values of $D_4(\ell, h)$ are
{\setlength{\tabcolsep}{9pt}
\begin{center}
    \begin{tabular}{c|c c c c c}
        \hline\hline
        $h \backslash \ell$ & 0 & 1 & 2 & 3 & 4 \\
        \hline
        2 & & & 1 & 0 & 0 \\
        1 & & 5 & 2 & 1 & 0 \\
        0 & 9 & 4 & 2 & 1 & 1 \\
        \hline\hline
    \end{tabular}
\end{center}}
The entries with $h > \ell$ are omitted, since no prefix of length $\ell$ can reach such heights.

A table like this can be efficiently evaluated for all possible values of $\ell$ and $h$ via dynamic programming in $O(N^2)$ time and space.

Next, we calculate the fraction of the probability of the current $(\ell, h)$-prefix that will go into the $(\ell + 1, h + 1)$-continuation of this prefix (up), which can be calculated as
\begin{equation}
    \label{eq:prob-frac-up}
    p_u(\ell, h) = \frac{D_N(\ell + 1, h + 1)}{D_N(\ell, h)}.
\end{equation}
The corresponding up rotation angles are
\begin{equation}
    \label{eq:rotation-angles}
    A_{u}(\ell, h) = 2 \arcsin \left( \sqrt{p_u(\ell, h)} \right).
\end{equation}
The dependence of $A_u$ on $\ell$ is realized via its position in the circuit, so it does not need a separate multiplexer.

For example, the first iteration ($\ell = 1$) needs only the angle $A_u(0, 0) = 2 \arcsin \sqrt{5/9} \approx 1.682$, since the 0-prefix always has height 0, so its up rotation is not multiplexed at all.
Similarly, the values of $A_u$ needed in the second iteration ($\ell = 2$) are $A_u(1, 0)$ and $A_u(1, 1)$, and can be evaluated to $A_u(1, :) = [\pi/2, 2 \arcsin \sqrt{1/5}] \approx [1.571, 0.927]$.

The size of $A_u$ has to be padded to $2^{a_{\ell - 1}}$, since a UCR rotates its target qubit by an angle that depends on the value of its multiplexing register, so this angle has to be defined for all $2^{a_{\ell - 1}}$ possible values of the register.
The extra angles corresponding to the heights greater than $h_{\max}(\ell - 1)$ do not affect the state since the corresponding prefixes do not exist, and thus can be chosen arbitrarily without compromising \textit{correctness} per se.
However, the \textit{cost} of the UCR implementation can still depend on how those extra angles are chosen (see Appendix~\ref{app:ucr}).

The down rotation angles are calculated similarly.
First, we calculate the fraction of the remaining (after the up rotation) prefix probability that will go into the $(\ell + 1, h - 1)$-continuation of this prefix (down) as
\begin{equation}
    \label{eq:prob-frac-down}
    p_d(\ell, h) = \frac{D_N(\ell + 1, h - 1)}{D_N(\ell + 1, h - 1) + D_N(\ell + 1, h)}.
\end{equation}
Then the corresponding rotation angles are once again obtained from Eq.~\eqref{eq:rotation-angles}, using $p_d$ in place of $p_u$.

\addvspace{10pt}
\textbf{Counter update.}
Now let us consider the internal structure of the ``update counter" operators in Fig.~\ref{fig:overview}.
Following the implementation of the ``prepare site" operators, we now have an $\ell$-prefix state in the data qubits, but the values in the height counter are still inherited from the original $(\ell - 1)$-prefixes.
As such, they need to be updated by $+1$ for the prefixes that made an up step, or by $-1$ for the prefixes that made a down step in site $\ell$.

To implement this efficiently, we employ the following binary arithmetic identity:
\begin{equation}
    B - 1 = \overline{\overline{B} + 1}.
\end{equation}
That is, for any binary value $B$, subtraction of 1 is equivalent to the bitwise negation of the result of adding 1 to the bitwise negated value of $B$.
This allows us to handle both $+1$ and $-1$ branches with the same $+1$ operator.

\begin{figure}
    \includegraphics[width=0.75\columnwidth]{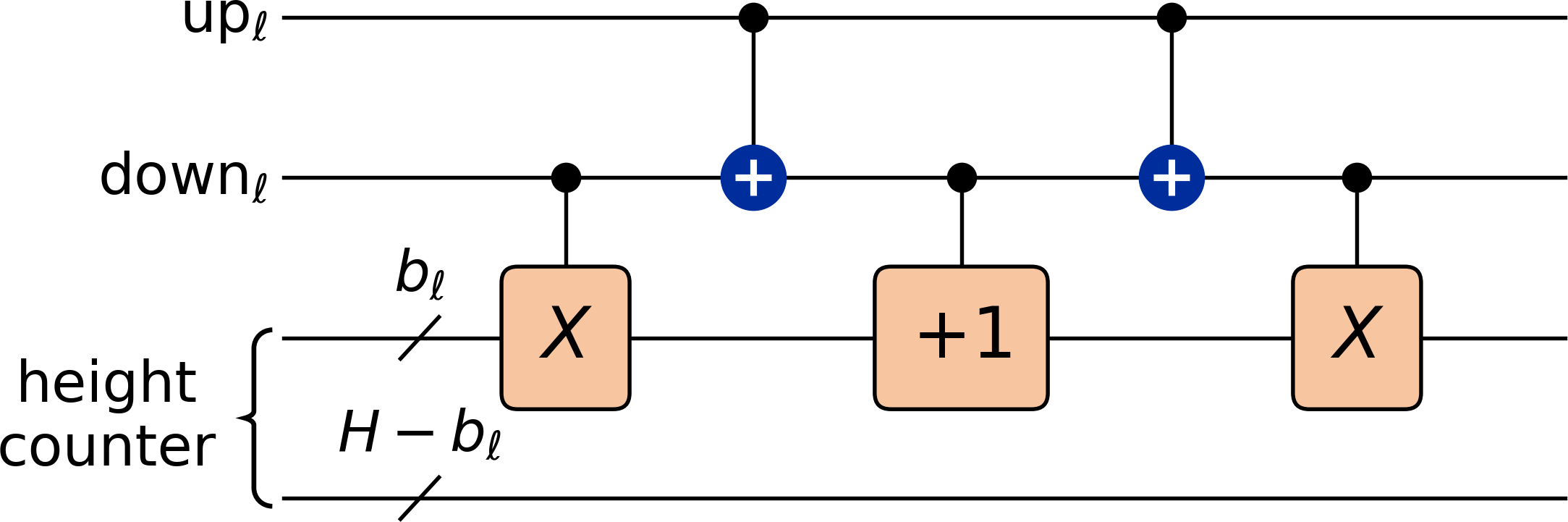}
    \caption{Implementation of the update counter operator. The height counter is incremented for an up step and decremented for a down step. Both operations are implemented using the same increment operator, with conditional bitwise inversion of the counter for down steps.}
    \label{fig:update-counter}
\end{figure}

The corresponding circuit is shown in Fig.~\ref{fig:update-counter}.
Since the number of active qubits of the height register changes from $a_{\ell - 1}$ to $a_\ell$ during this step, the circuit needs to act on whichever is larger, i.e. the first
\begin{equation}
    b_\ell = \max(a_{\ell - 1}, a_\ell)
\end{equation}
qubits of the height counter.

First, we apply the $X^{\otimes b_\ell}$ controlled on $\mathrm{down}_\ell$ to invert the height counter for the down branch (the prefixes that made a down step in the last site).
Next, since both up and down branches need to increment their counter, we temporarily save the result of ($\mathrm{up}_\ell\ \mathrm{xor}\ \mathrm{down}_\ell$) in qubit $\mathrm{down}_\ell$.
Because the two qubits cannot both be 1, their XOR is equivalent to OR, which is used as a control for the following +1 operator, after which we uncompute the XOR operation and the counter inversion.

\begin{figure}
    \includegraphics[width=\columnwidth]{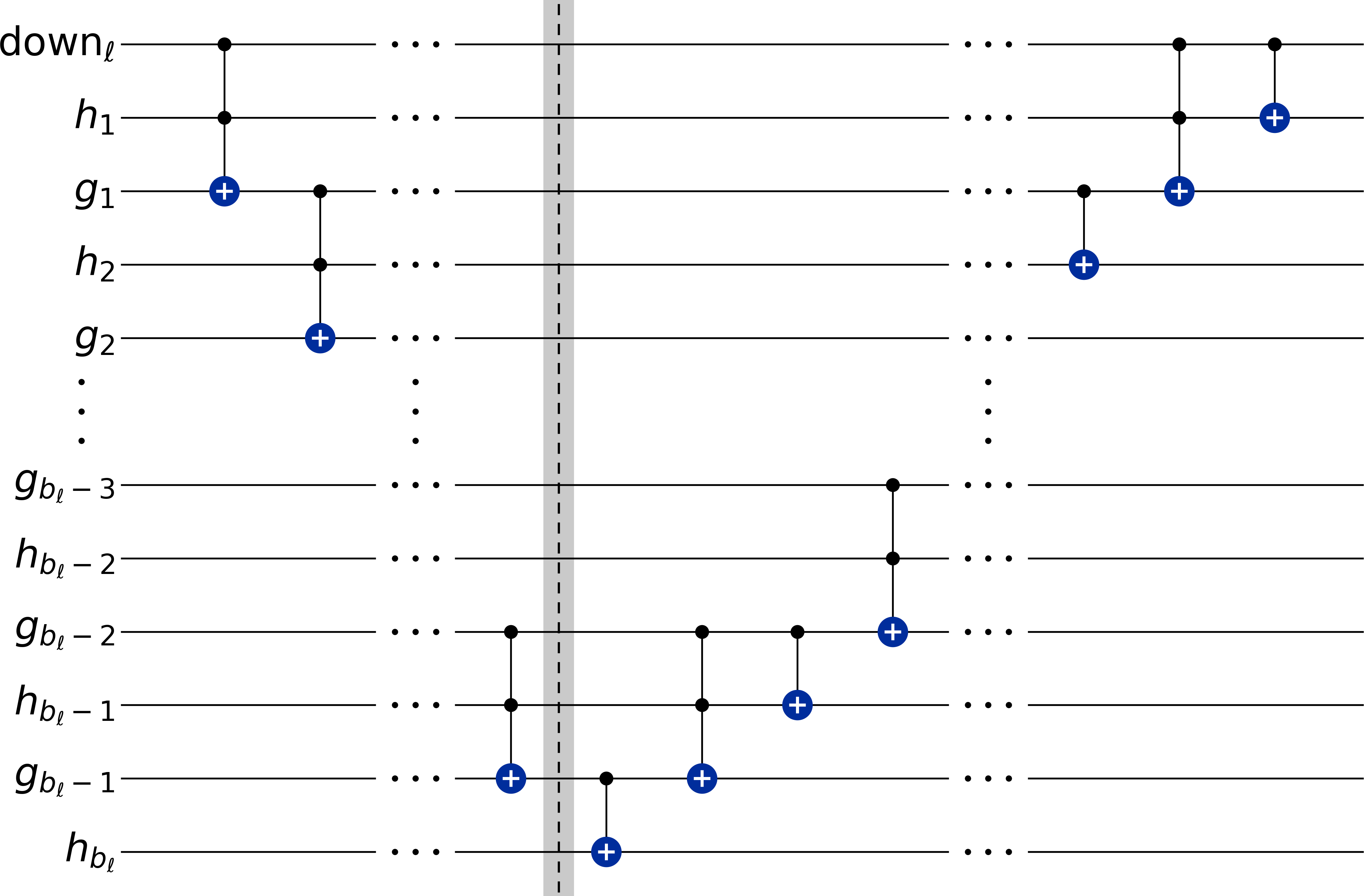}
    \caption{Implementation of the controlled +1 (increment) operator. The first part (before the barrier) of the circuit computes carry bits into temporary ancillas $g_i$. The second part (after the barrier) updates the height counter using these carry bits and uncomputes the ancillas.}
    \label{fig:increment}
\end{figure}

Implementation of the +1 operator is shown in Fig.~\ref{fig:increment}.
This is a standard binary increment circuit~\cite{vedral1996quantum, cuccaro2004new, gidney2018halving}.
It acts on the counter qubits $h_1 \dots h_{b_\ell}$, starting from the least significant one.
In the first part of the circuit (to the left of the barrier), we use a cascade of Toffoli gates to evaluate carry bits and store them in temporary ancilla qubits, labeled $g_1 \dots g_{b_\ell - 1}$.
In the second part we go backwards: the carry bits are used to invert the affected bits of the height counter, and then they are uncomputed.

The ancilla qubits $g_1 \dots g_{b_\ell - 1}$ do not need a separate register because we can always temporarily borrow them from unprepared data sites.
While the number of remaining qubits in unprepared data sites decreases as we prepare more sites, the value of $b_\ell$ decreases as well, so they always fit in.

For example, the action of ``update counter" on the $\ket{M_4; \ell = 2}'$ state, obtained after the ``down rotation" of the second iteration, would transform it to the proper 2-prefix state $\ket{M_4; \ell = 2}$ from Fig.~\ref{fig:state-tree}.

\addvspace{10pt}
\textbf{Finalize.}
By the time we reach the ``finalize" operator, the $(N - 1)$-prefix state has already been prepared.
$(N - 1)$-prefixes can only have 1 of the 2 possible heights: either 0 or 1.
If an $(N - 1)$-prefix has height 0, then its last step can only be $f$.
Since the unprepared site's qubits already encode $f$, there is nothing we need to do about it.

\begin{figure}
    \includegraphics[width=0.3\columnwidth]{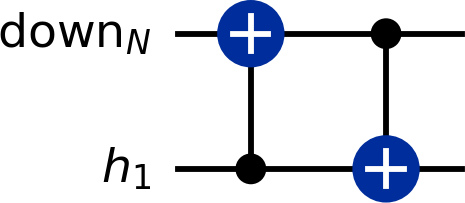}
    \caption{Implementation of the finalize operator. The last site is prepared by swapping its $\mathrm{down}_N$ qubit with the least significant bit of the height counter, which also resets the counter to zero. The swap requires only two CX gates because the $\mathrm{down}_N$ qubit is initially in the $\ket{0}$ state.}
    \label{fig:finalize}
\end{figure}

If an $(N - 1)$-prefix has height 1, then its last step can only be $d$.
In this case, we need to set its $\mathrm{down}_N$ qubit and decrease its height counter.
Both can be done simply by swapping the $\mathrm{down}_N$ and $h_1$ qubits (see Fig.~\ref{fig:finalize}).
This does not affect the height=0 branch, so no additional controls are needed.
Note that the swap can be implemented with 2 CX gates because the $\mathrm{down}_N$ qubit is guaranteed to be in the $\ket{0}$ state before the swap.

For example, in Fig.~\ref{fig:state-tree}, the ``finalize" operator transforms the 3-prefix state $\ket{M_4; \ell = 3}$ to the target state $\ket{M_4}$.

\subsection{Colored Motzkin States}
\label{sec:colored-motzkin}

\begin{figure*}
    \includegraphics[width=\linewidth]{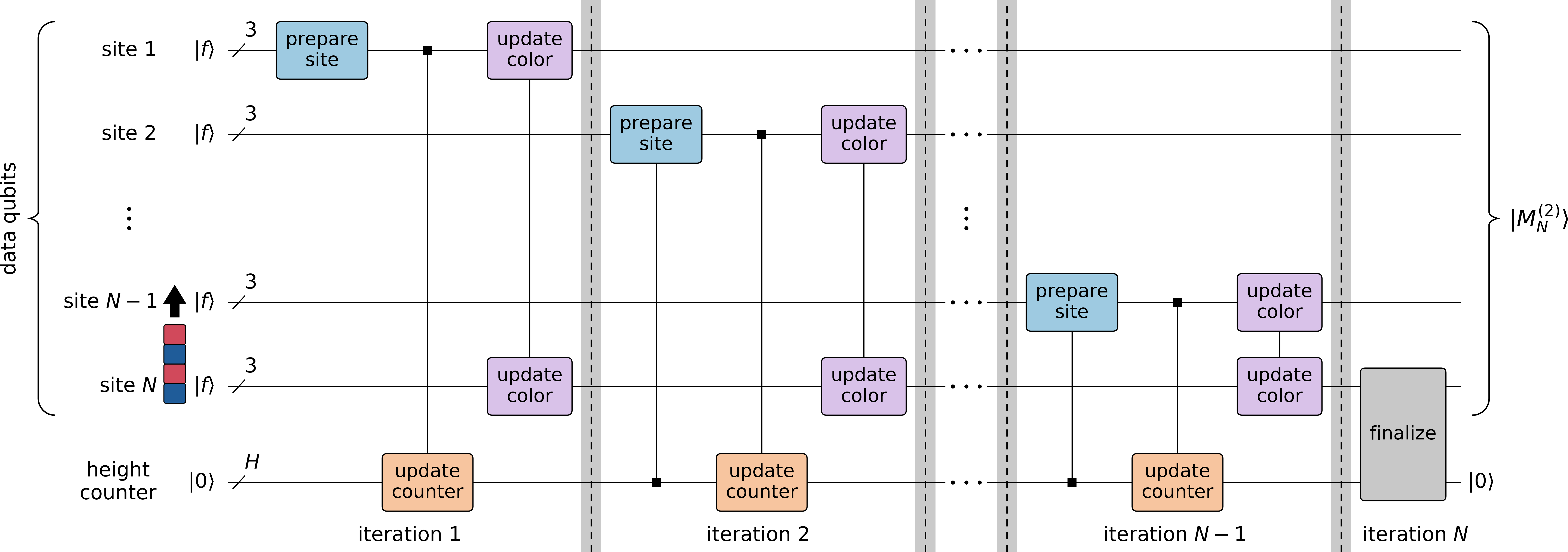}
    \caption{High-level overview of the SG-circuit preparing 2-colored $\ket{M_N^{(2)}}$. The color stack, shown with red and blue boxes, lives at the end of the data qubits, has variable size, and runs downward in qubit index. Note that the ``update color" operator interacts with the entire color stack, which fits within the last data site for the sites shown in this circuit, but can span up to $\lfloor N / 2 \rfloor$ tail data qubits in the worst case. As in Fig.~\ref{fig:overview}, kets on bundles of wires denote register states.}
    \label{fig:overview-colored}
\end{figure*}

Colored Motzkin states are prepared similarly to the uncolored Motzkin states, so here we explain only the differences relative to the uncolored state preparation, without restating the full procedure.
The high-level overview of the colored Motzkin state preparation circuit is shown in Fig.~\ref{fig:overview-colored}.

Specifically, we describe the preparation of 2-colored Motzkin states, although the procedure can be straightforwardly generalized to $k$-colored states.
The following list highlights the changes compared to the uncolored version.
\begin{enumerate}
    \item Since the up and down steps in the Motzkin paths can be of 2 colors now, we need an additional qubit in the site encoding -- the color qubit.
    So we use a total of 3 qubits per site now: up, down and color.
    As for the other two, we denote the color qubit of site $\ell$ by $\mathrm{color}_\ell$.
    The $\mathrm{color}_\ell$ qubit trivially encodes the color of the step taken in site $\ell$, i.e. 0 means color-0 and 1 means color-1.
    For flat steps the $\mathrm{color}_\ell$ qubit must be 0.

    \item Besides its height, each prefix now also needs to keep track of the colors of its unmatched up steps.
    This will be done via a stack of colors.
    Similarly to the carry ancillas for the increment operator of the height counter, the qubits for the color stack do not need dedicated ancillas and can be borrowed from the unprepared data qubits instead.
    Specifically, the stack qubits live in the tail of the data qubits, i.e. the top of the stack (the last pushed color) is stored in the last data qubit, and the colors below it run downward in qubit index.
    We denote the stack qubits by $s_1, s_2, \dots$, starting from the top of the stack, so that $s_1$ is the last data qubit.
    The total number of qubits involved in the color stack for a given $\ell$-prefix state is equal to the maximum height of its prefixes, i.e. $h_{\max}(\ell)$ (Eq.~\eqref{eq:hmax}).
    Even with the color stack, the number of unprepared data qubits at any iteration is always sufficient to accommodate both the color stack and the carry ancillas for the +1 operator.

    \item The recurrence relation of Eq.~\eqref{eq:uncolored-recurrence} acquires an additional factor of 2 in the up branch due to the fact that each step up can be of any of the 2 colors.
    The updated recurrence reads:
    \begin{equation}
        \label{eq:colored-recurrence}
        \begin{aligned}
            D_N(\ell, h) = &\ 2 D_N(\ell + 1, h + 1) \\
                           & + D_N(\ell + 1, h) + D_N(\ell + 1, h - 1),
        \end{aligned}
    \end{equation}
    with the same boundary conditions.

    Similarly, the same factor of 2 also shows up in the numerator of Eq.~\eqref{eq:prob-frac-up}, which in this case reads
    \begin{equation}
        \label{eq:prob-frac-up-colored}
        p_u(\ell, h) = \frac{2 D_N(\ell + 1, h + 1)}{D_N(\ell, h)}.
    \end{equation}

    \item The ``prepare site" operator remains exactly the same as in the uncolored case (Fig.~\ref{fig:prepare-site}), i.e. it prepares only the $\mathrm{up}_\ell$ and $\mathrm{down}_\ell$ qubits.
    Everything related to colors is done by a new ``update color" operator, which runs after the ``update counter" operator (Fig.~\ref{fig:overview-colored}).
    It prepares the $\mathrm{color}_\ell$ qubit and updates the color stack accordingly, as shown in Fig.~\ref{fig:update-color}.
    Since the maximum number of colors in the stack changes from $h_{\max}(\ell - 1)$ to $h_{\max}(\ell)$ during this step, the operator needs to act on whichever is larger, i.e. on the first
    \begin{equation}
        w_\ell = \max(h_{\max}(\ell - 1), h_{\max}(\ell))
    \end{equation}
    stack qubits.
    This operator is built around the reverse operator, which reverses the order of its $n$ input qubits as shown in Fig.~\ref{fig:reverse}.

    \begin{figure}
        \includegraphics[width=\columnwidth]{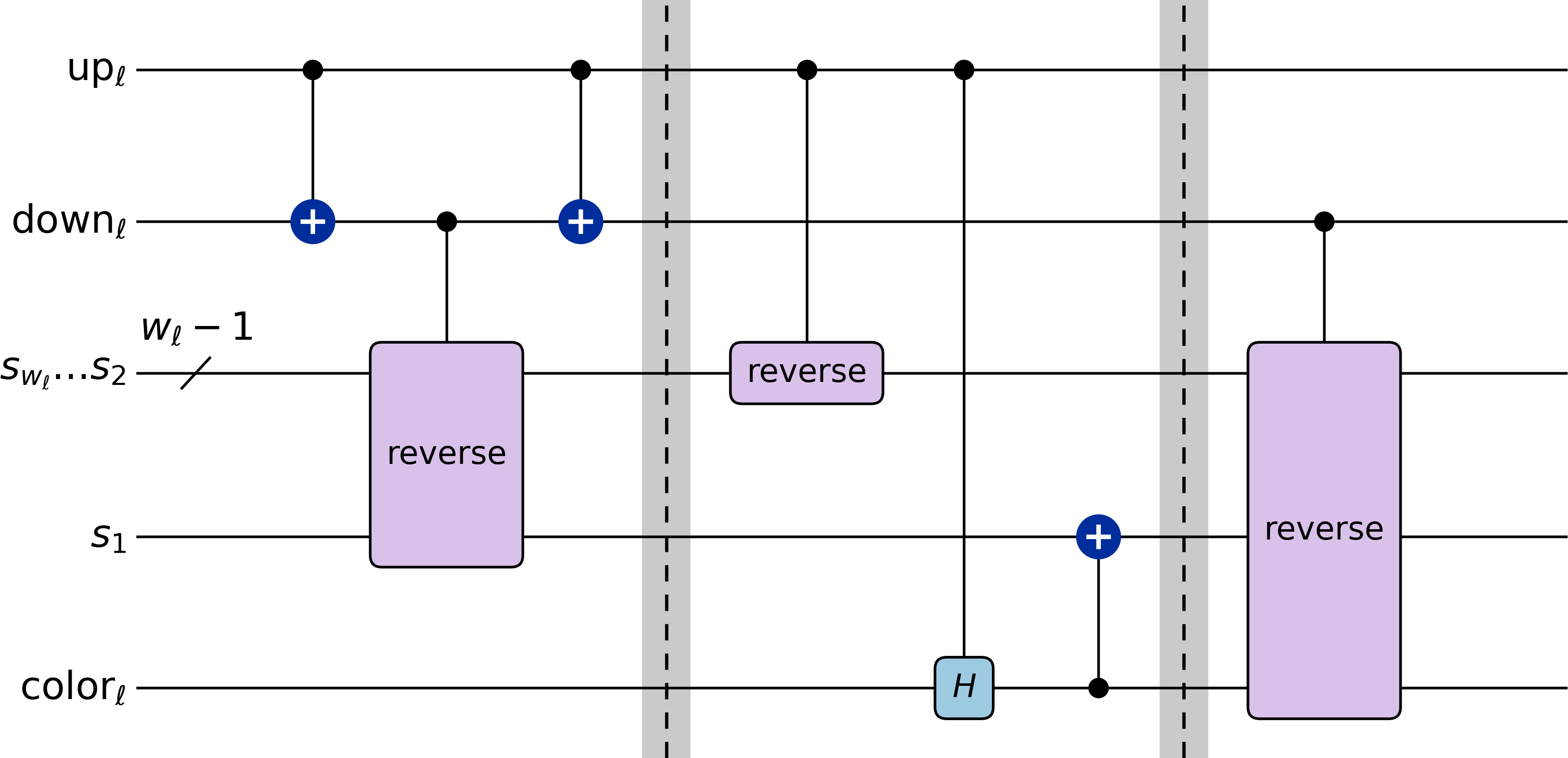}
        \caption{Implementation of the ``update color" operator. The first reversal is shared by all prefixes stepping up or down, the second reversal, the controlled Hadamard and the CX gate draw the color of the prefixes stepping up and push it onto the stack, and the third reversal pops the color of the last unmatched up step into the $\mathrm{color}_\ell$ qubit of the prefixes stepping down. Here, $s_1, \ldots, s_{w_\ell}$ denote the color stack qubits.}
        \label{fig:update-color}
    \end{figure}

    \begin{figure}
        \centering
        \includegraphics[width=0.5\columnwidth]{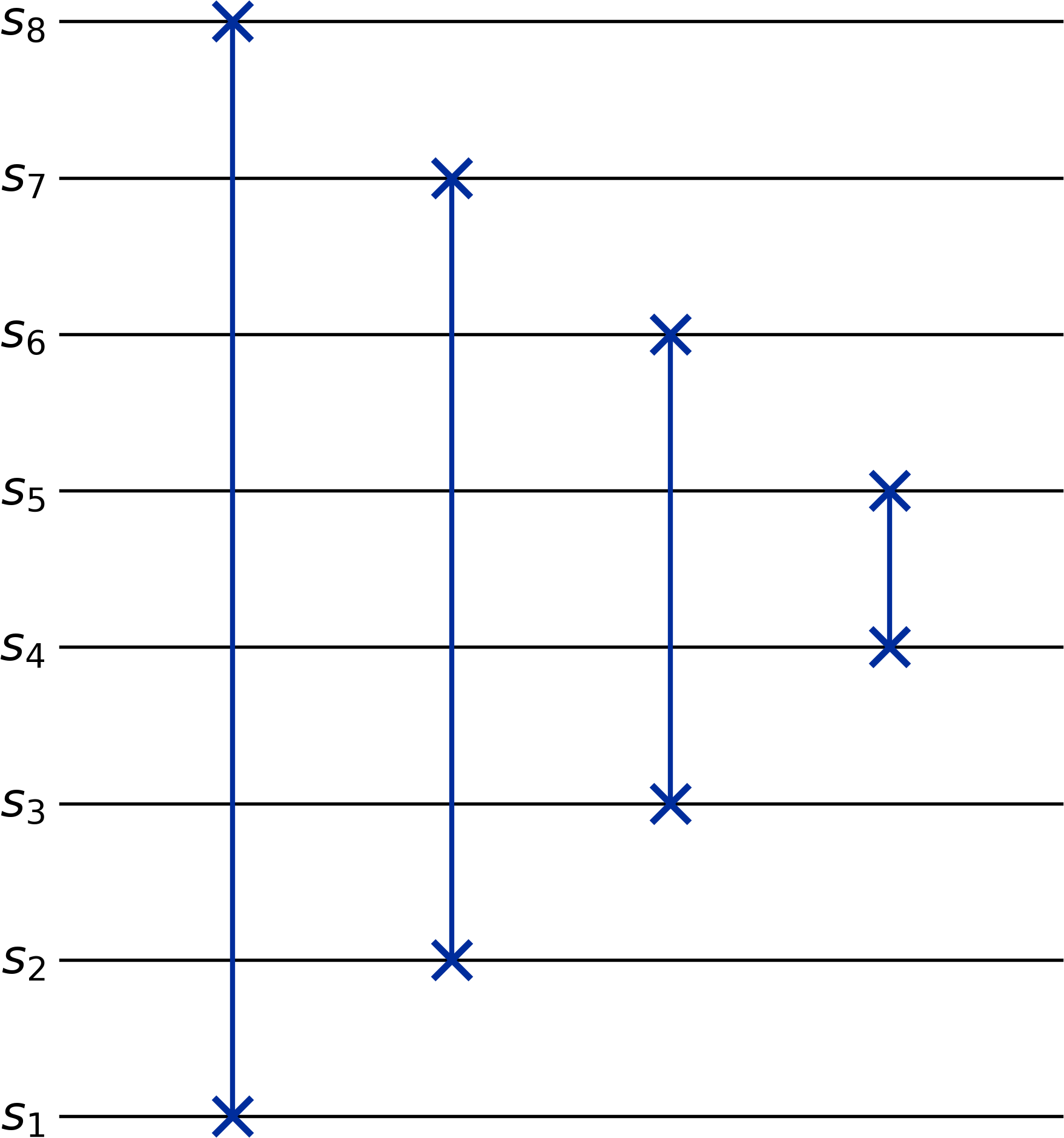}
        \caption{Implementation of the reverse operator for 8 qubits. The same construction extends to an arbitrary number of qubits.}
        \label{fig:reverse}
    \end{figure}

    The first block of Fig.~\ref{fig:update-color} reverses the entire color stack, which transforms the colors in it as $(s_{w_\ell}, \dots, s_2, s_1) \rightarrow (s_1, s_2, \dots, s_{w_\ell})$.
    This needs to be done for all prefixes that made either a step up or down in site $\ell$.
    Thus, the first reversal is controlled on $\mathrm{up}_\ell \oplus \mathrm{down}_\ell$, which is temporarily stored in the $\mathrm{down}_\ell$ qubit, similarly to the ``update counter" operator (Fig.~\ref{fig:update-counter}).

    The second block is specific to the prefixes stepping up and reverses all stack qubits other than $s_1$, transforming the stack colors as $(s_1, s_2, \dots, s_{w_\ell - 1}, s_{w_\ell}) \rightarrow (s_{w_\ell - 1}, \dots, s_2, s_1, s_{w_\ell})$.
    The height of these prefixes after this step is at most $h_{\max}(\ell)$, and we did not write the color of the last up step to the stack yet, so their stacks currently hold at most $h_{\max}(\ell) - 1 < w_\ell$ colors.
    Therefore, $s_{w_\ell}$ is guaranteed to be in the $\ket{0}$ state, and the overall effect of both reverse gates is to shift the stack down and put $\ket{0}$ on its top.
    The following Hadamard gate, controlled on the $\mathrm{up}_\ell$ qubit, creates a uniform superposition of the 2 colors in the $\mathrm{color}_\ell$ qubit, and the CX gate then writes that color to the top of the stack.
    The CX gate does not need to be controlled on the $\mathrm{up}_\ell$ qubit, since the $\mathrm{color}_\ell$ qubit is still in the $\ket{0}$ state for all other prefixes.

    The third block is specific to the prefixes stepping down and reverses the entire color stack together with the $\mathrm{color}_\ell$ qubit, which is still in the $\ket{0}$ state for these prefixes.
    This transforms the stack colors as $(s_1, s_2, \dots, s_{w_\ell} | c = 0) \rightarrow (0, s_{w_\ell}, \dots, s_2 | c = s_1)$, where $c$ is the value of the $\mathrm{color}_\ell$ qubit.
    In other words, it moves the color of the last unmatched up step to the $\mathrm{color}_\ell$ qubit, and shifts the rest of the stack up.
    After this, the color stack is up to date for all prefixes and we can start preparing the next site.

    Note that the construction of Fig.~\ref{fig:update-color} is optimized for circuit depth.
    An alternative construction that requires fewer CX gates, but results in a deeper circuit, is described in Appendix~\ref{app:opt-targets}.
\end{enumerate}

\subsection{Area-Weighted States}

The only difference for area-weighted states, compared to uniform weights, lies in the values of the classically calculated rotation angles for the ``prepare site" operator.

Specifically, since the superposition is no longer uniform, the meaning of $D_N(\ell, h)$ now generalizes from ``the number of ways in which a given prefix of length=$\ell$ and height=$h$ can be completed into a Motzkin path on $N$ sites" to the total probability weights (unnormalized) of those completions, i.e.
\begin{equation}
    D_N(\ell, h) = \sum_s t^{2A(s)},
\end{equation}
where the summation is over all valid completions $s$ of prefixes of length=$\ell$ and height=$h$ into $N$ sites.
For $t = 1$ this quantity reduces back to its original interpretation.

The corresponding recurrence relation acquires height-dependent multiplicative factors and becomes
\begin{equation}
    \label{eq:area-recurrence}
    \begin{aligned}
        D_N(\ell, h) &= t^{2h + 1} D_N(\ell + 1, h + 1) \\
                     &+ t^{2h} D_N(\ell + 1, h) \\
                     &+ t^{2h - 1} D_N(\ell + 1, h - 1),
    \end{aligned}
\end{equation}
for the uncolored case, with the same boundary conditions as in Eq.~\eqref{eq:uncolored-recurrence}.
For the 2-colored case there will also be an extra factor of 2 on the first term, similarly to Eq.~\eqref{eq:colored-recurrence}.

The explanation of the height-dependent factors is the following.
A given completion for a prefix of length=$\ell$ and height=$h$ has to start with one of the 3 possible steps: $u$, $f$, or $d$.
For the step up, the area under that step from height $h$ is $h + 0.5$, so the corresponding probability weight factor is $t^{2h + 1}$.
The total probability of all completions from site $\ell + 1$ and height $h + 1$ is $D_N(\ell + 1, h + 1)$, and extending the area of those completions by $h + 0.5$ changes their probability weight by a factor of $t^{2h + 1}$.
Similar considerations apply to the flat and down steps.

Because the corresponding probability equations are defined as ratios of recurrence terms, their expressions acquire the corresponding factors introduced by the modified recurrence relation as well.
For example, Eq.~\eqref{eq:prob-frac-up} becomes
\begin{equation}
    p_u(\ell, h) = \frac{t^{2h + 1} D_N(\ell + 1, h + 1)}{D_N(\ell, h)},
\end{equation}
and similarly for the down rotations and in the colored case.

\subsection{Fredkin States}
\label{sec:fredkin}

Fredkin states are Motzkin states minus the flat step, so each site has to step either up or down (and $N$ has to be even).
Preparation of the Fredkin states mostly follows that of the Motzkin states, except for the following differences.
\begin{enumerate}
    \item Since we only have 2 kinds of steps (up or down), we can encode the choice of the step at each site with only 1 qubit, $\mathrm{up}_\ell$ for site $\ell$, where the value of 1 encodes an up step, and the value of 0 encodes a down step.
    So in the uncolored case we only need 1 qubit per site, and in the 2-colored case, we need 2 qubits per site.
    Even with fewer qubits per site, the number of unprepared data qubits at any iteration is still sufficient to accommodate the color stack and the increment carry ancillas.

    \item The ``prepare site" operator (Fig.~\ref{fig:prepare-site}) loses the down rotation, so the probability amplitude remaining after the up rotation now encodes a down step.

    \item The recurrence relations of Eqs.~\eqref{eq:uncolored-recurrence}, \eqref{eq:colored-recurrence} and \eqref{eq:area-recurrence} lose the flat step terms ($D_N(\ell + 1, h)$).

    \item Because there is no flat step, the parity of the prefix heights is always the same as the parity of their lengths, so we can store only half of the height in the height counter, i.e. $\tilde{h} = \lfloor h / 2 \rfloor$, reducing the required number of qubits for it by 1 (and removing the gates acting on it).

    \item Storing only half of the height also changes the height arithmetic.
    If $\ell - 1$ is even, then an up step does not change $\tilde{h}$, while a down step decreases it by 1.
    If $\ell - 1$ is odd, then the opposite is true, i.e. an up step increases $\tilde{h}$ by 1, while a down step does not change it.
    Thus, at each iteration only one of the two branches needs to update the height counter, so the ``update counter" operator reduces to a single increment controlled on $\mathrm{up}_\ell = 1$ (for even $\ell$), or a single decrement (increment run in reverse) controlled on $\mathrm{up}_\ell = 0$ (for odd $\ell$), without the conditional inversions and the XOR of Fig.~\ref{fig:update-counter}.

    \item Because all $(N - 1)$-prefixes have height 1, their half height is already 0, and the down step in the last site is encoded by the default $\ket{0}$ state of its $\mathrm{up}_N$ qubit.
    Thus, the ``finalize" operator is not needed.

    \item In the colored case, the ``update color" operator (Fig.~\ref{fig:update-color}) also takes advantage of the absence of flat steps.
    Since every prefix steps either up or down, the first reversal does not need a control (and the XOR of the $\mathrm{up}_\ell$ and $\mathrm{down}_\ell$ qubits), and the third reversal is controlled on $\mathrm{up}_\ell = 0$ instead of the $\mathrm{down}_\ell$ qubit.
    Without the first control, this construction is cheaper for Fredkin states than the alternative of Appendix~\ref{app:opt-targets} in both CX count and depth.
\end{enumerate}

\subsection{Other States}
\label{sec:other-states}

What other states can we prepare with the same approach?
All circuits above follow the same pattern.
Each iteration applies an \textit{extension operator}, which extends the $\ell$-prefix state to the $(\ell + 1)$-prefix state and updates any information stored about the prefixes, such as their heights and color stacks.
The extension operator does not extend each prefix individually.
Instead, it partitions the prefixes into \textit{prefix groups}, and applies the extension to each group, multiplexed on a register holding the group label. 
Importantly, the complexity of the extension operator depends (among other things) on the number of prefix groups, but not on their sizes, since all prefixes from the same group are extended in the same way under the action of the same operator.

For example, for the Motzkin and Fredkin states considered here, the group label was the prefix height, since all prefixes of the same height have the same total weight of their completions, so they extend in the same way.
This is the key property that makes our construction efficient: even though the number of distinct prefixes grows exponentially with $N$, the number of their prefix groups grows only polynomially.

Any other state with real non-negative amplitudes can be prepared in much the same way, provided that
\begin{enumerate}
    \item The group labels of that state can be evaluated (or updated between iterations) efficiently.
    \item The extension info, i.e. the information describing how each group is extended, can be evaluated efficiently.
\end{enumerate}

In our case, the extension info was evaluated partially classically (UCR angles) and partially quantumly (color stack).
The reason for this split is that the UCR angles are shared by the entire group. 
The total number of groups at each iteration is equal to the number of different heights, so the number of angles is polynomial, and the angles themselves can be efficiently evaluated classically by exploiting the combinatorial structure of the states, so there is no reason to delegate this calculation to the quantum computer.
On the other hand, the color stack is different for each individual prefix.
Since the number of prefixes is exponential in $N$, it would not be efficient to evaluate this information classically.

The fully quantum SG approach of Ref.~\cite{grover2002creating} evaluates the extension info entirely quantumly, which removes the need for a polynomial number of prefix groups, at the cost of additional circuitry for the quantum angle evaluation.

Note that sparse states satisfy both requirements trivially, since the total number of prefixes is bounded by the number of non-zero amplitudes, consistent with their known efficient preparation~\cite{gleinig2021efficient, gonzales2025efficient, malvetti2021quantum, gaidai2026decomposition}.

\section{Circuit Complexity}
\label{sec:circuit_complexity}

\subsection{Analytical Complexity Estimate}
\label{sec:complexity_analysis}

In this section we count the CX gates in each operator of the SG-circuits and derive the leading terms of the overall CX counts and circuit depth.

\addvspace{10pt}
\textbf{Uncolored Motzkin states.}
As shown in Fig.~\ref{fig:overview}, the circuit consists of $N$ iterations.
Each of the first $N - 1$ iterations is made of a ``prepare site" and an ``update counter" operator, and the last one of a single ``finalize" operator.

The ``prepare site" operator (Fig.~\ref{fig:prepare-site}) consists of two UCRs.
A simplified implementation of a UCR is shown in Fig.~\ref{fig:rotation-recursion}.
In it, each recursion step consists of 2 CX gates and 2 more UCRs multiplexed on one fewer qubit.
Thus, for that version the number of CX gates from all recursion levels is given by $\sum_{i = 1}^{H} 2^i = 2^{H + 1} - 2 \approx 2^{H + 1}$ CX gates, i.e. approximately 2 CX gates per angle.

The UCR implementation used in our code is more efficient and requires approximately 1 CX gate per angle, counting only the angles reachable by $h_{\max}(\ell - 1)$.
The up rotation uses one angle per height of the $(\ell - 1)$-prefixes, so it requires $\approx h_{\max}(\ell - 1)$ CX gates.
The down rotation is additionally multiplexed on the $\mathrm{up}_\ell$ qubit, which doubles the number of its angles, so it requires $\approx 2 h_{\max}(\ell - 1)$ CX gates.
Together, the ``prepare site" operator requires $\approx 3 h_{\max}(\ell - 1) \approx 3 h_{\max}(\ell) = O(N)$ CX gates at iteration $\ell$.

The ``update counter" operator (Figs.~\ref{fig:update-counter} and \ref{fig:increment}) is made up of $O(b_\ell) = O(\log N)$ CX and CCX gates, so its contribution is negligible compared to the ``prepare site" operator.
The ``finalize" operator (Fig.~\ref{fig:finalize}) adds only 2 CX gates, so its cost is also negligible.

The total CX count is obtained by summing the dominant iteration cost (from ``prepare site") over the first $N - 1$ iterations.
Since $h_{\max}(\ell)$ grows linearly up to $N / 2$ in the middle of the chain and then decreases back to 0, its average over the iterations is approximately $N / 4$, i.e.
\begin{equation}
    \sum_{\ell = 1}^{N - 1} h_{\max}(\ell) = \sum_{\ell = 1}^{N - 1} \min(\ell, N - \ell) = \left\lfloor \frac{N^2}{4} \right\rfloor,
    \label{eq:hmax-sum}
\end{equation}
so the total CX count is $\approx 3 N^2 / 4$.

Unlike the CX count, the circuit depth depends on the choice of the gate set.
We count it in the gate set of CX, $R_z$, $\sqrt{X}$ and $X$ gates, i.e. the native single-qubit gates of IBM superconducting processors~\cite{ibm2026backend} with CX as the two-qubit gate (their native ECR and CZ gates are equivalent to CX up to single-qubit gates).
In this gate set, each $R_y$ rotation becomes 5 single-qubit gates.
All CX gates of a UCR act on the same target qubit, alternating with its $R_y$ rotations, so they cannot be parallelized, and each angle adds 6 layers to the depth.
The down rotation is multiplexed on the $\mathrm{up}_\ell$ qubit, so it has to wait for the up rotation, and the ``prepare site" operator of the next iteration has to wait for the updated counter.
Thus, the rotations of all iterations are executed sequentially, which gives $\approx 6 \cdot 3 h_{\max}(\ell) = 18 h_{\max}(\ell)$ layers per iteration and a total depth of $\approx 9 N^2 / 2$.

\addvspace{10pt}
\textbf{Colored Motzkin states.}
As shown in Fig.~\ref{fig:overview-colored}, each of the first $N - 1$ iterations is made up of the same ``prepare site" and ``update counter" operators, followed by the new ``update color" operator.

The ``update color" operator (Fig.~\ref{fig:update-color}) is made up of 3 conditional reverse operators, and a constant number of other 2-qubit gates, whose contribution is negligible.

The reverse operators act on $w_\ell$, $w_\ell - 1$ and $w_\ell + 1$ qubits, where $w_\ell \approx h_{\max}(\ell)$, and each needs half as many swaps as qubits, so they require $\approx 3 h_{\max}(\ell) / 2$ controlled swaps in total.
Each controlled swap requires a Toffoli gate between 2 CX gates, i.e. 8 CX gates per swap.
Thus the ``update color" operator requires a total of $\approx 12 h_{\max}(\ell)$ CX gates.

Together with the $3 h_{\max}(\ell)$ from the site preparation, one iteration in the 2-colored case therefore requires $\approx 15 h_{\max}(\ell)$ CX gates.
Using Eq.~\eqref{eq:hmax-sum}, the total CX count is therefore $\approx 15 N^2 / 4$.
Thus, maintenance of the color stack is responsible for about $12/15 = 80\%$ of the overall CX count, asymptotically.

The circuit depth, in contrast, is the same as in the uncolored case, to leading order.
The ``update color" operator of iteration $\ell$ acts on different qubits than the ``prepare site" operator of iteration $\ell + 1$, so the two run in parallel, and the overall depth is set by the deeper of the two sequences of operators.

All swaps of a reversal act on disjoint pairs of qubits, so only their gates acting on the shared control qubit have to be executed sequentially.
In the standard decomposition of the Toffoli gate, the gates acting on its control span 8 layers, so each controlled swap adds 8 layers to the depth of a reversal, and the 3 reversals take at most $\approx 8 \cdot 3 h_{\max}(\ell) / 2 = 12 h_{\max}(\ell)$ layers per iteration.
This is less than the $18 h_{\max}(\ell)$ layers of the ``prepare site" operator, so the color stack updates do not contribute to the leading-order depth at all, which remains $\approx 9 N^2 / 2$.

\addvspace{10pt}
\textbf{Area-weighted states.}
Area-weighted circuits differ from the uniform ones only in the rotation angles, so they have the same structure, CX count and depth, up to angle-dependent pruning of the UCRs.

\addvspace{10pt}
\textbf{Fredkin states.}
The Fredkin circuits consist of the same operators, with the modifications listed in Sec.~\ref{sec:fredkin}.
The ``prepare site" operator loses the down rotation and its associated $2 h_{\max}(\ell - 1)$ CX cost.
In addition, the height of a prefix has the same parity as its length, so only every second height is reachable, and the height counter stores only half of the height.
The up rotation then uses one angle per half height, so it requires only $\approx h_{\max}(\ell) / 2$ CX gates.

The ``update counter" operator remains lower-order, as in the Motzkin case, and the ``finalize" operator is not needed at all.
One uncolored iteration therefore requires $\approx h_{\max}(\ell) / 2$ CX gates, and the total CX count of the uncolored Fredkin circuit is $\approx N^2 / 8$, i.e. $\approx 6$ times smaller than that of the uncolored Motzkin circuit.
Similarly, each iteration adds $\approx 6 \cdot h_{\max}(\ell) / 2 = 3 h_{\max}(\ell)$ layers to the depth, so the total depth is $\approx 3 N^2 / 4$.

The color stack, in contrast, is not reduced, since it has to store the colors of all unmatched up steps, whose number can still reach $h_{\max}(\ell)$.
The colored Fredkin circuit uses the same ``update color" operator (Fig.~\ref{fig:update-color}), except that its first reversal does not need a control.
The uncontrolled reversal acts on the $w_\ell \approx h_{\max}(\ell)$ qubits of the color stack, so it requires $\approx h_{\max}(\ell) / 2$ swaps of 3 CX gates each.
The two controlled reversals require $\approx h_{\max}(\ell)$ controlled swaps of 8 CX gates each, as in the Motzkin case.
Together they require $\approx 3 h_{\max}(\ell) / 2 + 8 h_{\max}(\ell) = 19 h_{\max}(\ell) / 2$ CX gates.
With the $h_{\max}(\ell) / 2$ CX from site preparation, one iteration therefore requires $\approx 10 h_{\max}(\ell)$ CX gates, and the total CX count is $\approx 5 N^2 / 2$.
As one can see, 95\% of those gates are spent on color stack updates, which makes it by far the heaviest operation for both state families.
Because of this, the two colored circuits are much closer to each other, despite large differences in the uncolored circuits.

Unlike in the Motzkin case, the color stack updates also set the depth of the colored Fredkin circuit.
The swaps of the uncontrolled reversal run in parallel, but the two controlled reversals share the same control (the $\mathrm{up}_\ell$ qubit), so they take $\approx 8 h_{\max}(\ell)$ layers per iteration, which exceeds the $3 h_{\max}(\ell)$ layers of the ``prepare site" operator.
The total depth is therefore $\approx 2 N^2$.

\addvspace{10pt}
\textbf{Summary.}
Overall, the complexity of all state preparation circuits considered here scales as $O(N^2)$ asymptotically, which offers a significant advantage over MPS-based preparation of the colored chains, where the circuit depth and CX gate count are lower-bounded by the Schmidt rank (for generic site tensors~\cite{schon2005sequential}), which grows exponentially with system size for colored chains.

The prefactors of $N^2$ derived here are summarized in the table below.

{\setlength{\tabcolsep}{6pt}
\renewcommand{\arraystretch}{1.2}
\begin{center}
    \begin{tabular}{|c|c|c|c|c|}
        \hline
        \multirow{2}{*}{State}& \multicolumn{2}{c|}{Motzkin}& \multicolumn{2}{c|}{Fredkin} \\
        \cline{2-5}&   CX&   Depth& CX&    Depth \\\hline
        Uncolored&     0.75& 4.5&   0.125& 0.75  \\\hline
        2-colored&     3.75& 4.5&   2.5&   2     \\\hline
        Area-weighted& 3.75& 4.5&   2.5&   2     \\\hline
    \end{tabular}
\end{center}}

Note that these prefactors are derived for the unoptimized circuits.
Transpiler optimizations reduce these numbers further, but the exact degree of reduction is transpiler-dependent and not analyzed here.
However, the overall scaling and relative order are not expected to be affected by the optimizations.

In addition, the circuits require
\begin{itemize}
    \item $2N$ (uncolored) or $3N$ (colored) data qubits for the Motzkin states, and $N$ or $2N$ for the Fredkin states.
    \item $\lfloor \log_2 \lfloor N/2 \rfloor \rfloor + 1$ (Motzkin) or $\lfloor \log_2 \lfloor N/4 \rfloor \rfloor + 1$ (Fredkin) ancilla qubits for the height register.
    An ancilla-free version is also possible, but requires more CX gates and larger depth (see Appendix~\ref{app:opt-targets}).
    \item $O(N^2)$ classical time and space for the evaluation of the UCR angles.
\end{itemize}

\addvspace{10pt}
\textbf{Other states.}
The same accounting applies to any state prepared with the SG approach.
Let $G_\ell$ denote the number of prefix groups at iteration $\ell$.
In general, each prefix group requires its own rotation angle.
If the groups are identified by a label register of $\lceil \log_2 G_\ell \rceil$ qubits, the extension of each group can be implemented with a constant number of UCRs multiplexed on this register, which requires $O(G_\ell)$ CX gates.
This cost is difficult to avoid within the SG approach, which makes the total number of prefix groups $\sum_\ell G_\ell$ a natural lower bound on the overall complexity.

In addition, each iteration has to update the label register and the quantum part of the extension info, at some cost $U_\ell$ that depends on how much information about a prefix is stored and how difficult it is to update it.
Not much can be said about this cost for a general state.

The total CX count is therefore given by
\begin{equation}
    \sum_{\ell = 1}^{N} \bigl[ O(G_\ell) + U_\ell \bigr],
\end{equation}
which is polynomial whenever both the number of prefix groups and the cost of the updates are polynomial in $N$.

\subsection{Numerical Benchmarks}
\label{sec:complexity}

\begin{figure}
    \includegraphics[width=\columnwidth]{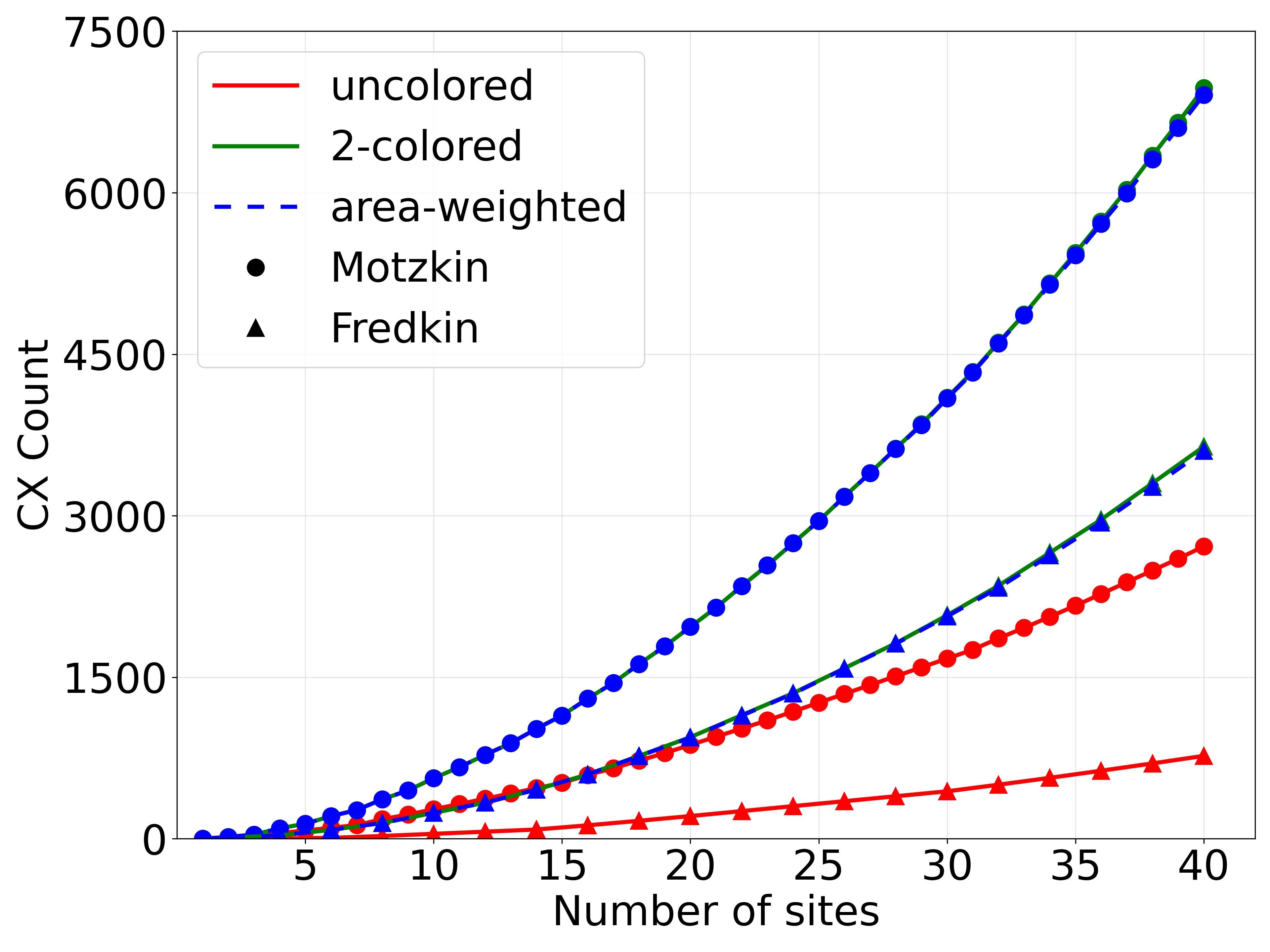}
    \includegraphics[width=\columnwidth]{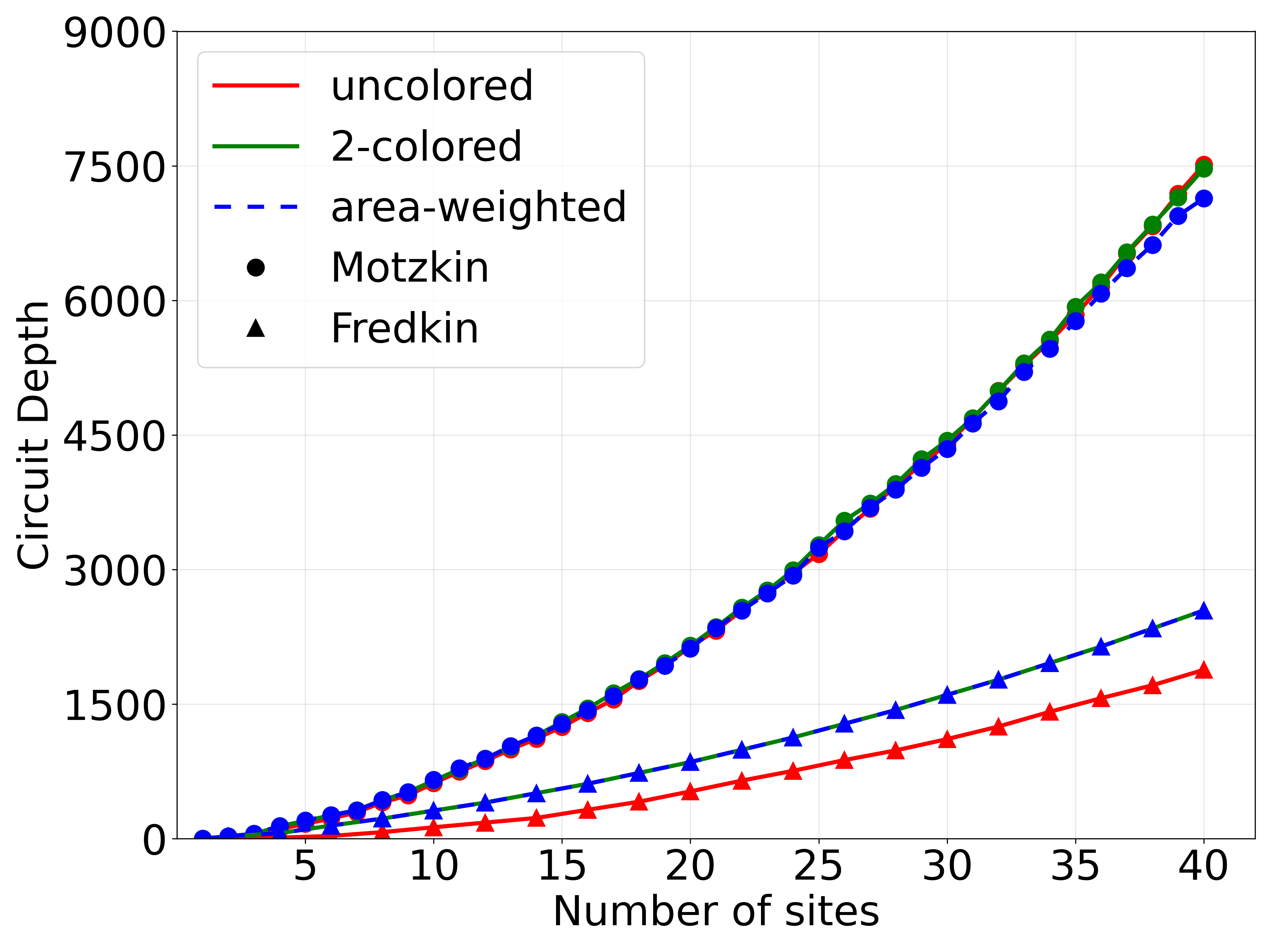}
    \caption{CX count (top) and depth (bottom) of the transpiled circuits preparing uncolored (red), 2-colored (green) and area-weighted 2-colored with $t = 2$ (blue) Motzkin and Fredkin states vs the number of sites $N$. The 2-colored and area-weighted curves nearly overlap. In the depth plot, the 2-colored and area-weighted Motzkin curves also nearly overlap with the uncolored one, since the color updates run in parallel with the rotations (Sec.~\ref{sec:complexity_analysis}).}
    \label{fig:resources}
\end{figure}

The exact number of CX gates and depth of the \textit{transpiled} circuits preparing uncolored, 2-colored and area-weighted 2-colored (with $t = 2$) Motzkin and Fredkin states for $N \le 40$ are shown in Fig.~\ref{fig:resources}.
Transpilation has been performed with \textit{qiskit} 2.5.0 at optimization level 3, to the gate set of CX, $R_z$, $\sqrt{X}$ and $X$ gates.

The area-weighted circuit differs only in the values of the rotation angles.
As such, it follows the results of the 2-colored circuit very closely.
Slight differences that can be noticed around $N \approx 40$ are due to some angle-dependent circuit optimizations.

\subsection{Experimental Feasibility}
The numbers shown in Fig.~\ref{fig:resources} assume all-to-all connectivity, which is needed because during the construction all sites access both the height counter and the color stack.
In practice, if the connectivity of a given quantum hardware is limited, there will be an additional overhead due to the swap networks implementing long-range interactions, which can increase the required resources roughly by a factor of 2--4 for such hardware architectures.
For this reason, all-to-all connectivity will be desirable for practical implementations of Motzkin chains, as in the trapped-ion architectures of Quantinuum~\cite{ransford202698} and IonQ~\cite{wright2019benchmarking}, or in the neutral-atom architectures of QuEra~\cite{bluvstein2022quantum, bluvstein2024logical} and Infleqtion~\cite{rines2025demonstration}, where it is provided by the coherent transport of atoms.

\section{Conclusions and Outlook}

We have presented an algorithm for the exact preparation of Motzkin and Fredkin states, in both their uncolored and colored variants, on a digital quantum computer with circuit depth and CX gate count that scale polynomially with system size.
The underlying principle is not specific to Motzkin and Fredkin walks, and other families of constrained states with similar structure can be prepared efficiently in the same way.
Our algorithm therefore provides access to a regime of strongly entangled ground states whose dynamics lies beyond the reach of classical tensor-network methods~\cite{schuch2008entropy, orus2014practical}, enabling quantum simulation of this dynamics~\cite{lloyd1996universal, georgescu2014quantum, atallah2026matching}, Haldane phases~\cite{barbiero2017haldane}, and holographic correspondence~\cite{alexander2019exact, alexander2021exact}.
A natural next step would be to investigate the fault-tolerant implementation of our protocol on quantum hardware, including the associated error-correction requirements, resource overheads, and the possibility of optimizing for different targets, such as the T-gate count~\cite{bravyi2005universal, gosset2026quantum}.

Finally, the states prepared here offer a natural testbed for entanglement measurement.
A central challenge in measuring entanglement lies in accessing the von Neumann entropy $S_A$, which for pure bipartite states equals the rate at which Bell pairs can be distilled by local operations and classical communication~\cite{Bennett1996, horodecki2009quantum}.
While $S_A$ generally requires full state tomography, the second R\'enyi entropy $S_2$ can be efficiently measured via two-copy interference~\cite{Daley2012, Islam2015} or randomized measurements~\cite{elben2018renyi, brydges2019probing}.
However, for some states, $S_2$ provides a poor proxy for $S_A$.
For instance, in the case of the colored Motzkin chain, $S_A$ grows as $\sqrt{N}$ while all R\'enyi entropies with $\alpha > 1$ grow only logarithmically~\cite{movassagh2016supercritical, Sugino2018}.
Recent work~\cite{Barghathi2026} showed that symmetry-resolved R\'enyi entropies, measurable by combining magnetization-resolved readout with two-copy or randomized protocols~\cite{Lukin2019}, yield a lower bound on $S_A$ that detects this anomalous scaling.
Together with the polynomial-depth preparation protocol presented here, these developments open a path toward the experimental realization of Motzkin and Fredkin states exhibiting power-law violations of the entanglement area law, starting with small chains on current quantum hardware.

\section*{Declarations}

\subsection*{Availability of Data and Materials}
For the exact code, finer technical details, and the raw data supporting the findings of this study, readers are invited to inspect our code archived on Zenodo~\cite{gaidai2026motzkin}.

\subsection*{Use of Generative Artificial Intelligence}
The authors used generative artificial intelligence (AI) tools to assist with code development and debugging, study design and methodology, literature-review organization, and manuscript drafting and editing.
All scientific decisions, code, analyses, interpretations, conclusions, and final manuscript text were reviewed, verified, and approved by the authors.

\subsection*{Acknowledgments}

I.G., K.M. and R.M. acknowledge support from the U.S. National Institute of Standards and Technology (NIST) through the CIPP program under Award No. 60NANB24D218.
R.M. also acknowledges support from the U.S. National Science Foundation (NSF) through the NSF TIP program under Award No. 2534232.
T.S. and P.S. acknowledge funding from the Federal Ministry of Research, Technology and Space under Project No. 13N17157 (Q-ROM).
This research was partially supported by the National Science Foundation Materials Research Science and Engineering Center program through the UT Knoxville Center for Advanced Materials and Manufacturing (DMR-2309083).

\clearpage
\appendix
\renewcommand{\thefigure}{\thesection\arabic{figure}}
\counterwithin*{figure}{section}
\renewcommand{\thetable}{\thesection\arabic{table}}
\counterwithin*{table}{section}

\section{Construction of the Reduced Density Matrix from Constrained Paths}
\label{app:rdm_construction}

\begin{figure*}
    \centering
    \includegraphics[width=1\linewidth]{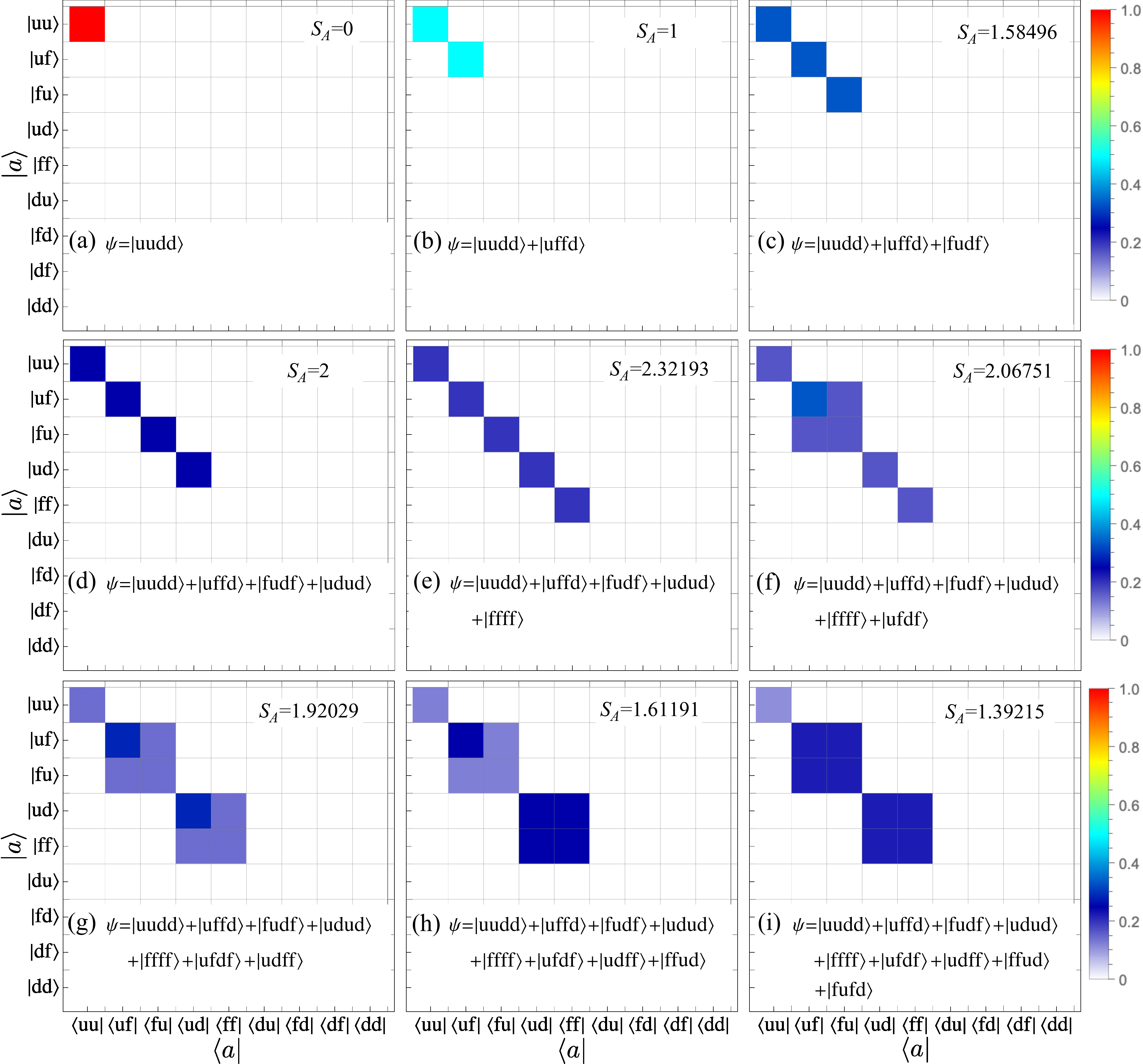}
    \caption{Reduced density matrix $\rho_A$ of equal-weight superpositions of constrained paths on an $N=4$ chain, for the bipartition into sites 1--2 ($A$) and sites 3--4 ($B$). Rows and columns are the nine basis states $\ket{a}$ of subsystem $A$. Shaded cells are non-zero matrix elements, and color intensity gives their magnitude. The entanglement entropy $S_A$ is given in each panel. Panels (a)--(i) add the Motzkin paths one at a time, from the product state $\ket{uudd}$ in (a) to the full Motzkin state in (i).}
    \label{fig:rdm}
\end{figure*}

In this section we describe how the reduced density matrix (RDM) of a Motzkin state follows directly from the set of allowed path configurations, and how adding or removing configurations modifies its block structure and entanglement entropy.
We illustrate the construction with the $N=4$ chain bipartitioned into a left half $A$ (sites 1--2) and a right half $B$ (sites 3--4), as shown in Fig.~\ref{fig:rdm}.

\subsection{General Construction}

Each site carries a local state $\ket{u}$, $\ket{f}$, or $\ket{d}$, corresponding to an up, flat, or down step of a lattice path.
A configuration $p$ is an allowed Motzkin path if its heights (see Eq.~\eqref{eq:area}) satisfy $h_j(p)\ge 0$ for all $j$ and $h_N(p)=0$.
We consider equal-weight superpositions over a set $\mathcal{P}$ of allowed paths,
\begin{equation}
    \ket{\psi}=\frac{1}{\sqrt{Z}}\sum_{p\in\mathcal{P}}\ket{p} ,
    \qquad Z=|\mathcal{P}| ,
\end{equation}
so that the Motzkin state corresponds to $\mathcal{P}=M_N$ being the full set of $|M_N|$ uncolored Motzkin paths.

Writing each path as a concatenation $p=a\,b$ of an $N/2$-prefix $a$ on $A$ and an $N/2$-suffix $b$ on $B$, we define for every prefix $a$ the set of suffixes that complete it,
\begin{equation}
    \mathcal{B}(a)=\{\,b \;:\; a\,b\in\mathcal{P}\,\}.
\end{equation}
Tracing out $B$ then gives
\begin{equation}
    \left(\rho_A\right)_{a a'}
    =\frac{1}{Z}\sum_{b}\bra{a b}\,\sum_{p,p'}\ket{p}\!\bra{p'}\,\ket{a' b}
    =\frac{1}{Z}\,\bigl|\mathcal{B}(a)\cap\mathcal{B}(a')\bigr| .
    \label{eq:rdm_overlap}
\end{equation}
Equation~\eqref{eq:rdm_overlap} contains the entire construction: a diagonal element counts how many allowed paths share the prefix $a$, and an off-diagonal element counts how many suffixes are shared by two different prefixes $a$ and $a'$.
Two properties follow immediately.
\begin{enumerate}
    \item Forbidden rows and columns.
    If $a$ cannot be the beginning of any allowed path, $\mathcal{B}(a)=\emptyset$ and the corresponding row and column of $\rho_A$ vanish.
    For $N=4$ the prefixes $\ket{du},\ket{fd},\ket{df},\ket{dd}$ take the height below zero and are therefore always absent, so only the five states $\ket{uu},\ket{uf},\ket{fu},\ket{ud},\ket{ff}$ can appear.

    \item Height-sector block structure.
    A suffix $b$ that completes $a$ must start at height $h_{N/2}(a)$ and return to zero.
    Hence $\mathcal{B}(a)\cap\mathcal{B}(a')=\emptyset$ whenever $h_{N/2}(a)\neq h_{N/2}(a')$, and $\rho_A$ is block diagonal in the height at the cut,
    \begin{equation}
        \rho_A=\bigoplus_{h}\rho_A^{(h)} .
    \end{equation}
    For $N=4$ the sectors are $h=2$: $\{\ket{uu}\}$, $h=1$: $\{\ket{uf},\ket{fu}\}$, and $h=0$: $\{\ket{ud},\ket{ff}\}$, which fixes the $1\times1$, $2\times2$, $2\times2$ block pattern seen in Fig.~\ref{fig:rdm}.
\end{enumerate}

Within a sector, the block is the Gram matrix of the sets $\mathcal{B}(a)$.
For the complete Motzkin (or Fredkin) state, the height is the only information that couples $A$ and $B$: every prefix ending at height $h$ can be completed by every suffix starting at $h$.
All sets $\mathcal{B}(a)$ within a sector are then identical, each block has all entries equal, and it is rank one,
\begin{equation}
    \begin{aligned}
        \rho_A^{(h)}&=p_h\,\proj{\chi_h} ,\\
        \ket{\chi_h}&=\frac{1}{\sqrt{W_h}}\sum_{a:\,h_{N/2}(a)=h}\ket{a} ,\\
        p_h&=\frac{W_h^2}{|M_N|} ,
    \end{aligned}
    \label{eq:rank_one}
\end{equation}
where $W_h$ is the number of $N/2$-prefixes that stay non-negative and end at height $h$.
By reflection, the number of $N/2$-suffixes from height $h$ back to zero is also $W_h$.
The Schmidt rank therefore equals the number of accessible heights, and the entanglement entropy is $S_A=-\sum_h p_h\log_2 p_h$.

\subsection{Building the RDM Path by Path}
\label{app:building_rdm}
Figure~\ref{fig:rdm} shows how $\rho_A$ changes as the nine Motzkin paths of length four are added one at a time, from the product state $\ket{uudd}$ in panel~(a) to the full Motzkin state in panel~(i).
The prefixes and suffixes of each path are listed in Table~\ref{tab:rdm_paths}.

\begin{table}
    \centering
    \begin{tabular}{c c c c c}
        \hline\hline
        Panel & Added path & $a$ (on $A$) & $b$ (on $B$) & $h_2$ \\
        \hline
        (a) & $\ket{uudd}$ & $uu$ & $dd$ & 2 \\
        (b) & $\ket{uffd}$ & $uf$ & $fd$ & 1 \\
        (c) & $\ket{fudf}$ & $fu$ & $df$ & 1 \\
        (d) & $\ket{udud}$ & $ud$ & $ud$ & 0 \\
        (e) & $\ket{ffff}$ & $ff$ & $ff$ & 0 \\
        (f) & $\ket{ufdf}$ & $uf$ & $df$ & 1 \\
        (g) & $\ket{udff}$ & $ud$ & $ff$ & 0 \\
        (h) & $\ket{ffud}$ & $ff$ & $ud$ & 0 \\
        (i) & $\ket{fufd}$ & $fu$ & $fd$ & 1 \\
        \hline\hline
    \end{tabular}
    \caption{Decomposition of the $N=4$ Motzkin paths into prefixes and suffixes for the bipartition of Fig.~\ref{fig:rdm}, in the order in which they are added. $h_2$ is the height at the cut.}
    \label{tab:rdm_paths}
\end{table}

\emph{Panels (a)--(e): diagonal growth.}
In the first five steps every new path brings both a new prefix and a new suffix that has not appeared before.
All sets $\mathcal{B}(a)$ are then disjoint, so Eq.~\eqref{eq:rdm_overlap} gives no off-diagonal elements, and $\rho_A$ is diagonal with $Z$ equal entries $1/Z$ after $Z$ paths.
The entropy grows as $S_A=\log_2 Z$: $0$, $1$, $\log_2 3\simeq1.585$, $2$, and $\log_2 5\simeq2.322$.
At this stage the state is maximally entangled within the five-dimensional support of $\rho_A$, even though it contains only a subset of the Motzkin paths.

\emph{Panel (f): first coherence in the $h=1$ sector.}
Adding $\ket{ufdf}$ reuses the suffix $df$, already paired with $fu$ in $\ket{fudf}$.
The overlap $\mathcal{B}(uf)\cap\mathcal{B}(fu)=\{df\}$ produces an off-diagonal element between $\ket{uf}$ and $\ket{fu}$, and because $\ket{uf}$ now has two partners ($\mathcal{B}(uf)=\{fd,df\}$) its diagonal weight is doubled,
\begin{equation}
    \rho_A^{(1)}=\frac{1}{6}\begin{pmatrix}2&1\\1&1\end{pmatrix}.
\end{equation}
This block has eigenvalues $(3\pm\sqrt5)/12$.
The spectrum becomes non-uniform and the entropy drops to $S_A\simeq2.068$ even though the number of paths has grown.
This is the key mechanism: an added path whose prefix and suffix have both appeared before adds no new row or column to $\rho_A$.
Instead, it adds coherence between existing prefixes.

\emph{Panel (g): first coherence in the $h=0$ sector.}
Adding $\ket{udff}$ pairs $ud$ with the suffix $ff$ already used by $\ket{ffff}$, which creates the analogous non-uniform block $\tfrac17\left(\begin{smallmatrix}2&1\\1&1\end{smallmatrix}\right)$ in the $\{\ket{ud},\ket{ff}\}$ sector, and $S_A\simeq1.920$.

\emph{Panels (h) and (i): completion of the blocks.}
Adding $\ket{ffud}$ gives $\mathcal{B}(ud)=\mathcal{B}(ff)=\{ud,ff\}$.
The $h=0$ block becomes $\tfrac{2}{8}\left(\begin{smallmatrix}1&1\\1&1\end{smallmatrix}\right)$, which is rank one with eigenvalue $1/2$, while the $h=1$ block is still incomplete, and $S_A\simeq1.612$.
Finally, $\ket{fufd}$ supplies the missing combination, $\mathcal{B}(uf)=\mathcal{B}(fu)=\{fd,df\}$, so the $h=1$ block also becomes uniform and rank one.
Panel~(i) reproduces Eq.~\eqref{eq:rank_one} with $W_0=W_1=2$, $W_2=1$, and $|M_4|=9$:
\begin{equation}
    \begin{gathered}
        \{p_0,p_1,p_2\}=\left\{\tfrac49,\tfrac49,\tfrac19\right\},\\
        S_A=\tfrac89\log_2\tfrac94+\tfrac19\log_2 9\simeq1.392 .
    \end{gathered}
\end{equation}

The intermediate panels therefore interpolate between two limits.
When no two paths share a prefix or a suffix, the RDM is diagonal and each path contributes an independent Schmidt vector.
When the path set is closed under exchanging suffixes of equal height, as for the Motzkin state, each height sector collapses to a single Schmidt vector.
Removing a configuration from the Motzkin set breaks this closure: it lowers the diagonal weight of the affected prefixes and reduces the corresponding off-diagonal elements, which, in a sector with more than one prefix, raises the rank of that height block and makes its spectrum non-uniform.

\section{Estimating Preparation Cost from Reduced Density Matrix (RDM) Structure}
\label{app:rdm_complexity}
\begin{figure}[b]
    \includegraphics[width=0.99\columnwidth]{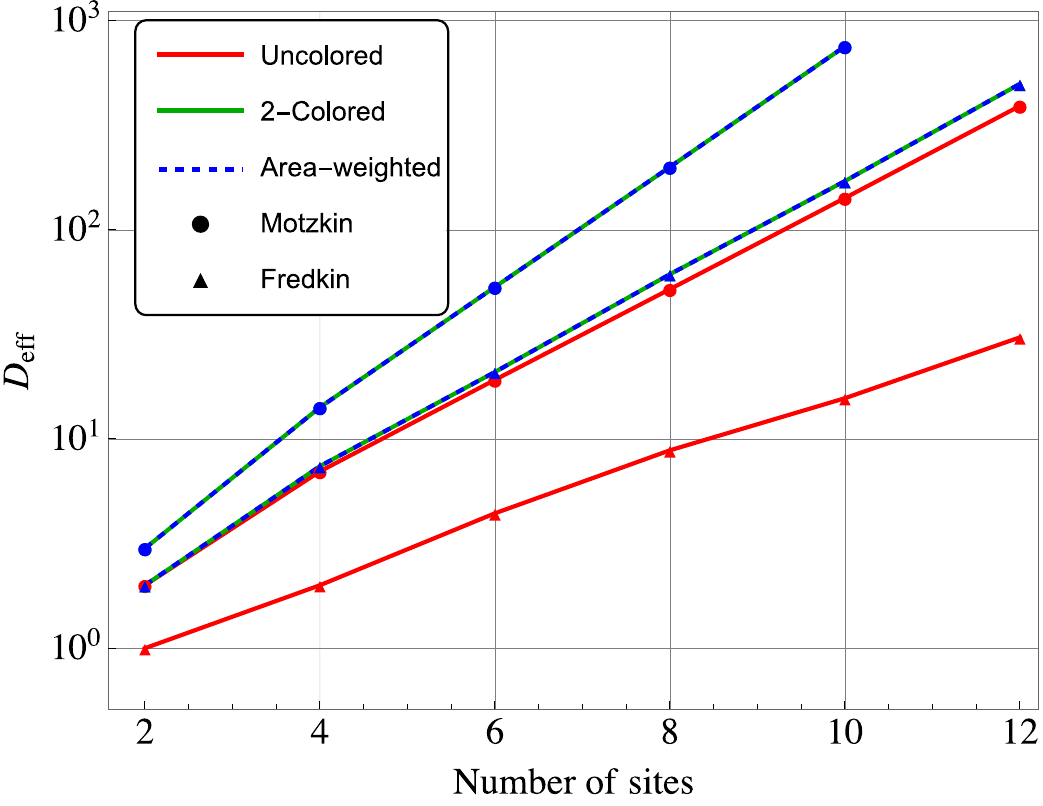}\\
    \caption{Effective number of non-zero matrix elements, $D_{\rm eff}$ (see Eq.~(\ref{eq:Deff})), from the reduced density matrix of the Fredkin (triangle) and Motzkin (circle) families of states, as the system size grows. Each family has an uncolored variant (red-solid lines), a 2-colored one (green-solid lines), and an area-weighted one (blue-dashed lines). The vertical axis is logarithmic.}
    \label{fig:Paths_EE_CX_CD}
\end{figure}

In Sec.~\ref{sec:circuit_complexity}, we have established how the total CX gate count and circuit depth of the SG protocol emerge from the gates required at each iteration, identifying the specific operations they implement.
When extending the SG approach to other state families, such as ground states of constrained-walk Hamiltonians, reproducing this detailed gate-level analysis for each new state may be impractical.
We show that a preliminary estimate of the relative preparation cost can instead be obtained from the reduced density matrix (RDM) structure.
Specifically, we compare the RDM structures of Motzkin and Fredkin states, using the effective number of non-zero RDM matrix elements $D_{\mathrm{eff}}$, defined as
\begin{equation}
    \label{eq:Deff}
    D_{\mathrm{eff}}=n_{\mathrm d}+\sqrt{n_{\mathrm{od}}},
\end{equation}
where $n_{\mathrm d}$ ($n_{\mathrm{od}}$) is the number of non-zero diagonal (off-diagonal) elements of the RDM (see Figs.~\ref{fig:EE_Motzkin_Fredkin}(c)--(f)).
The square root accounts for the quadratic growth of off-diagonal elements, allowing diagonal and off-diagonal contributions to enter the measure on comparable scales.
While $D_{\mathrm{eff}}$ grows exponentially with $N$, its growth rate is set by the number of allowed local steps and color choices per site, the same quantities that influence the implementation cost of our construction.
Consequently, $D_{\mathrm{eff}}$ reproduces the hierarchy of CX counts across the states considered here (Figs.~\ref{fig:Paths_EE_CX_CD} and~\ref{fig:resources}): colored Motzkin $>$ colored Fredkin $>$ uncolored Motzkin $>$ uncolored Fredkin, including the relative closeness of the uncolored Motzkin and colored Fredkin chains and the coincidence of area-weighted and colored variants, whose reduced density matrices share the same structure.

The structural analysis presented here has also been checked against the additional states with block-diagonal RDM structure introduced in Appendix~\ref{app:symmetric_states} (Figs.~\ref{fig:Deff_SI} and~\ref{fig:symmetric-resources}), although its applicability to arbitrary quantum states remains unexplored.
For states with similar RDM structures, the approach can provide a preliminary estimate of the relative CX gate count through \(D_{\mathrm{eff}}\), even using small-system RDMs (e.g., \(N=4\)), with the Motzkin and Fredkin states serving as reference points for comparison.
This analysis further motivates the construction of alternative states from the RDM structure that show volume-law entanglement scaling but have lower preparation costs than the Motzkin and Fredkin states considered here (see Appendix~\ref{app:symmetric_states}).
Thus, \(D_{\mathrm{eff}}\) serves as a qualitative indicator of relative preparation cost, while the polynomial scaling of the circuit cost follows from the number of prefix groups, which is set by the maximum height \(h_{\mathrm{max}}(\ell)\), as discussed in Secs.~\ref{sec:other-states} and~\ref{sec:complexity_analysis}.

\section{Uniformly Controlled Rotations}
\label{app:ucr}

Suppose we have some quantum state $\ket{\psi}$ with a register $C$ that has $H$ qubits in it, and stores some binary number $B_i$ for each non-zero CBS $i$ of $\ket{\psi}$ in LSB-first order.
In other words, $\ket{\psi}$ has the form $\ket{\psi} = \sum_i c_i \ket{i}\ket{B_i}$.
Suppose also that we have some classical array of angles $A$ of size $2^{H}$.
The goal is to apply an $R_y(A)$ rotation multiplexed by $C$ to some chosen qubit $t$, i.e. to rotate $t$ in each CBS $i$ by an angle whose index in $A$ is equal to $B_i$.

\begin{figure}
    \includegraphics[width=0.75\linewidth]{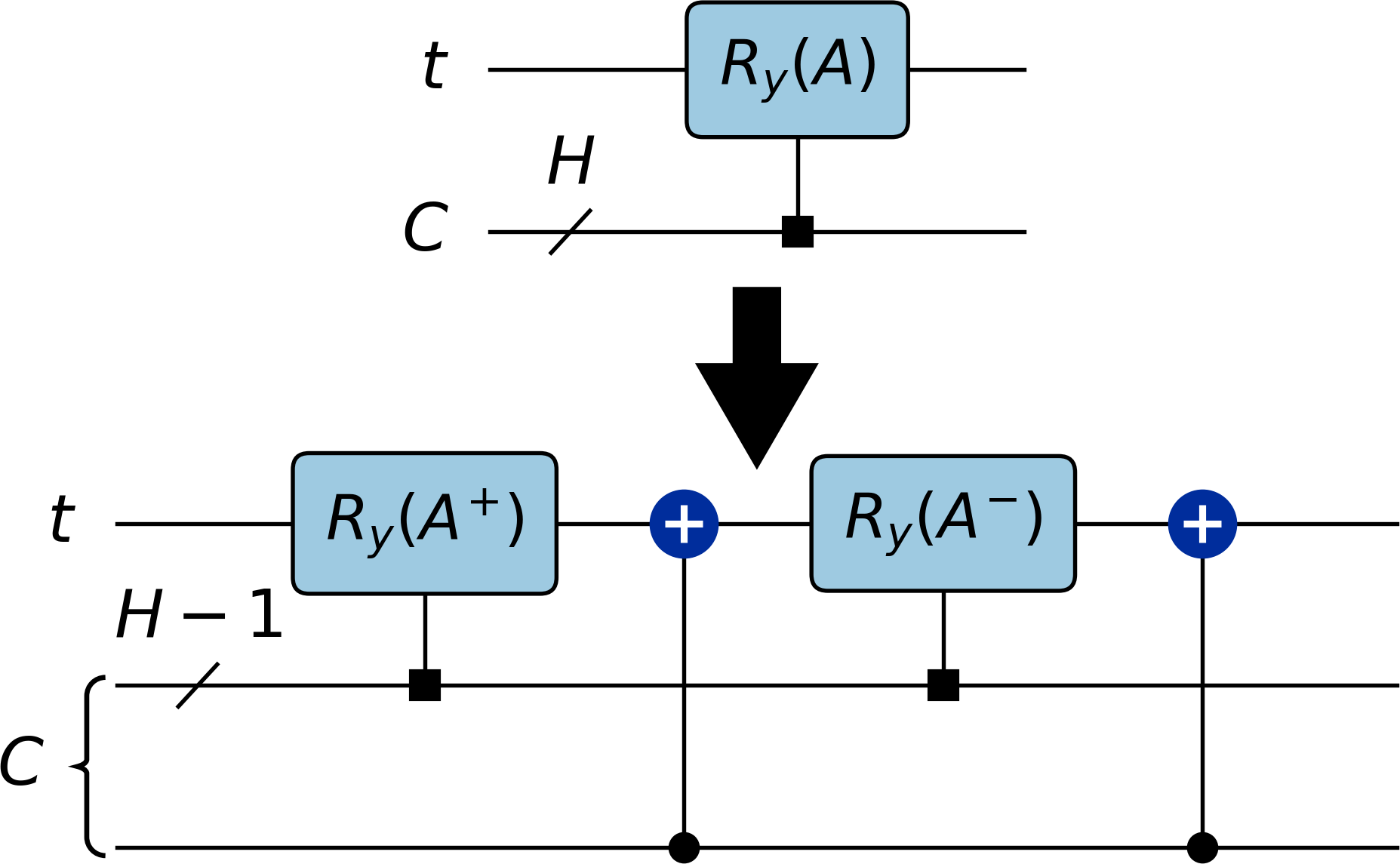}
    \caption{One recursion step of the uniformly controlled rotations. A rotation multiplexed on an $H$-qubit register $C$ is decomposed into two rotations multiplexed on $H - 1$ qubits, interleaved with CX gates controlled on the remaining qubit. The new angle arrays are $A^\pm = (A_L \pm A_R)/2$, where $A_L$ and $A_R$ are the two halves of the original angle array $A$.}
    \label{fig:rotation-recursion}
\end{figure}

This can be done recursively as follows~\cite{mottonen2005transformation}.
Take the array $A$ and split it into two halves: left ($A_L$) and right ($A_R$).
Notice that any angle in $A_L$ is applicable only when the most significant bit of $C$ ($C_{H - 1}$) is 0.
Similarly, $A_R$ is only applicable if $C_{H - 1} = 1$.
Define 2 new angle arrays: $A^+ = (A_L + A_R) / 2$ and $A^- = (A_L - A_R) / 2$.
If we know how to implement a UCR by angles $A^+$ and $A^-$ multiplexed by the first $H - 1$ qubits of $C$, then the overall UCR by the original angles can be implemented as shown in Fig.~\ref{fig:rotation-recursion}.

The above works because $X R_y(\theta) X = R_y(-\theta)$.
Thus, if $C_{H - 1} = 0$, the rotation applied to $t$ is $A^+ + A^- = A_L$, and if $C_{H - 1} = 1$, the rotation is $A^+ - A^- = A_R$, which is exactly what is needed.
$R_y(A^+)$ and $R_y(A^-)$ can themselves be implemented by applying the same recursion step again.
Each application reduces the number of qubits in the multiplexor register by 1, so after $H$ recursion steps we will get down to a plain uncontrolled rotation which can be implemented directly and finish the recursion.

Recall that the angle arrays of the ``prepare site" operators are padded with extra angles that do not affect any of the non-zero CBS (Sec.~\ref{sec:uncolored-motzkin}).
Choosing them in a way that makes $A^- = 0$ allows one to prune the corresponding $A^-$ branch with its CX gates.
Therefore, in practice one tries to choose them in a way that allows for pruning of more $A^-$ branches.

The actual implementation of the UCR in the code is somewhat more complicated and switches to a different, non-recursive (dense) implementation (not explained here) at some level of recursion, depending on the angle structure, instead of recursing through all $H$ levels, but fundamentally it is not required.
The version shown here would work as well, just less efficiently.
See the associated code~\cite{gaidai2026motzkin} for details.

\section{Optimization Targets}
\label{app:opt-targets}

In Sec.~\ref{sec:complexity} we showed the number of CX gates and depth of the circuits preparing uncolored and 2-colored Motzkin and Fredkin states.
That section uses constructions optimized for circuit depth.
However, other optimization targets could be considered as well, which quantitatively (but not qualitatively) change the results shown in Fig.~\ref{fig:resources}.
In this section we consider 2 other optimization targets: CX gate count and number of qubits.

\begin{figure}
    \includegraphics[width=\columnwidth]{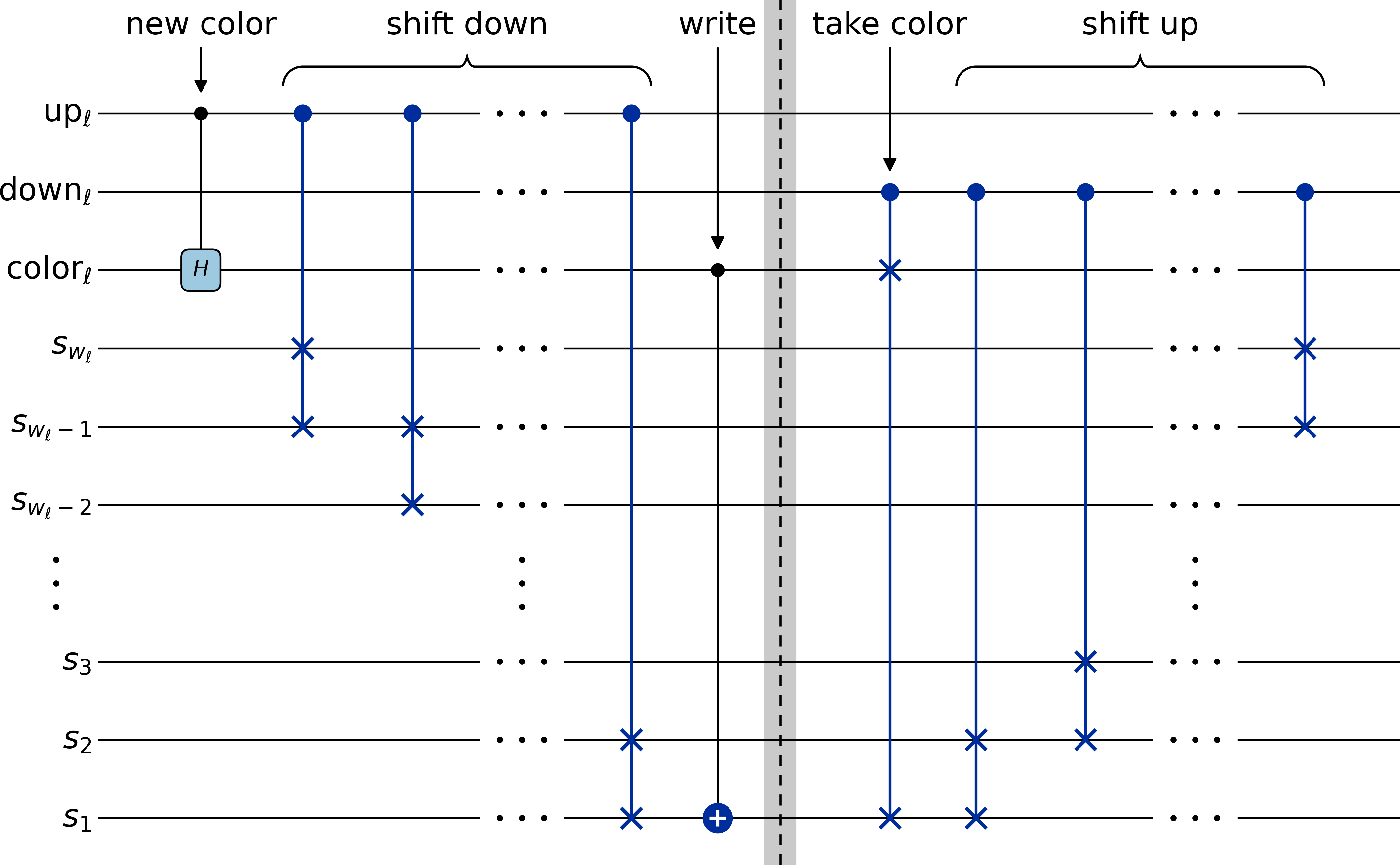}
    \caption{CX-optimized implementation of the ``update color" operator for Motzkin states. The left part of the circuit draws the color of the new up step, shifts the color stack down and writes that color to the top of the stack. The right part takes the color of the last unmatched up step off the stack for a down step and shifts the stack up. Here, $s_1, \ldots, s_{w_\ell}$ denote the color stack qubits.}
    \label{fig:update-color-cx}
\end{figure}

The CX gate count of the Motzkin states can be reduced (at the cost of circuit depth) by implementing the ``update color" operator with chains of controlled swaps of neighboring stack qubits instead of reversals, as shown in Fig.~\ref{fig:update-color-cx}.

The operator is made of 2 parts.
In the left half (before the barrier) we implement the stack push operation.
As in Fig.~\ref{fig:update-color}, we first apply a Hadamard gate to the $\mathrm{color}_\ell$ qubit, controlled on the $\mathrm{up}_\ell$ qubit, which creates a uniform superposition of the 2 colors for the prefixes stepping up.
Then we have a chain of swap gates controlled on the $\mathrm{up}_\ell$ qubit that perform pairwise swaps of all color stack qubits starting from the bottom of the stack (qubit $s_{w_\ell}$).
As in Fig.~\ref{fig:update-color}, the stacks of these prefixes currently hold at most $h_{\max}(\ell) - 1 < w_\ell$ colors.
As such, qubit $s_{w_\ell}$ is guaranteed to be in the $\ket{0}$ state for them, and this chain of swaps shifts the entire stack down (in terms of data qubit index), while pushing that $\ket{0}$ all the way to the top of the stack.
This $\ket{0}$ serves as a blank slate for the following CX gate that writes the color of the last up step into it (which, as in Fig.~\ref{fig:update-color}, does not need to be controlled on the $\mathrm{up}_\ell$ qubit).

In the right half (after the barrier), we take care of the prefixes stepping down.
First, we swap the $\mathrm{color}_\ell$ qubit with the top of the color stack, controlled on the $\mathrm{down}_\ell$ qubit.
This ensures that the color of the last down step matches the color of the last unmatched up step, and puts the $\ket{0}$ state from the $\mathrm{color}_\ell$ qubit on the top of the stack.
Then we propagate that $\ket{0}$ to the bottom of the stack and shift the stack up along the way, which is accomplished by the same chain of controlled swaps run in reverse.

Since one of the swapped qubits is always in the $\ket{0}$ state, each controlled swap can be implemented as a relative-phase Toffoli gate (3 CX gates)~\cite{maslov2016advantages} between 2 CX gates, i.e. with 5 CX gates.
Each half contains $\approx h_{\max}(\ell)$ controlled swaps, so this construction requires $\approx 10 h_{\max}(\ell)$ CX gates instead of $\approx 12 h_{\max}(\ell)$ of Fig.~\ref{fig:update-color}, which reduces the total CX count from $\approx 15 N^2 / 4$ to $\approx 13 N^2 / 4$.
However, swaps of neighboring stack qubits share a qubit and cannot overlap at all, so each of the $\approx 2 h_{\max}(\ell)$ controlled swaps adds its full depth of 17 layers (in the gate set of Sec.~\ref{sec:complexity_analysis}).
The resulting $\approx 34 h_{\max}(\ell)$ layers per iteration exceed the $18 h_{\max}(\ell)$ layers of the rotations, so the color stack updates set the overall depth instead of the rotations, which grows from $\approx 9 N^2 / 2$ to $\approx 17 N^2 / 2$ for the unoptimized circuits.

The qubit-optimized target prioritizes using fewer qubits first, then optimizes depth as a secondary target, i.e. it uses the ``update color" operator of Fig.~\ref{fig:update-color}.
The reduction in the number of qubits is achieved by removing ancilla qubits, which used to store the height counter in the CX- and depth-optimized versions.
Even without dedicated ancilla qubits, we can still largely follow the same preparation pattern by placing the counter in the uninitialized data qubits, like we did with the color stack and the ripple carry ancillas (at the cost of higher CX count and depth).
Specifically, the height counter is allocated between the initialized data qubits and the color stack, and is shifted at every iteration because the free gap where it can be allocated closes on the left due to initialized data sites, and on the right due to the color stack qubits.
Despite this, the height counter still fits in the remaining space for most of the sites, except for the last few sites (2 for Motzkin and 4 for Fredkin).
Before deallocating the counter, we copy the height it stores into the color stack, by placing a sentinel $\ket{1}$ in the stack qubit right below the colors (i.e. in $s_{h + 1}$ for a prefix of height $h$), which then moves together with them in the following stack updates.
The sentinel is needed because color-0 is stored as $\ket{0}$, so without it, the height could not be read off a stack whose deepest colors are color-0.
Since the height is now stored twice, we can use the sentinel to return the counter to the $\ket{0}$ state, and control the UCR rotations of the last few sites directly on the color stack instead, where the height is given by the position of the sentinel.
This remains cheap, since $h_{\max}(\ell)$ (Eq.~\eqref{eq:hmax}), and with it the part of the color stack that the rotations are controlled on, is small in the last few sites.

\begin{figure}
    \includegraphics[width=\columnwidth]{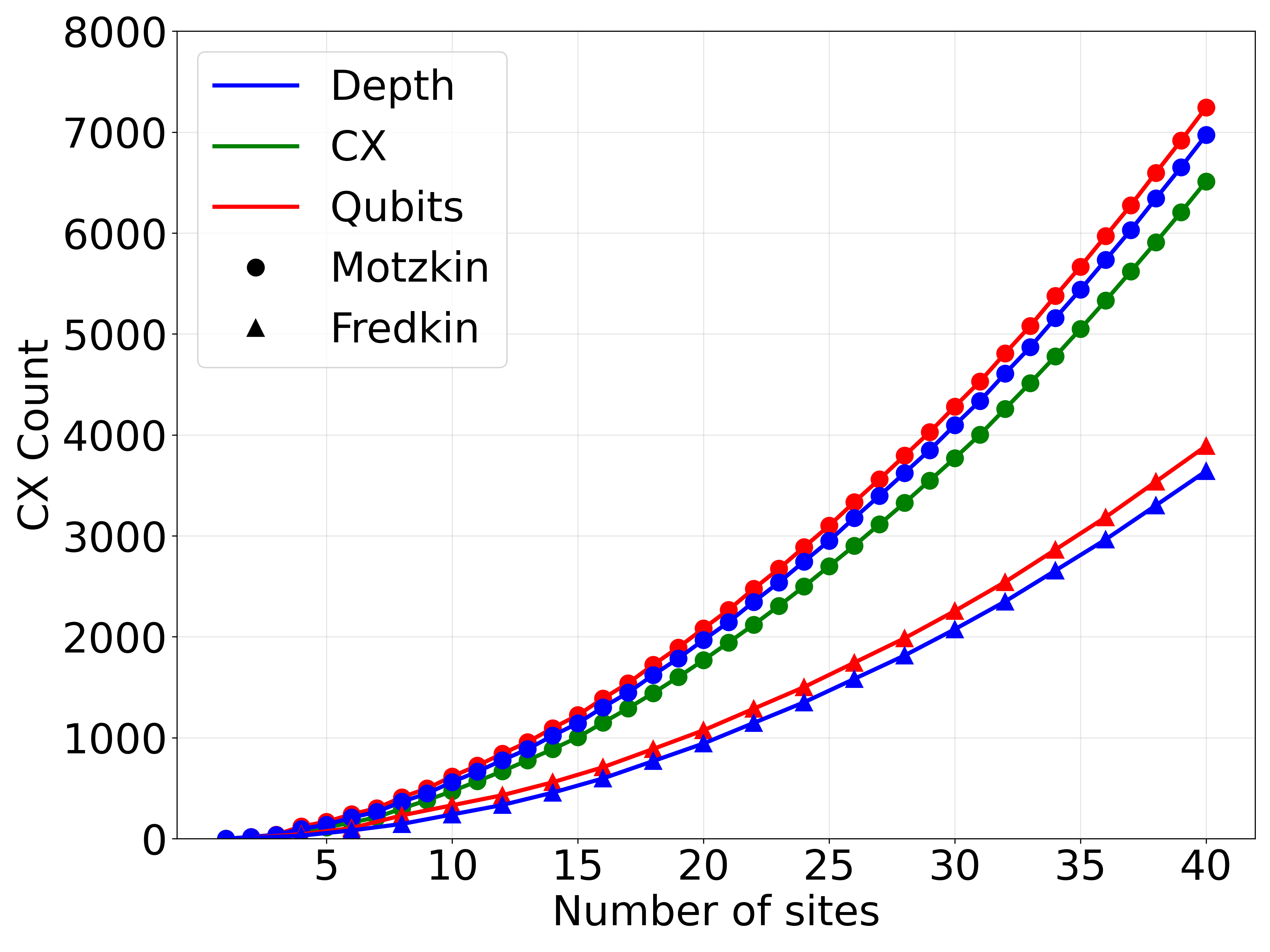}

    \includegraphics[width=\columnwidth]{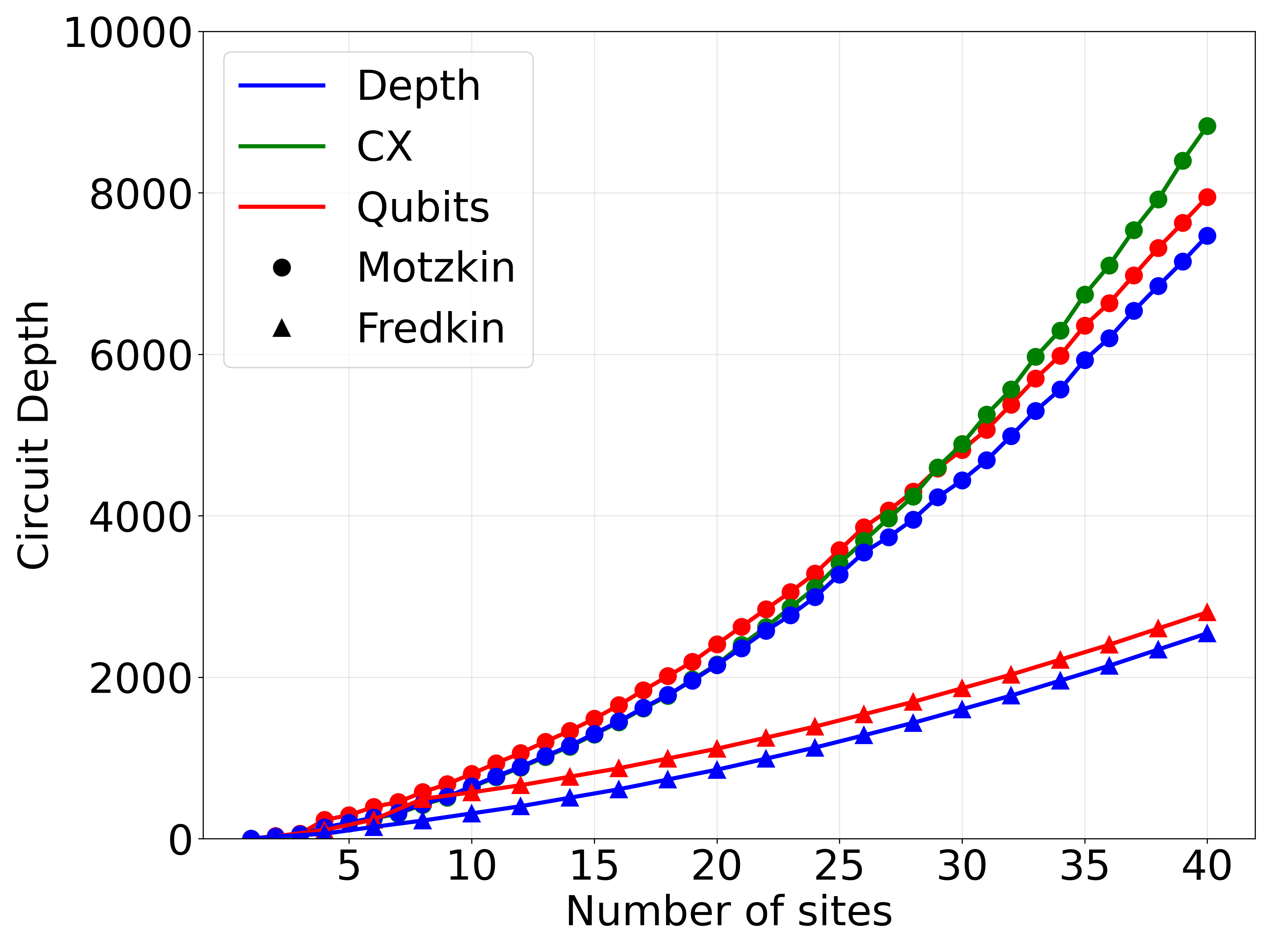}

    \caption{CX count and depth of the circuits preparing 2-colored Motzkin (circles) and Fredkin (triangles) states vs the number of sites $N$. Different colors show different optimization targets for the circuits. The main text uses the depth-optimized target.}
    \label{fig:targets-resources}
\end{figure}

The resulting CX gate count and depth of the transpiled circuits, optimized for each target, are shown in Fig.~\ref{fig:targets-resources} for 2-colored Motzkin and Fredkin states.
As before, transpilation for all targets has been done with \textit{qiskit}.

For Fredkin states the CX- and depth-optimized targets happen to coincide, which is why there are only 2 Fredkin lines in Fig.~\ref{fig:targets-resources}.
This happens because the first reversal of the Fredkin ``update color" operator (Fig.~\ref{fig:update-color}) does not need a control, which makes the reversal-based construction favorable for both CX count and depth simultaneously.

Qualitatively, all curves in Fig.~\ref{fig:targets-resources} scale as $O(N^2)$.
For all targets, the Fredkin circuits are about 2 times cheaper than the Motzkin ones in CX count and about 3 times cheaper in depth at $N = 40$, while the prefactors of $N^2$ for the unoptimized circuits (Sec.~\ref{sec:complexity_analysis}) give asymptotic ratios of $1.3$--$1.5$ in CX count and $2.25$--$4.25$ in depth.
The CX count of each iteration consists of the rotations of ``prepare site", whose cost is proportional to the number of prefix groups, and the updates of the height counter and the color stack.
Fredkin prefixes reach only every second height, so they form about half as many prefix groups, and ``prepare site" needs only the up rotation, which makes the rotations about 6 times cheaper.
However, the color stack updates dominate the CX count of both colored circuits and are only slightly cheaper for Fredkin states, due to the uncontrolled first reversal, which keeps the CX ratio small.
The depth ratio is larger, since the depth of the Motzkin circuits is set by their rotations (or by the swap chains for the CX-optimized target), while the shallower Fredkin rotations leave the depth to the color updates.
The qubit-optimized targets follow the same logic, since they use the same ``update color" operator as the depth-optimized ones, and moving the height counter and controlling the last rotations on the color stack add only 4--10\% to their CX count and depth at $N = 40$.

\section{Extension of Motzkin and Fredkin Families of States}
\label{app:symmetric_states}

\begin{figure}
    \centering
    \makebox[\linewidth][l]{\hspace{0.1193\linewidth}\includegraphics[width=0.8414\linewidth, trim=53 276.74 5.27 0, clip]{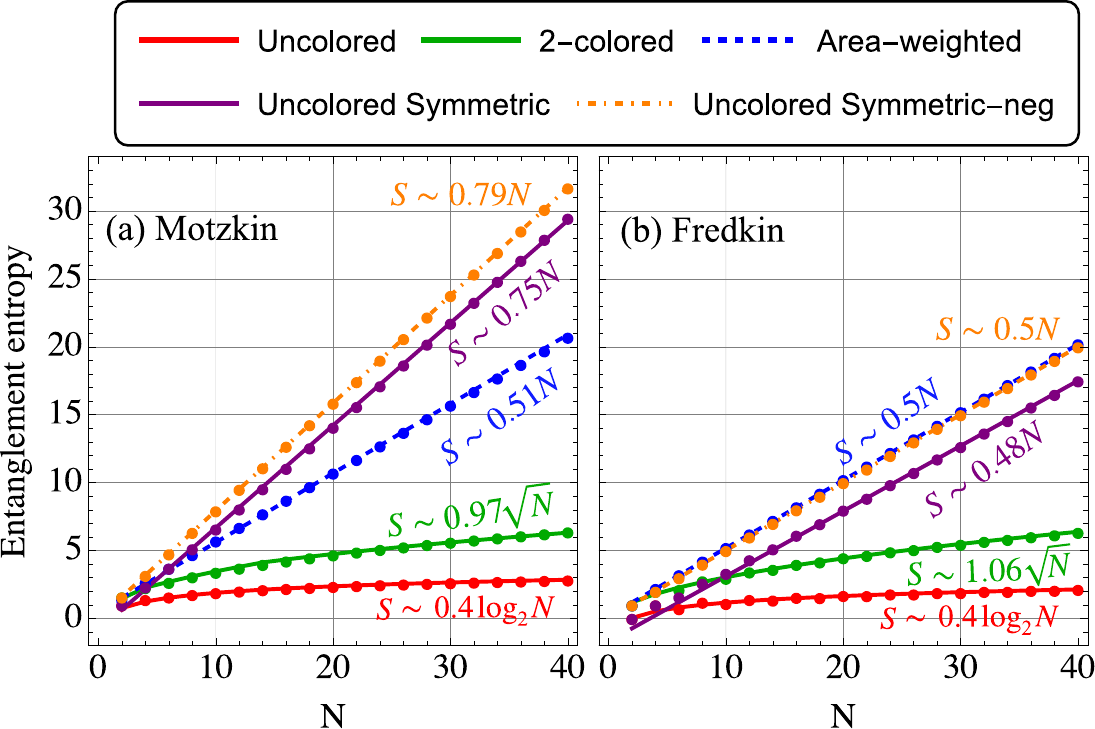}}\\
    \makebox[\linewidth][l]{\hspace{0.0094\linewidth}\includegraphics[width=0.5115\linewidth, trim=0 0 241.37 72.7, clip]{figures/rdm/RDM_symmetric_plots_v2a.pdf}\hspace{0.0442\linewidth}\includegraphics[width=0.4349\linewidth, trim=283.9 0 0 72.7, clip]{figures/rdm/RDM_symmetric_plots_v2a.pdf}}\\
    \includegraphics[width=\linewidth, trim=0 0 17.30 13, clip]{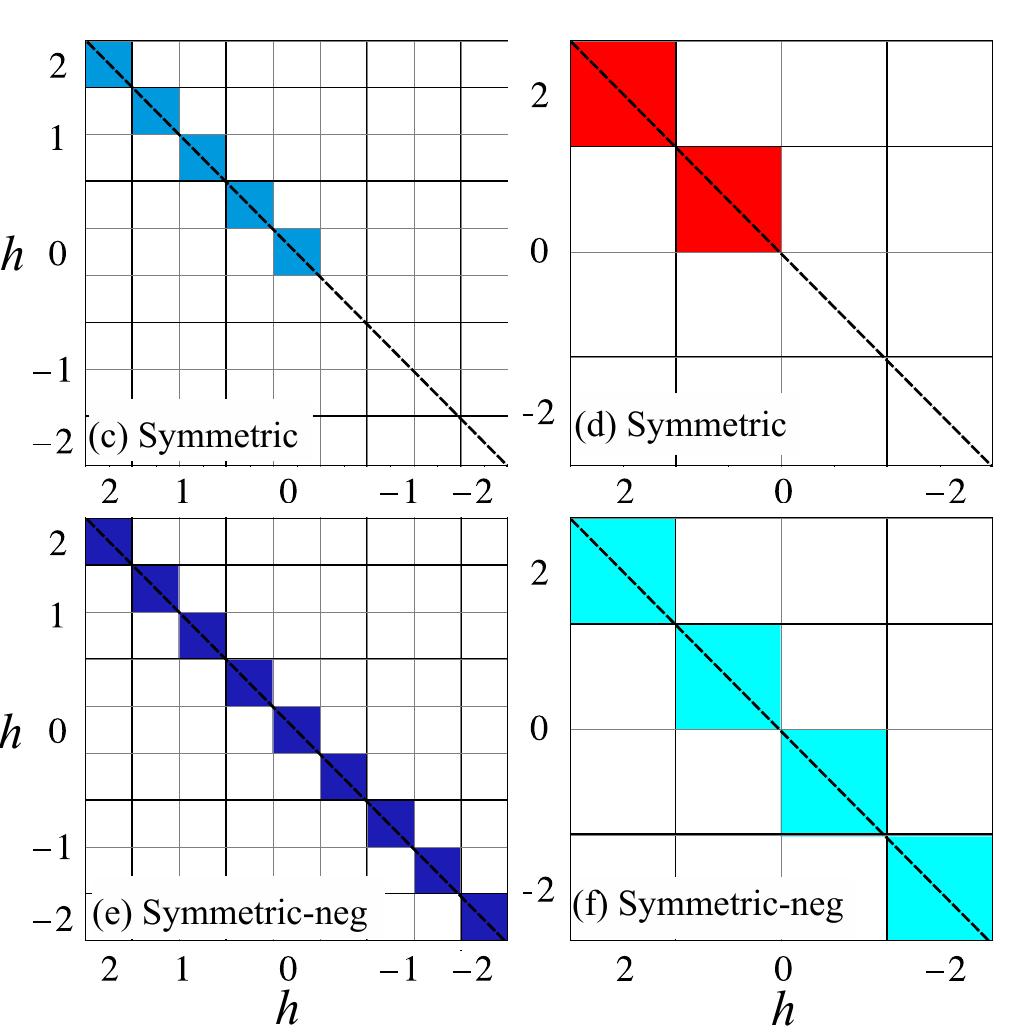}\\
    \makebox[\linewidth][l]{\hspace{0.3459\linewidth}\includegraphics[width=0.4029\linewidth]{figures/rdm/RDM_colorbar_horizontal.pdf}}
    \caption{Half-chain entanglement entropy $S_A$ (in bits) versus the number of sites $N$ for (a) the Motzkin and (b) Fredkin states. Red, green, blue, purple, and orange markers represent the uncolored, 2-colored, area-weighted 2-colored ($t=2$), uncolored symmetric, and uncolored symmetric-negative states, respectively. Dashed and solid lines are fits, labeled by their leading terms (with $S \equiv S_A$). Reduced density matrix (RDM) of the (c) uncolored symmetric Motzkin, (d) uncolored symmetric Fredkin, (e) uncolored symmetric-negative Motzkin, and (f) uncolored symmetric-negative Fredkin states for a chain of $N = 4$ sites, obtained by tracing out the right half of the chain. Rows and columns are labeled by the height $h$ of the left half ($N/2$-prefix), and the color scale indicates the magnitude of each matrix element.}
    \label{fig:EE_sym_states}
\end{figure}

The reduced density matrix (RDM) analysis presented in Appendix~\ref{app:rdm_complexity} motivates the construction of two additional families of states, which we refer to as uncolored symmetric Motzkin and uncolored symmetric Fredkin states.
These states are constructed by dividing the chain into two halves, with the first half following the conventional Motzkin or Fredkin path constraints and the second half forming its mirror image.
This construction selectively retains paths corresponding to the diagonal elements of the RDM, as illustrated in Fig.~\ref{fig:EE_sym_states}(c,d).
As discussed in Sec.~\ref{app:building_rdm}, the resulting states are maximally entangled within their allowed subspace of the full Hilbert space, exhibiting volume-law entanglement despite being uncolored.
A parent Hamiltonian for which the uncolored symmetric Motzkin state is the unique ground state is derived in Sec.~\ref{app:symmetric-motzkin-hamiltonian}.

\begin{figure}
    \centering
    \includegraphics[width=\linewidth]{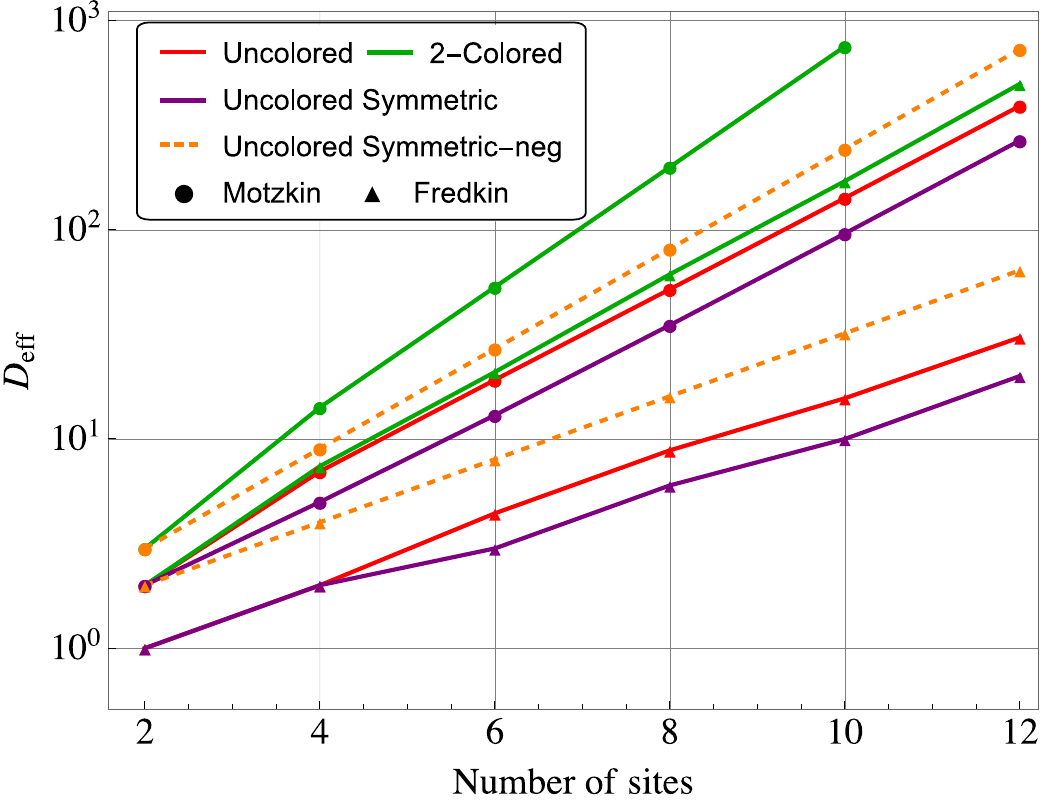}
    \caption{Effective number of non-zero matrix elements, $D_{\rm eff}$, from the reduced density matrix of the Fredkin (triangle) and Motzkin (circle) families of states, as the system size grows. Each family has an uncolored variant (red-solid lines), a 2-colored one (green-solid lines), an uncolored symmetric one (purple lines), and an uncolored symmetric-negative one (orange-dashed lines). The vertical axis is logarithmic.}
    \label{fig:Deff_SI}
\end{figure}

\begin{figure}
    \includegraphics[width=\columnwidth]{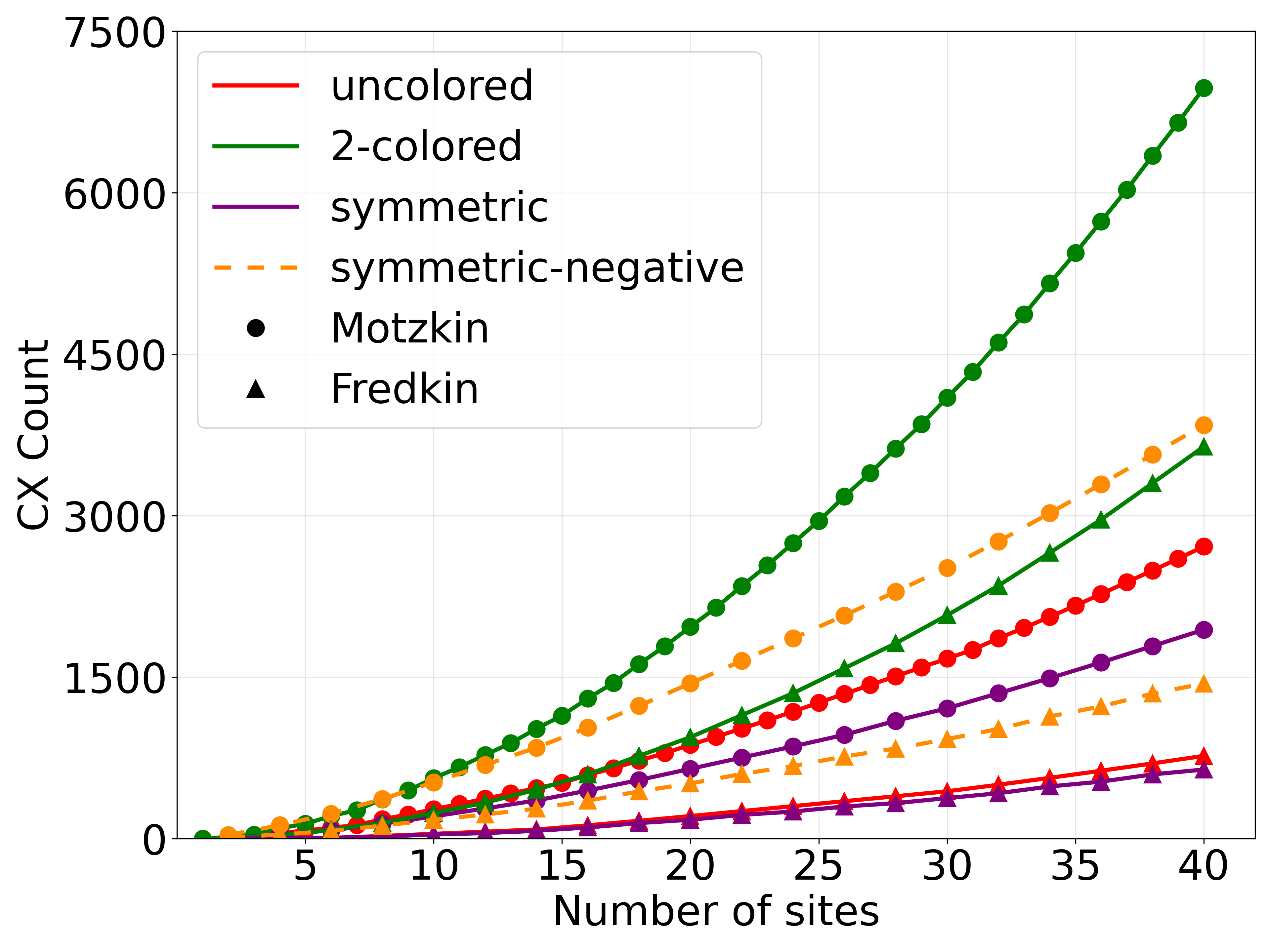}
    \includegraphics[width=\columnwidth]{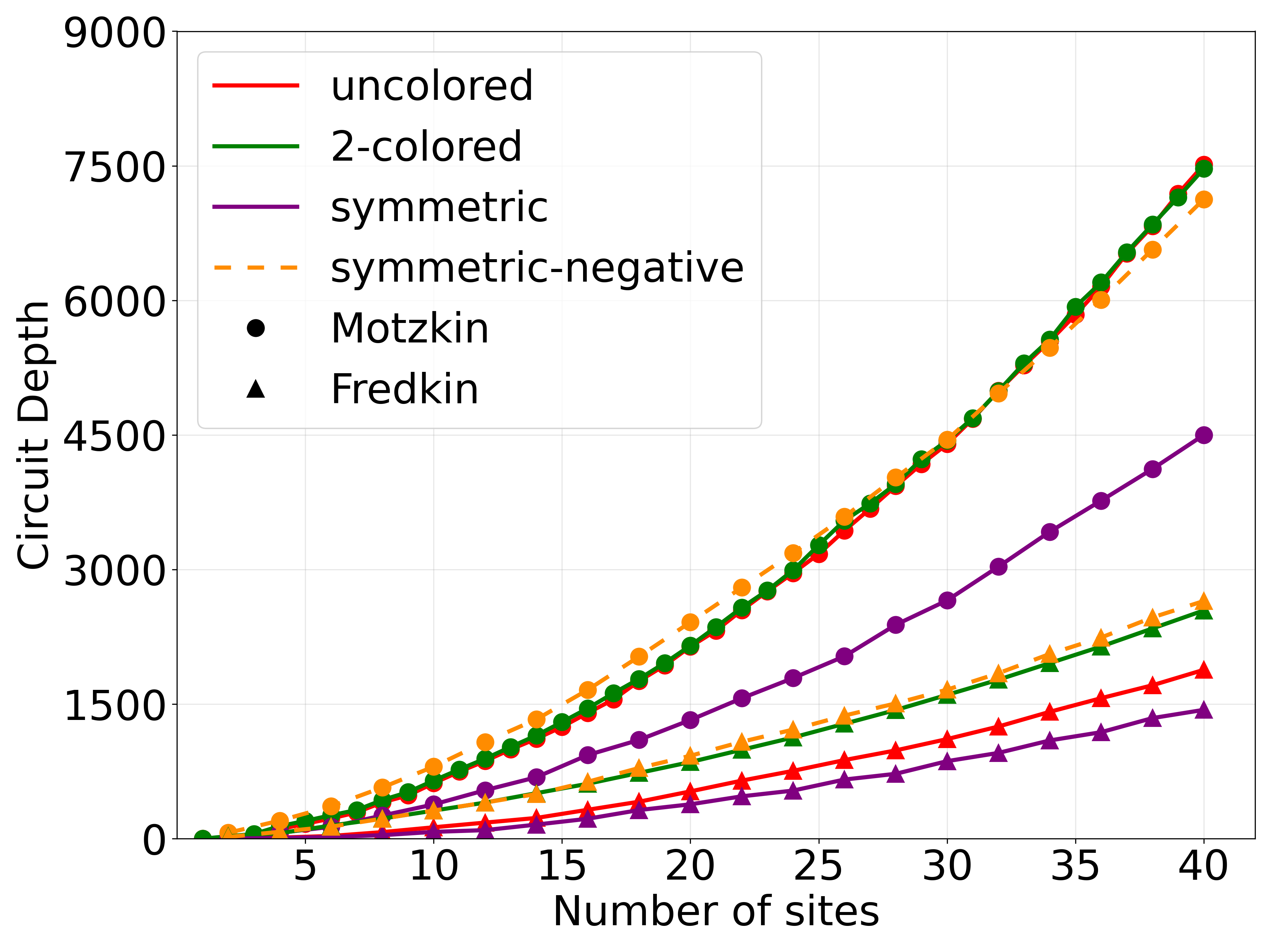}
    \caption{CX count (top) and depth (bottom) of the transpiled SG circuits preparing uncolored (red), 2-colored (green), uncolored symmetric (purple) and uncolored symmetric-negative (orange, dashed) Motzkin (circles) and Fredkin (triangles) states vs the number of sites $N$. The uncolored and 2-colored curves are the same as in Fig.~\ref{fig:resources}.}
    \label{fig:symmetric-resources}
\end{figure}

The effective number of non-zero RDM matrix elements, $D_{\mathrm{eff}}$, provides a useful indicator of the CX gate requirements for preparing these states.
As shown in Fig.~\ref{fig:Deff_SI}, the uncolored symmetric Motzkin and Fredkin states have smaller $D_{\mathrm{eff}}$ than their conventional uncolored counterparts, suggesting lower CX gate counts.
Remarkably, while the conventional uncolored Motzkin and Fredkin states exhibit logarithmic entanglement scaling, their symmetric counterparts achieve volume-law scaling with fewer CX gates and a lower circuit depth (Fig.~\ref{fig:symmetric-resources}).
Although the conventional and symmetric states have the same number of distinct prefixes, the mirror constraint substantially reduces the number of compatible suffixes for each prefix, simplifying the corresponding extension operators.
This observation applies to both the Motzkin and Fredkin families.

We further introduce symmetric-negative states (abbreviated as ``symmetric-neg'' in Figs.~\ref{fig:EE_sym_states} and~\ref{fig:Deff_SI}), which extend the symmetric construction by allowing paths to attain negative heights.
These states are designed to be maximally entangled within the full Hilbert space of the corresponding uncolored spins.
They are products of maximally entangled states of the symmetric pairs of sites $(i, N + 1 - i)$, similar to the rainbow states of spin-1/2 chains, in which each such pair forms a singlet~\cite{vitagliano2010volume, ramirez2014conformal}.
Their RDM structures are shown in Fig.~\ref{fig:EE_sym_states}(e) and (f) for the Motzkin and Fredkin types, respectively.
Parent Hamiltonians for which the symmetric-negative Motzkin and Fredkin states are the unique ground states are derived in Sec.~\ref{app:symmetric-negative-hamiltonian}.

As shown in Fig.~\ref{fig:Deff_SI}, the symmetric-negative states have larger $D_{\mathrm{eff}}$ than their conventional uncolored counterparts but smaller values than the corresponding $2$-colored states.
This suggests intermediate resource requirements for both CX gate count and circuit depth.
The actual circuit depths and CX gate counts obtained from our preparation protocol for these state families are presented in Fig.~\ref{fig:symmetric-resources}.
They follow these estimates, except for the depth of the Motzkin circuits, where the uncolored, 2-colored and symmetric-negative curves nearly overlap.
However, since the symmetric-negative states are products of pair states (Sec.~\ref{app:symmetric-negative-hamiltonian}), they can also be prepared directly, one pair at a time, which after the same transpilation takes only $2N$ CX gates and depth 10 for Motzkin, and $N/2$ CX gates and depth 5 for Fredkin.

\subsection{Parent Hamiltonian for the Motzkin State}
\label{app:motzkin-hamiltonian}

Here we restate the parent Hamiltonian of the uncolored Motzkin state and its proof, both given by Bravyi et al.~\cite{bravyi2012criticality}.
We include it since the proof for the symmetric Motzkin state in Sec.~\ref{app:symmetric-motzkin-hamiltonian} reuses most of its steps.
Table~\ref{tab:notation-appendix} lists the notation that this and the following sections use in addition to Table~\ref{tab:notation}.

\begin{table}
    \centering
    \renewcommand{\arraystretch}{1.2}
    \begin{tabular}{| l | >{\raggedright\arraybackslash}p{0.72\columnwidth} |}
        \hline
        Symbol & Meaning \\
        \hline
        $M_{N, h}^{(k)}$ & Set of $k$-colored Motzkin prefixes of length $N$ with final height $h$ \\
        $M_{N, *}$ & Set of all Motzkin prefixes of length $N$ and any final height, i.e. $\bigcup_h M_{N, h}$ \\
        $M_N^{\mathrm{sym}}$ & Set of mirror-symmetric Motzkin paths of length $N$, see Sec.~\ref{app:symmetric-motzkin-hamiltonian} \\
        $M_N^{\mathrm{neg}}$ & Set of all mirror-symmetric paths of length $N$, including those that go below height $0$, see Sec.~\ref{app:symmetric-negative-hamiltonian} \\
        \hline
    \end{tabular}
    \caption{Notation used in the appendices in addition to Table~\ref{tab:notation}. As there, qualifiers and parameters at their default values can be omitted.}
    \label{tab:notation-appendix}
\end{table}

\begin{theorem}
    \label{thm:motzkin}
    The uncolored Motzkin state $\ket{M_N}$, i.e. Eq.~\eqref{eq:weighted-motzkin-state} with $k = 1$ and $t = 1$, is the unique ground state of its parent Hamiltonian
    \begin{equation}
        H_M = \sum_{i = 1}^{N - 1} \Pi_i + \proj{d}_1 + \proj{u}_N,
        \label{eq:H_M}
    \end{equation}
    where the subscripts denote the sites the operators act on, and
    \begin{subequations}
        \label{eq:H_M_terms}
        \begin{align}
            \Pi_i &= (\proj{U} + \proj{D} + \proj{F})_{i, i + 1},
            \label{eq:Pi}\\
            \ket{U} &= \frac{1}{\sqrt{2}} \bigl(\ket{fu} - \ket{uf}\bigr),
            \label{eq:U}\\
            \ket{D} &= \frac{1}{\sqrt{2}} \bigl(\ket{fd} - \ket{df}\bigr),
            \label{eq:D}\\
            \ket{F} &= \frac{1}{\sqrt{2}} \bigl(\ket{ff} - \ket{ud}\bigr).
            \label{eq:F}
        \end{align}
    \end{subequations}
\end{theorem}
\begin{proof}
    \begin{claim}
        \label{clm:nonnegative}
        $H_M$ has no negative eigenvalues.
    \end{claim}
    \begin{proof}
        $H_M$ is a sum of projectors: the three projectors in each $\Pi_i$, called bulk terms, and the two boundary terms.
        Projectors are positive semidefinite, so $H_M$ is positive semidefinite too.
        Therefore, it has no negative eigenvalues.
    \end{proof}

    \begin{claim}
        \label{clm:motzkin-only}
        If an eigenstate of $H_M$ with eigenvalue $0$ exists, then it must be a multiple of $\ket{M_N}$.
    \end{claim}
    \begin{proof}
        Let $\ket{\psi}$ be an arbitrary eigenstate of $H_M$ with eigenvalue $0$, so
        \begin{equation}
            H_M \ket{\psi} = 0.
        \end{equation}
        Each term in $H_M$ is positive semidefinite.
        Therefore, this is only possible if every term annihilates $\ket{\psi}$ separately.

        Let us define a \emph{local move} as one of these replacements $\{fu \leftrightarrow uf, fd \leftrightarrow df, ff \leftrightarrow ud\}$ in any two adjacent sites.

        \begin{subclaim}
            \label{clm:local-moves}
            Local moves do not change the minimum height $h_{\min}(p) = \min_{0 \le \ell \le N} h_\ell(p)$ and the final height $h_N(p)$ of a path $p$.
            Therefore, the set of Motzkin paths is closed under local moves.
        \end{subclaim}
        \begin{proof}
            Consider a local move in sites $(i, i + 1)$, where $1 \le i < N$.
            The heights $(h_{i - 1}, h_i, h_{i + 1})$ before and after the move are
            \begin{equation}
                \begin{aligned}
                    fu \leftrightarrow uf &: \quad (h, h, h + 1) \leftrightarrow (h, h + 1, h + 1), \\
                    fd \leftrightarrow df &: \quad (h, h, h - 1) \leftrightarrow (h, h - 1, h - 1), \\
                    ff \leftrightarrow ud &: \quad (h, h, h) \leftrightarrow (h, h + 1, h),
                \end{aligned}
            \end{equation}
            where $h = h_{i - 1}$ and $h_0 = 0$.
            As one can see, none of the moves changes the height before site $i$ or after site $i + 1$.
            Only the height after site $i$ ($h_i$) changes.
            Since $i < N$, $h_N$ cannot be changed by the local moves.
            Additionally, $h_i \ge \min(h_{i - 1}, h_{i + 1})$ both before and after all moves.
            Therefore, $h_i$ cannot become a new minimum.
            Motzkin paths are the paths with minimum and final heights =$0$, so local moves cannot turn a Motzkin path into a non-Motzkin path or the other way around.
        \end{proof}

        Two paths (and the corresponding basis states) are called \emph{connected} if a sequence of local moves turns one into the other.

        \begin{subclaim}
            \label{clm:connectivity}
            Any path $p$ is connected to $d^a f^{N - a - b} u^b$, where $a = -h_{\min}(p)$ and $b = h_N(p) - h_{\min}(p)$.
        \end{subclaim}
        \begin{proof}
            Local moves $fu \leftrightarrow uf$ and $fd \leftrightarrow df$ can move any $f$ step to any position.
            First, we move all $f$ steps of $p$ to the end.
            Then, while the remaining $u$ and $d$ steps contain $ud$, we replace it with $ff$ and move the two new $f$ steps to the end too.
            Each replacement removes one $u$ and one $d$ step, so this process ends with $d^a u^b f^{N - a - b}$ for some $a$ and $b$.
            Finally, we move all $f$ steps back between $d^a$ and $u^b$, which gives $d^a f^{N - a - b} u^b$.
            This path has the minimum height $-a$ and the final height $b - a$.
            Since, by Claim~\ref{clm:local-moves}, local moves do not change the minimum and final heights, the original path $p$ must have the same minimum and final heights, so $a = -h_{\min}(p)$ and $b = h_N(p) - h_{\min}(p)$.
        \end{proof}

        \begin{subclaim}
            \label{clm:bulk}
            Connected basis states have equal amplitudes in $\ket{\psi}$.
        \end{subclaim}
        \begin{proof}
            Consider the term $\proj{U}$, applied at some sites $(i, i + 1)$.
            Each basis state $\ket{p}$ can be written as $\ket{p} = \ket{xyz}$, where $y$ are the 2 steps taken in sites $(i, i + 1)$, and $x$ and $z$ are the steps taken before and after $y$, respectively.
            Then
            \begin{equation}
                \label{eq:xyz}
                \ket{\psi} = \sum_{x, y, z} c_{xyz} \ket{xyz}.
            \end{equation}
            Since $\proj{U}$ annihilates $\ket{\psi}$,
            \begin{equation}
                \braket{U | \psi} = \frac{1}{\sqrt{2}} \sum_{x, z} \bigl(c_{x f u z} - c_{x u f z}\bigr) \ket{x z} = 0.
                \label{eq:U_overlap}
            \end{equation}
            This is only possible if $c_{x f u z} = c_{x u f z}$ for all $x$ and $z$.
            Similarly, from the other bulk terms $\proj{D}$ and $\proj{F}$, we get $c_{x f d z} = c_{x d f z}$ and $c_{x f f z} = c_{x u d z}$.
            Hence, any two basis states connected by a single move in sites $(i, i + 1)$ have equal amplitudes.
            Since this holds for every $i$, the same is true for any two basis states connected by a sequence of moves as well.
        \end{proof}

        \begin{subclaim}
            \label{clm:left-boundary}
            $\ket{\psi}$ must have the form
            \begin{equation}
                \ket{\psi} = \sum_{h = 0}^{N} A_h \sum_{p \in M_{N, h}} \ket{p},
                \label{eq:prefix_sectors}
            \end{equation}
            for some amplitudes $A_h$.
        \end{subclaim}
        \begin{proof}
            Since $\proj{d}_1$ annihilates $\ket{\psi}$, $c_{d z} = 0$ for all $z$, i.e. every basis state starting with $d$ has zero amplitude.
            By Claim~\ref{clm:connectivity}, for every path $p$ with $h_{\min}(p) < 0$, $\ket{p}$ is connected to a basis state starting with $d$, so by Claim~\ref{clm:bulk}, $c_p = 0$ too.
            The remaining paths, with $h_{\min}(p) = 0$, are the Motzkin prefixes.
            By Claim~\ref{clm:connectivity}, all Motzkin prefixes with final height $h$ are connected to $\ket{f^{N - h} u^h}$, so by Claim~\ref{clm:bulk}, they have equal amplitudes $A_h$, which gives Eq.~\eqref{eq:prefix_sectors}.
        \end{proof}

        \begin{subclaim}
            \label{clm:right-boundary}
            $A_h = 0$ for all $h > 0$.
        \end{subclaim}
        \begin{proof}
            Since $\proj{u}_N$ annihilates $\ket{\psi}$, $c_{x u} = 0$ for all $x$, i.e. every basis state ending with $u$ has zero amplitude.
            For $h > 0$, $f^{N - h} u^h \in M_{N, h}$ ends with $u$, so $A_h = c_{f^{N - h} u^h} = 0$.
        \end{proof}

        By Claims~\ref{clm:left-boundary} and~\ref{clm:right-boundary}, $\ket{\psi} = A_0 \sum_{p \in M_N} \ket{p}$, so $\ket{\psi}$ is a multiple of $\ket{M_N}$.
    \end{proof}

    \begin{claim}
        \label{clm:motzkin-annihilated}
        $\ket{M_N}$ is an eigenstate of $H_M$ with eigenvalue $0$.
    \end{claim}
    \begin{proof}
        $H_M \ket{M_N}$ is the sum of all bulk and boundary terms applied to $\ket{M_N}$, so it suffices to show that each of them annihilates $\ket{M_N}$ separately.
        Since no Motzkin path starts with $d$ or ends with $u$, the boundary terms annihilate $\ket{M_N}$.

        Consider the application of the term $\proj{U}$ to some sites $(i, i + 1)$ of $\ket{M_N}$.
        $\ket{M_N}$ can still be written in the form of Eq.~\eqref{eq:xyz}, and the first equation of Eq.~\eqref{eq:U_overlap} still holds.
        However, since $\ket{M_N}$ is a uniform superposition of all Motzkin paths, the coefficients $c_{xyz}$ can only take 2 values: either some constant $c$, the same for all $xyz$ that are Motzkin paths, or 0 for all $xyz$ that are not Motzkin paths.
        $fu \leftrightarrow uf$ is a local move.
        Therefore, by Claim~\ref{clm:local-moves}, $xfuz$ and $xufz$ are either both Motzkin paths, or both non-Motzkin paths.
        In both cases, $c_{xfuz} - c_{xufz} = 0$.
        Therefore, $\proj{U}$ annihilates $\ket{M_N}$ for all $i$.

        Similarly, $\braket{D | M_N} = \braket{F | M_N} = 0$, so the other bulk terms annihilate $\ket{M_N}$ too.
        Therefore $H_M$ annihilates $\ket{M_N}$.
    \end{proof}

    By Claims~\ref{clm:nonnegative} and~\ref{clm:motzkin-annihilated}, the lowest eigenvalue of $H_M$ is $0$, and $\ket{M_N}$ is a ground state of $H_M$.
    By Claim~\ref{clm:motzkin-only}, every ground state of $H_M$ is a multiple of $\ket{M_N}$, but only $\ket{M_N}$ (up to a global phase) has the correct normalization.
    Therefore $\ket{M_N}$ is the unique ground state of $H_M$.
\end{proof}

\subsection{Parent Hamiltonian for the Symmetric Motzkin State}
\label{app:symmetric-motzkin-hamiltonian}

We now extend the Motzkin Hamiltonian of Sec.~\ref{app:motzkin-hamiltonian} to the symmetric Motzkin state.
Consider a chain of an even number $N$ of sites, and let $n = N / 2$.
For each $i = 1, \dots, n$, we pair site $i$ with its symmetric site $i' = N + 1 - i$ and define the three-dimensional symmetric-pair subspace spanned by
\begin{equation}
    \ket{\tu}_i = \ket{u_i d_{i'}}, \quad
    \ket{\tf}_i = \ket{f_i f_{i'}}, \quad
    \ket{\td}_i = \ket{d_i u_{i'}},
    \label{eq:symmetric_basis}
\end{equation}
where the subscript $i$ denotes the pair $(i, i')$.
The symmetric Motzkin state is
\begin{equation}
    \ket{\Msym} = \frac{1}{\sqrt{|M_{n, *}|}} \sum_{p \in M_{n, *}} \ket{\tilde{p}},
    \label{eq:Psi_sym}
\end{equation}
where $M_{n, *} = \bigcup_h M_{n, h}$ is the set of all Motzkin prefixes of length $n$, and $\ket{\tilde{p}} = \ket{\tilde{p}_1}_1 \cdots \ket{\tilde{p}_n}_n$.
Each $\tilde{p}_i$ is $\tu$, $\tf$ or $\td$ if the $i$-th step of $p$ is $u$, $f$ or $d$, respectively.
The basis states $\ket{\tilde{p}}$ correspond to the mirror-symmetric Motzkin paths of length $N$, i.e. those whose step at site $i'$ mirrors the step at site $i$, which form the set $\Msym$.

\begin{theorem}
    \label{thm:symmetric}
    $\ket{\Msym}$ is the unique ground state of its parent Hamiltonian
    \begin{equation}
        \Hsym = \sum_{i = 1}^{n} Q_i + \sum_{i = 1}^{n - 1} \tilde{\Pi}_i + \proj{\td}_1 + \proj{R}_n,
        \label{eq:H_sym}
    \end{equation}
    where $\tilde{\Pi}_i$ is $\Pi_i$ of Eq.~\eqref{eq:Pi} with $u$, $f$ and $d$ replaced by $\tu$, $\tf$ and $\td$, acting on the pairs $i$ and $i + 1$, and
    \begin{subequations}\label{eq:H_sym_terms}
        \begin{align}
            Q_i &= I - (\proj{\tu} + \proj{\tf} + \proj{\td})_i,
            \label{eq:Qi}\\
            \ket{R} &= \frac{1}{\sqrt{2}} \bigl(\ket{\tf} - \ket{\tu}\bigr).
            \label{eq:R}
        \end{align}
    \end{subequations}
\end{theorem}
\begin{proof}
    $\Hsym$ is again a sum of projectors.
    Therefore, Claim~\ref{clm:nonnegative} holds for $\Hsym$ with the same proof.

    \begin{claim}
        \label{clm:symmetric-only}
        If an eigenstate of $\Hsym$ with eigenvalue $0$ exists, then it must be a multiple of $\ket{\Msym}$.
    \end{claim}
    \begin{proof}
        Let $\ket{\psi}$ be an arbitrary eigenstate of $\Hsym$ with eigenvalue $0$, so
        \begin{equation}
            \Hsym \ket{\psi} = 0.
        \end{equation}
        Each term in $\Hsym$ is positive semidefinite.
        Therefore, this is only possible if every term annihilates $\ket{\psi}$ separately.

        \begin{subclaim}
            \label{clm:symmetric-subspace}
            $\ket{\psi}$ lies in the symmetric-pair subspace.
        \end{subclaim}
        \begin{proof}
            $Q_i$ is the projector onto the orthogonal complement of the symmetric-pair subspace of the pair $i$, and all $Q_i$ annihilate $\ket{\psi}$.
            Therefore, no component of $\ket{\psi}$ lies outside of the symmetric-pair subspace.
        \end{proof}

        \begin{subclaim}
            \label{clm:symmetric-left}
            $\ket{\psi}$ must have the form
            \begin{equation}
                \ket{\psi} = \sum_{h = 0}^{n} A_h \sum_{p \in M_{n, h}} \ket{\tilde{p}},
                \label{eq:symmetric_prefix_sectors}
            \end{equation}
            for some amplitudes $A_h$.
        \end{subclaim}
        \begin{proof}
            By Claim~\ref{clm:symmetric-subspace}, $\ket{\psi}$ can be written as
            \begin{equation}
                \ket{\psi} = \sum_{p \in \{u, f, d\}^n} c_p \ket{\tilde{p}}.
                \label{eq:symmetric_expansion}
            \end{equation}
            From here, the proof follows the exact same steps as in Claims~\ref{clm:local-moves}--\ref{clm:left-boundary} (starting from the definition of local moves), except that individual path steps $p_i \in \{u, f, d\}$ are replaced with their symmetric 2-site counterparts $\tilde{p}_i \in \{\tilde{u}, \tilde{f}, \tilde{d}\}$, and $N$ with $n$.
            At the end of it, one can arrive at the same result as in Claim~\ref{clm:left-boundary}, except with $\tilde{p}$ instead of $p$.
        \end{proof}

        \begin{subclaim}
            \label{clm:symmetric-right}
            $A_0 = A_1 = \dots = A_n$.
        \end{subclaim}
        \begin{proof}
            Each basis state $\ket{\tilde{p}}$ in Eq.~\eqref{eq:symmetric_expansion} can be written as $\ket{\tilde{p}} = \ket{\tilde{x} \tilde{y}}$, where $y$ is the step taken in the last pair $n$, and $x$ are the steps taken before $y$.
            Then
            \begin{equation}
                \ket{\psi} = \sum_{x, y} c_{xy} \ket{\tilde{x} \tilde{y}}.
                \label{eq:xy}
            \end{equation}
            Since $\proj{R}_n$ annihilates $\ket{\psi}$, by Eq.~\eqref{eq:R},
            \begin{equation}
                {}_n\!\braket{R | \psi} = \frac{1}{\sqrt{2}} \sum_x \bigl(c_{x f} - c_{x u}\bigr) \ket{\tilde{x}} = 0.
                \label{eq:right_overlap}
            \end{equation}
            This is only possible if $c_{x f} = c_{x u}$ for all $x$, i.e. replacing the last step $f$ with $u$ does not change the amplitude.
            For $h < n$, $f^{n - 1 - h} u^h f \in M_{n, h}$ ends with $f$, and replacing it with $u$ gives $f^{n - 1 - h} u^{h + 1} \in M_{n, h + 1}$, so $A_h = c_{f^{n - 1 - h} u^h f} = c_{f^{n - 1 - h} u^{h + 1}} = A_{h + 1}$.
        \end{proof}

        By Claims~\ref{clm:symmetric-left} and~\ref{clm:symmetric-right}, $\ket{\psi} = A_0 \sum_{p \in M_{n, *}} \ket{\tilde{p}}$, so $\ket{\psi}$ is a multiple of $\ket{\Msym}$.
    \end{proof}

    \begin{claim}
        \label{clm:symmetric-annihilated}
        $\ket{\Msym}$ is an eigenstate of $\Hsym$ with eigenvalue $0$.
    \end{claim}
    \begin{proof}
        $\Hsym \ket{\Msym}$ is the sum of all terms of $\Hsym$ applied to $\ket{\Msym}$, so it suffices to show that each of them annihilates $\ket{\Msym}$ separately.
        $\ket{\Msym}$ lies in the symmetric-pair subspace, so all $Q_i$ annihilate it.
        Since no Motzkin prefix starts with $d$, the left boundary term annihilates $\ket{\Msym}$.

        The bulk terms in $\tilde{\Pi}_i$ annihilate $\ket{\Msym}$ by the same steps as in the proof of Claim~\ref{clm:motzkin-annihilated}, with $p_i$ replaced with $\tilde{p}_i$, $N$ with $n$, and Motzkin paths with Motzkin prefixes, which are also closed under local moves by Claim~\ref{clm:local-moves}.
        Similarly, $\proj{R}_n$ annihilates $\ket{\Msym}$ by the same steps with Eqs.~\eqref{eq:xy} and~\eqref{eq:right_overlap} in place of Eqs.~\eqref{eq:xyz} and~\eqref{eq:U_overlap}, since appending $f$ or $u$ to $x$ does not change its minimum height, so $x f$ and $x u$ are either both Motzkin prefixes, or both non-Motzkin prefixes.
        Therefore $\Hsym$ annihilates $\ket{\Msym}$.
    \end{proof}

    By Claims~\ref{clm:nonnegative} and~\ref{clm:symmetric-annihilated}, the lowest eigenvalue of $\Hsym$ is $0$, and $\ket{\Msym}$ is a ground state of $\Hsym$.
    By Claim~\ref{clm:symmetric-only}, every ground state of $\Hsym$ is a multiple of $\ket{\Msym}$, but only $\ket{\Msym}$ (up to a global phase) has the correct normalization.
    Therefore $\ket{\Msym}$ is the unique ground state of $\Hsym$.
\end{proof}

Unlike $H_M$, $\Hsym$ is no longer local in 1D, but it can be local in 2D (on a chain folded around its midpoint).

\subsection{Parent Hamiltonians for the Symmetric-Negative States}
\label{app:symmetric-negative-hamiltonian}

We now find the parent Hamiltonians of the symmetric-negative states, with $N$, $n$ and the symmetric pairs as in Sec.~\ref{app:symmetric-motzkin-hamiltonian}.
The symmetric-negative Motzkin state is the uniform superposition of all mirror-symmetric paths of length $N$, including those that go below height $0$.
These paths form the set $\Mneg$, and their basis states are $\ket{\tilde{p}}$ for all sequences of steps $p \in \{u, f, d\}^n$.
Since $\ket{\tilde{p}} = \ket{\tilde{p}_1}_1 \cdots \ket{\tilde{p}_n}_n$, the state is a product of identical pair states
\begin{equation}
    \ket{\Mneg} = \frac{1}{\sqrt{3^n}} \sum_{p \in \{u, f, d\}^n} \ket{\tilde{p}} = \ket{\Phi}_1 \cdots \ket{\Phi}_n,
    \label{eq:Psi_neg}
\end{equation}
where
\begin{equation}
    \ket{\Phi} = \frac{1}{\sqrt{3}} \bigl(\ket{\tu} + \ket{\tf} + \ket{\td}\bigr).
    \label{eq:Phi}
\end{equation}

\begin{theorem}
    \label{thm:negative}
    $\ket{\Mneg}$ is the unique ground state of its parent Hamiltonian
    \begin{equation}
        \Hneg = \sum_{i = 1}^{n} \bigl(I - \proj{\Phi}\bigr)_i.
        \label{eq:H_neg}
    \end{equation}
\end{theorem}
\begin{proof}
    $\Hneg$ is again a sum of projectors.
    Therefore, Claim~\ref{clm:nonnegative} holds for $\Hneg$ with the same proof.
    Each term $(I - \proj{\Phi})_i$ annihilates $\ket{\Phi}_i$, so $\Hneg$ annihilates $\ket{\Mneg}$, and the lowest eigenvalue of $\Hneg$ is $0$.

    Let $\ket{\psi}$ be an arbitrary eigenstate of $\Hneg$ with eigenvalue $0$.
    As in the proof of Claim~\ref{clm:motzkin-only}, every term of $\Hneg$ annihilates $\ket{\psi}$ separately, so $\proj{\Phi}_i \ket{\psi} = \ket{\psi}$ for all $i$.
    Applying this for $i = 1, \dots, n$ gives $\ket{\psi} = \proj{\Phi}_1 \cdots \proj{\Phi}_n \ket{\psi} = \braket{\Mneg | \psi} \ket{\Mneg}$, so $\ket{\psi}$ is a multiple of $\ket{\Mneg}$, but only $\ket{\Mneg}$ (up to a global phase) has the correct normalization.
    Therefore $\ket{\Mneg}$ is the unique ground state of $\Hneg$.
\end{proof}

Theorem~\ref{thm:negative} and its proof also hold for the symmetric-negative Fredkin state, i.e. the uniform superposition of all mirror-symmetric paths of length $N$ without flat steps, with $\{u, f, d\}$ replaced with $\{u, d\}$, $3^n$ with $2^n$, and $\ket{\Phi}$ with $\frac{1}{\sqrt{2}} \bigl(\ket{\tu} + \ket{\td}\bigr)$.
Like $\Hsym$, $\Hneg$ is local on the folded chain.

\clearpage
\bibliography{refs}

\end{document}